 %This is the correct arxiv file as of 1/27/15. Since 3/22/09, this file was corrected to change operator to operations where appropriate and other stuff. There may still be typos. Corrected and added to as of 12/30/12. Two slight typos corrected 1/6/13 in added material. Does not yet contain in appendix the 27 JAN 2013 ultra-logic-system paper of additions. 
\def\real{{\tt I\kern-.2em{R}}}
\def\nat{{\tt I\kern-.2em{N}}}

\def\realp#1{{\tt I\kern-.2em{R}}^#1} 
\def\natp#1{{\tt I\kern-.2em{N}}^#1}
\def\hyper#1{\ ^*\kern-.2em{#1}}
\def\monad#1{\mu (#1)}

\def\st#1{{\tt st}(#1)}
\def\hyperreal{{^*{\real}}}
\def\hyperrealp#1{{\tt ^*{I\kern-.2em{R}}}^#1} 
\def\hypernat{{^*{\nat }}}
\def\hypernatp#1{{{^*{{\tt I\kern-.2em{N}}}}}^#1} 
\def\eskip{\hskip.25em\relax}

\def\Hyper#1{\hyper {\eskip #1}}
\def\leaderfill{\leaders\hbox to 1em{\hss.\hss}\hfill}
\def\srealp#1{{\rm I\kern-.2em{R}}^#1}
\def\sp{\vert\vert\vert} 

\def\power#1{{{\cal P}(#1)}}
\def\iff{\leftrightarrow}
\def\qed{{\vrule height6pt width3pt depth2pt}\par\medskip}
\def\pars{\par\smallskip}
\def\parm{\par\medskip}
\def\r#1{{\rm #1}}
\def\b#1{{\bf #1}}
\def\ref#1{$^{#1}$}

\def\sig{{^\sigma}}
\def\m@th{\mathsurround=0pt}
\def\rightarrowfill{$\m@th \mathord- \mkern-6mu \cleaders\hbox{$\mkern-2mu 
\mathord- \mkern-2mu$}\hfil \mkern-6mu \mathord\rightarrow$}
\def\leftarrowfill{$\mathord\leftarrow
\mkern -6mu \m@th \mathord- \mkern-6mu \cleaders\hbox{$\mkern-2mu 
\mathord- \mkern-2mu$}\hfil $}
\def\noarrowfill{$\m@th \mathord- \mkern-6mu \cleaders\hbox{$\mkern-2mu 
\mathord- \mkern-2mu$}\hfil$}
\def\orgate{$\bigcirc \kern-.80em \lor$}
\def\andgate{$\bigcirc \kern-.80em \land$}
\def\inverter{$\bigcirc \kern-.80em \neg$}
%\magnification=\magstep0

\def\id{\par\hangindent2\parindent\textindent}
\def\textindent#1{\indent\llap{#1}}
\tolerance 10000
\baselineskip  14pt
\hoffset=.25in
\hsize 6.00 true in
\vsize 8.75 true in
%\font\eightrm=amr9
%\font\bfs=ambx9
\centerline{\bf Solutions to the ``\underbar{General} Grand Unification Problem$,$''}
\centerline{\bf and the Questions ``How Did Our Universe Come Into Being?''}
\centerline{{\bf and ``Of What is Empty Space  Composed?''}\footnote*{rah@usna.edu or drrahgid@hotmail.com}}
\medskip 
\centerline{\sl Robert A. Herrmann}\pars
\centerline{Mathematics Department}
\centerline{U. S. Naval Academy}
\centerline{572 Holloway Rd.}
\centerline{Annapolis$,$ MD 21402-5002}
\centerline{Originally presented before the Mathematics Association of America$,$}
\centerline{Western Maryland College$,$ 12 NOV 1994. Last revision 27 JAN 2015}
\bigskip\bigskip
{\leftskip 0.5in \rightskip 0.5in \noindent {\it Abstract}:
 Using mathematical techniques to model one of the most simplistic of 
human linguistic processes$,$ it is rationally predicted that within the 
nonstandard physical world there exists a force-like (logical) operator
$\Hyper {\b S}$ and an entity $w'$ such that $\Hyper {\b S}\{w'\}$ sequentially 
generates each of the Natural (i.e. physical) systems that comprise a Universe. This scientific model 
shows specifically that within the nonstandard physical world the behavior of 
each Natural world Natural-system is related logically. Further$,$ the model 
predicts the rational existence of a single type of entity within the 
nonstandard physical world's substratum that can be used to construct$,$ by 
means of an exceptionally simple process$,$ all
of the fundamental Natural (i.e. physial) world particles used within particle physics. In important section 11.2$,$ it is shown how (Natural (i.e. physical) law) allowable perturbations in Natural-system behavior are also included within this mathematical model. These 
results solve the pre-geometry problem of Wheeler. In general$,$ the model predicts that when the 
behavior of these 
Universe creating processes are viewed globally$,$ they can be interpreted as the behavior associated with a powerful intelligent agent. (An extensive refinement has been added to the original and further refinements are find in the (2014, 2013) references.)\par}\par 
\vskip 0.75in
\centerline{\bf CONTENTS}
\bigskip
\line{\indent 1. \ \ History\leaderfill 3}
\line{\indent 2. \ \ Two General Discussions\leaderfill 7}
\line{\indent 3. \ \ The Mathematics\leaderfill 15}
\line{\indent 3. \ \ Appendix\leaderfill 62}

\vfil
\eject
{\quad}
\vfil
\eject

%=========== FONTS =====================

\noindent {\bf 1. History.}
\medskip 
[Note: References for this section appear on pages 4 -- 5.] 
 In August  1979$,$ I was to give a paper before a gathering of 
mathematicians at the Summer Meeting of The American Mathematical Society
[1]. This summer meeting is a 
joint meeting with the Mathematical Association of America and was being
held at the University of Minnesota at Duluth. I boarded 
an airplane at Chicago for the last leg of my trip$,$ and who should be setting 
next to me at the window seat but John Wheeler$,$ the Joseph Henry Professor of 
Physics emeritus at Princeton.
Professor Wheeler was to deliver an invited talk before the Mathematical 
Association of America on what he believed was a fundamental difficulty with 
the philosophy of science associated with physical theories.\pars
 We discussed 
various things as we passed over some magnificent Minnesota thunderheads. 
Looking out of 
the window Professor Wheeler made the curious but simple statement$,$ ``We 
can't do that.'' What does this statement mean and who are the ``we''?\pars
After more discussion$,$ the meaning of his comment was clear. First$,$ the 
``we'' are physicists and what  they can't do was illustrated by the 
thunderhead. The thunderhead is assumed to be a collection of Natural (i.e. physical) systems and, as a collection, can also display emerging properties. 
You have the water droplets$,$ or crystals. Then their paths of motion caused by 
internal forces. The various electric potential differences produced by such 
motion and hundreds of other factors that scientists claim contribute to the 
overall behavior we observed from our window. \pars
A Natural-system is a set or arrangement of physical entities that are so 
related or connected as to form an identifiable whole. 
Many Natural-systems have an associated physical theory. These theories 
are used to predict behavior for each Natural-system. BUT many of these 
theories use different methods to predict behavior. Indeed$,$ one often has to
restrict a particular theory to specific areas of application. When this is 
done you often have a boundary between two theories -- an overlapping region 
-- where one or the other but not both apply 
for the methods used in one of the theories 
are 
inconsistent with the methods used in the other theory. The hundreds of 
Natural-systems (now called subnatural-systems) that comprise the Natural 
system called a thunderhead are then put together in some 
manner to obtain the entire thunderhead one observes. With respect to the 
universe as a whole$,$ the differences between various predicting theories are  
profound. What Professor Wheeler meant by the  ``. . . can't do that'' 
with respect 
to our universe is the acknowledgment that physicists had not found a 
general unifying theory that predicts all of the behavior of all subnatural 
systems that comprise the entire Natural-system called our  universe. 
Science needed a theory that would be totally consistent$,$ and a theory that would replace the piecemeal approach. Of course$,$ such
a unifying theory did not even exist for the behavior of the thunderhead 
we observed. \pars
I restated this problem in the following manner. \pars
{\leftskip=0.25in \rightskip=0.25in \noindent {\bf Does Nature really combine      
subnatural-systems together$,$ in a way that we can understand$,$ to produce an 
entire Natural-system or does it 
use an entirely different method so that inconsistencies are somehow 
avoided? Is there something else required$,$ something more basic than science 
has yet described and
that's needed to combine all the Natural-systems together that comprise 
our universe? Indeed$,$ how can a universe that's perceived by humans to 
have order and harmony really be a product of chaos?}\par}\pars 
\noindent By the way$,$ I added that last question to this problem 
since it is certainly relevant.\pars
After our discussion$,$ it occurred to me that something discovered 
in October 1978$,$ and now called ultralogics [2]$,$   
might possibly lead to an answer to these questions. But I also knew that the 
answer might be rather startling. \pars
  The standard approach is a piecemeal  ``bottom-up''
approach beginning with the  ``bottom$,$'' so to speak$,$ of the scientific 
hierarchy (or ``domain of explanation''). Supposedly$,$ if you can find a unification of the fundamental 
forces (or interactions)$,$ then this would lead to an ``upward'' process that would eventually 
unify all physical theories. As discussed above$,$ this is a doubtful assumption. 
Moreover$,$ many scientists and philosophers of science believe that there 
are emergent properties of 
organisms that cannot be fully understood as products of DNA or chemistry and$,$ 
thus$,$ do not follow directly from the four fundamental forces.  
\pars
On the other hand$,$ there may be a {\it top-down} approach that answers the 
Wheeler questions. This means to find some scientific theory that gives 
processes which yield cosmologies$,$ cosmologies that contain all of the 
processes that control the behavior of all Natural-systems. \pars

The  ``deductive-world model$,$'' (i.e. the D-world model) was the first 
constructed (1978) and is interpreted in a linguistic sense. D-world model  
properties will not be discussed directly within the following pages. 
Many of these properties can be found in sections 1 -- 5 of the book 
{\it Ultralogics and More} [2]. 
However$,$ some of what appears in the next sections does have the 
requisite linguistic 
interpretation associated with the D-world model. The actual model 
used to solve these problems is generally termed the {\it General Grand Unification Model (GGU-model).} (For a special purpose, a portion of the GGU-model, the 
{\it metamorphic-anamorphosis model} (i.e. the MA-model) is employed.) 
Further$,$ the intuitive 
concepts and mathematical methods that determine the GGU-model 
properties are 
very similar to those used for the additional D-world model conclusions. Thus 
becoming familiar with the GGU-model conclusions will aid in ones comprehension 
of the D-world model.\pars
The basic construction of the GGU-model uses the following philosophical stance 
as described by deBroglie.\parm 
{\leftskip 0.25in \rightskip 0.25in \noindent {\it . . . the structure of the 
material universe has something in common with the laws that govern the 
workings of the human mind} [3$,$ p. 143] \par}\parm
\noindent It was determined that an appropriate 
cosmogony would depend upon the properties of a logic-like operator and 
subsidiary concepts associated with the {\it nonstandard physical world}
 (i.e. NSP-world) [4$,$ part III].
 The construction of the GGU-model began in August 1979 and a series of 
announcements [5 -- 10] relative to its mathematical construction appeared in 
the Abstracts of papers presented to the  American Mathematical Society. In 
particular$,$ [5] mentions the logic-like operators.\par

In 1982$,$ papers by Bastin [11]$,$ and Wheeler and Patton [12] were discovered in 
{\it The Encyclopaedia of Ignorance.} I mention that [12] does not contain an 
important appendix that appears in the original 1975 paper. The paper by 
Bastin describes the {\it discreteness paradox} and the paper [13]$,$ using a few 
GGU-model procedures$,$ presents a solution to this paradox. Next the 
requirements for an acceptable cosmogony stated in [12] were compared with 
GGU-model properties. A major difficulty in showing that the GGU-model meets
all of these requirements is in the language used to give physical-like meaning  
to the abstract entities. After considerable reflection$,$ the prefix  ``ultra'' 
was decided upon. The use of this prefix is consistent with its use in the 
construction of the nonstandard structure. A relatively explicit 
construction of 
such a structure uses the concepts of the ultrafilter$,$ the ultraproduct and 
the ultralimit. In 1987$,$ a paper justifying the fundamental  
mathematical theory of 
nonstandard consequence operators was published [14]. \parm
Although a few technical aspects associated with the GGU-model were yet to be 
fully justified$,$ two papers were written in 1986 [15]$,$ [16] announcing the 
solution to the general grand unification problem and the pregeometry problem of 
Wheeler and Patton. To properly prepare paper [16]$,$ the original Wheeler and 
Patton paper [17] was utilized. An appendix in [17] describes how 
Wheeler and his colleagues at Princeton tried to construct a pregeometry from 
the statistics of very long propositions and very many propositions$,$ where the 
term  ``proposition'' refers to the propositional (i.e. the sentence) logical
calculus. They failed to achieve a solution$,$ but Wheeler left open the 
possibility that a solution could be obtained using concepts from the area of 
Mathematical Logic. This is exactly what has occurred. \parm
Beginning in about 1989$,$ a project was instituted to justify fully all of the 
GGU-model concepts$,$ among other aspects of the NSP-world$,$ and to present them 
in monograph form. 
The basic monograph 
[2] was completed in  1991 and the final result that completes 
the justification process 
was published in 1993 [18]. \parm  
In the next section is given a very brief$,$ general and mostly nonmathematical 
description of how the 
GGU-model and the language of the NSP-world solve the problems mentioned in the 
title.   
Presented in the third and technical section titled ``The Mathematics$,$'' with  
additional comments$,$ is  most of the actual mathematics that$,$ when interpreted$,$ 
describes 
the properties of the GGU-model [2] that correspond to the nonmathematical 
descriptions. \parm 
The following quotations are taken from [17] and are all relative to the 
Patton and Wheeler cosmogony requirements.\par
{(1) {\it Five bits of evidence argue 
that geometry is as far from giving an understanding of space as elasticity is 
from giving a understanding of a solid.} [17$,$ p. 539]\par}\parm
{(2) {\it They also suggest that 
the basic structure is something deeper than geometry$,$ that underlies both 
geometry and particles (``pregeometry'').} [17$,$ p. 539] \par}\parm
{(3) {\it For someday revealing 
this structure no perspective seems more promising than the view that it must 
provide the universe with a way to come into being.} [17$,$ p. 539] \par}\parm
{(4) {\it It brings us into closer 
confrontation than ever with the greatest questions on the book of physics: 
How did our universe come into being? And of what is it made?} [17$,$ p. 540] \par}\parm
{(5) {\it Tied to the paradox of 
the big bang and collapse is the question$,$ what is the \underbar{substance} out of which 
the universe is made?} [17$,$ p. 543] \par\bigskip}\par
{(6) {\it But is it really 
imaginable that this deeper structure of physics should govern how the 
universe came into being? Is it not more reasonable to believe the converse$,$ 
that the requirement that the universe should come into being governs the 
structure of physics? } [17$,$ p. 558]\par}
\indent {(7) {\it It is difficult to avoid 
the impression that every law of physics is  ``mutable'' under conditions 
sufficiently extreme$,$ . . . .} [17$,$ p. 568] \par}\parm
{(8) {\it It is difficult to 
believe that we can uncover this pregeometry except as we come to understand 
at the same time the necessity of the quantum principle$,$ with its  
``observer-participator$,$'' in the construction of the world.} [17$,$ p. 575\par}\parm
{(9) {\it . . . . a guiding 
principle$,$ is what we seek.} [17$,$ p. 575] \par}\parm
As the GGU-model pregeometry is discussed$,$ I will refer to these quotations 
where applicable. Relative to (9)$,$ the first half of the guiding principle 
is the deBroglie statement viewed from the NSP-world. The second half 
of the guiding principle is relative to the philosophy of realism and an
observation made by the originator of nonstandard analysis Abraham Robinson.
First recall that Newton believed that infinitesimal measures were real 
measures associated with objectively real entities. Berkeley and Leibniz did 
not accept Newton's belief.   
Robinson   
at the end of his very first 
published paper on nonstandard analysis made the 
following statement relative to the modern concepts of how mathematical models 
are used to predict indirectly Natural-system behavior.\par
{\leftskip 0.25in \rightskip 0.25in \noindent {\it For phenomena on a different 
scale$,$ such as are considered in Modern Physics$,$ the dimensions of a 
particular body or process may not be observable directly. Accordingly$,$ the 
question whether or not the scale of non-standard analysis is appropriate to 
the physical world really amounts to asking whether or not such a system 
provides a better explanation of certain observable phenomena than the 
standard system. . . . The possibility that this is the case should be borne 
in mind.}\parm
\rightline{Fine Hall$,$ \hskip 1.35in}
\rightline{Princeton University [19$,$ p. 440]}\par}\parm     
Robinson is referring to infinitesimal measures in this quotation. Since  
the publication of [19]$,$ nonstandard analysis has been applied to entities 
that are not 
infinitesimal in character. The second half of the guiding principle 
is an extension of this Robinson 
statement to the NSP-world. Thus the acceptance of the NSP-world 
as a viable realism  depends upon whether or not it provides a ``better'' 
rational  
explanation of certain observable phenomena than the standard world model.  
The philosophy of what constitutes  ``a better explanation'' is left to 
individual choice.  
\bigskip\bigskip
\centerline{\bf References}
\medskip
\id [1] Herrmann$,$ Robert. A. (1979). Perfect maps on convergence spaces I$,$ 
{\it Notices of the AMS$,$} (26)(5): A-475.\par
\id [2] Herrmann$,$ Robert. A. (1979-93). The Theory of Ultralogics$,$  See\hfil\break http://raherrmann.com/books.htm or\hfil\break http://arxiv.org/abs/math/9903081 and http://arxiv.org/abs/math/9903082 
\id [3] deBroglie$,$ L. (1963). In March$,$ Arthur and Ira M. Freeman$,$ The 
New World of Physics$,$ Vintage Books$,$ New York. 
\id [4] Herrmann$,$ Robert A. (1989). Fractals and ultrasmooth microeffects$,$ 
{\it J. Math. Physics$,$} 30(4) (April):805--808.
\id [5] Herrmann$,$ Robert A. (1981). Mathematical philosophy$,$ {\it
Abstracts of papers presented to the AMS$,$} Vol. 2(6). \#81T-03-529:527.
\id [6] Herrmann$,$ Robert A. (1983). A useful *-real valued function$,$ {\it 
Abstracts of papers presented to the AMS$,$} Vol. 4 (4)$,$ \#83T-26-280:318.
\id [7] Herrmann$,$ Robert A. (1984). Nonstandard consequence operators I$,$ 
{\it 
Abstracts of papers presented to the AMS$,$} Vol. 5 (1)$,$ \#84T-03-61:129.
\id [8] Herrmann$,$ Robert A. (1984). Nonstandard consequence operators II$,$ 
{\it Abstracts of papers presented to the AMS$,$} Vol. 5 (2)$,$ \#84T-03-93:195.
\id [9] Herrmann$,$ Robert A. (1984). D-world alphabets I$,$ {\it Abstracts of papers presented to the AMS$,$}
Vol. 5 (4)$,$ \#84T-03-320:269.
\id [10] Herrmann$,$ Robert A. (1984). D-world alphabets II$,$ 
{\it Abstracts of papers presented to the AMS$,$} Vol. 5 (5)$,$ \#84T-03-374:328. 
\id [11] Bastin$,$ T.$,$ (1977). A clash of paradigms$,$ in {\it The Encyclopaedia 
of Ignorance$,$} Duncan and Westen--Smith$,$ eds. Pergamon$,$ New York. 
\id [12] Wheeler$,$ J. A. and C. M. Patton. (1977). Is physics legislated by 
cosmogony? in {\it The Encyclopaedia 
of Ignorance$,$} Duncan and Westen--Smith$,$ eds. Pergamon$,$ New York.
\id [13] Herrmann$,$ Robert A. (1983). Mathematical philosophical and 
developmental processes$,$ {\it Nature and System$,$} 5(1/2):17--36. 
\id [14] Herrmann$,$ Robert A. (1987). Nonstandard consequence operators$,$ {\it Kobe 
J. Math.} 4(1):1--14.
\id [15] Herrmann$,$ Robert A. (1986). A solution to the grand unification 
problem$,$ {\it Abstracts of papers presented to the AMS$,$} Vol. 7 (2)$,$ 
\#86T-85-41:238.
\id [16] Herrmann$,$ Robert A. (1988) Physics is legislated by a cosmogony$,$ 
{\it Speculations in Science and Technology$,$} 11(1):17--24. 
\id [17] Patton$,$ C. M and J. A. Wheeler$,$ (1975). Is physics legislated 
by cosmogony? in {\it Quantum Gravity: an Oxford Symposium$,$} Ishan$,$ C.$,$ 
Penrose$,$ R.$,$ and D. Sciama$,$ eds$,$ Clarendon$,$ Oxford. 
\id [18] Herrmann$,$ Robert A. (1993). A special isomorphism between 
superstructures$,$ {\it Kobe J. Math.} 10(2):125-129. 
\id [19] Robinson$,$ A.$,$ (1961). Non-standard analysis$,$ {\it Nederl. Akad. 
Weimsch Proc.} Ser. A64 and {\it Indiag. Math.} 23:432 -- 440.  
\medskip
\centerline{\bf Important Reference}\pars
The above reference [2]$,$ from which the mathematical portions of the monograph 
are taken$,$  
contains {\bf all} of the fundamental concepts 
associated 
with the methods and processes used within the discipline of nonstandard 
analysis as they would apply to all aspects of the GGU-model and D-world 
model.\pars
\bigskip 
\leftline{\bf 2. Two General Discussions}
\medskip
[Note: This first discussion is a portion of a general audience non-technical 
and elementary talk I
gave on this subject. The numbers that appear in the double square brackets refer 
to the quotations from [17] as they appear on pages 5.]  
Let me point out that as long as scientists use 
mathematics to obtain their theories and written symbols$,$ diagrams$,$ 
photographs and the like to communicate their concepts then the GGU-model 
can't be eliminated. It will always be there lurking in the background.  
 \pars
Now to answer the question  ``How was our universe created?'' 
Let's start by considering a single geometric point a few feet in front of you. 
A geometric 
point in this sense is a {\it position} in our universe and$,$ for the present$,$ 
has 
no other meaning. Now$,$ I'm able to magnify this point for you by using a 
mathematical 
microscope with a power that's greater 
than any power that can ever be 
achieved by human means. \pars
Suddenly$,$ you see the point open up$,$ like the 
iris of your 
eyes. What's revealed to you is a {\it background 
universe}$,$ a {\it substratum}$,$ or whatever you might like to call it. 
Now$,$ the Natural-world is the world we can scientifically perceive$,$  
and this 
Natural-world 
point is still in your field
of view with a small portion of the background 
universe surrounding it. You can't make out much detail$,$ but there's 
definitely something there. The detail you see is sharper and clearer
 near 
to the single Natural-world point. Then clarity slowly fads as you proceed 
further 
from that one solitary Natural-world position within our universe. You can 
find no clear outer edge within 
your field of view.  This
background
universe forms a portion of what I called the nonstandard physical world - 
the NSP-world ([4] above). [Note: The entire collection of all possible standard world 
points coupled with all of the NSP-world points that are infinitesimally near 
to them is 
called the {\it finite} or {\it bounded} portion of the NSP-world.]
The term the NSP-world  
is also used for other applications. However$,$ for our purposes
I'll discuss a portion of this NSP-world$,$ the GGU-model.\pars

Our universe is inside and ``just as near to'' the NSP-world as I have 
described 
it. One might say that this background universe is scientifically omnipresent. 
Now the  GGU-model portion of this background universe specifically states that 
there will {\it never} be a human language 
that can give a completely detailed description for 
the mechanisms that may have produced our universe$,$ but there do exist such 
mechanisms within this background universe. Such mechanisms exist but$,$ no 
matter how hard we try$,$ the human mind can't comprehend {\it all} of the 
{\it details}. Well then$,$ 
what can be known about how universes$,$ such as ours$,$ can be created by
 NSP-world  
mechanisms? We can know general properties. \pars  
To begin with$,$ there exists within the NSP-world a set of  ``things.'' These
``things''
need not be considered as being within our Natural universe. But why do I call them 
``things''?  I do know a lot about these ``things.'' But$,$ unfortunately$,$ 
if I were to give you a basic description for their contents$,$ I would only use 
various mathematical terms from the fields of mathematical logic and 
set-theory.  The most difficult task I've faced is to give some  ``physical''
meaning to these  ``things.''  These objects could easily fall into the 
category 
of those objects within the  NSP-world that have NO human language 
physical descriptions 
at all. I know that these ``things'' behave like informational  
``superballs.'' \pars
After literally years of reflection$,$ it was determined that these  ``things''
can be described as  
containing all of the building plans$,$ the laws of ``Nature'' and even  
step-by-step images of how an ideal universe will appear as it develops. They also contain a great deal of information that is incomprehensible to the human mind. It's interesting to note that$,$ after I came to these 
conclusions$,$ I came across a 
quotation from 
one  of the greatest scientists of our time. Hermann Weyl is quoted as 
saying the following: \parm
 {\leftskip=0.25in \rightskip=0.25in \noindent{\it 
Is it conceivable that immaterial factors having the nature of images$,$ 
ideas$,$ `building plans' also intervene in the evolution of the world as a 
whole?}\par}\parm
We have  ``things'' in the background universe 
that are ready$,$ when the conditions are just right$,$ to {\it aid} in the 
production of  universes$,$ ours included. And different  ``things'' aid in the 
production of different universes. 
Actually$,$ these  ``things$,$'' which I 
have called {\it ultimate ultrawords}$,$ (or now {\it ultra-logic-systems}) don't work alone. Each is 
 just one 
piece of an entire process. All pieces of the ``puzzle'' must be put together 
before a universe is produced. Now the existence of these ultra-objects  
may not seem startling$,$ but a lot more is yet to come. \parm
What are the conditions that must exist in the NSP-world -- conditions needed 
to trigger the creation of a universe such as ours? At present$,$ there is no
human 
scientific language that can detail these conditions$,$ no human understanding 
of what these conditions are or were$,$ only that such conditions exist.  
\pars
I remind you that this cosmogony comes from a mathematical model$,$ a model that 
can't be eliminated from modern science. 
This cosmogony 
satisfies all of the basic theoretical requirements of the scientific method. 
 The universe in which we dwell$,$ our solar system$,$
the Earth$,$ or a virus are Natural-systems. (Now called physical-systems.) Natural-systems are studied 
by the physical scientist in a 
piecemeal fashion. They apply distinct 
procedures that   
seem to describe the moment-to-
moment behavior of each distinct Natural-system. 
Now within the NSP-world there's \underbar{one} special process called ``$\Hyper {\b S},$''
it's a hidden 
process$,$  
an {\it intrinsic} (hidden) {\it ultranatural process} (IUN-process). The $\Hyper {\b S}$-process is one 
of the entities know as an {\it ultralogic.} When the 
conditions are just right$,$ this force-like process takes one of these 
ultimate ultrawords and produces a universe. Indeed$,$ it 
combines together$,$ controls and coordinates all of the distinctly different  
Natural-systems that comprise a universe. The $\Hyper {\b S}$-process applied to an ultimate 
ultraword produces each Natural-system and also yields the moment-to-moment 
alterations in the behavior of each and every Natural-system 
$\lbrack\!\lbrack$2$,$ 3$,$ 4$,$ 6$\rbrack\!\rbrack$. 
But$,$ how does it do this? We 
can't know many details$,$ but a few simple mechanisms do present themselves. 
\pars
Within the NSP-world there are objects$,$ of a single type$,$ (that I term {\it ultimate subparticles} 
) that$,$ from the NSP-world viewpoint$,$ can be ``easily'' combined 
together to produce every material object$,$ electrons$,$ protons$,$ our earth$,$ 
and everything else that might be termed material as well as immaterial fields$,$ if 
such exist $\lbrack\!\lbrack$2$,$ 4$,$ 5$,$ 6$\rbrack\!\rbrack$. This combining process cannot be 
reproduced by Natural means within any laboratory
within our universe. (Note: 12/12/12 Subparticles are now called ``propertons'' and ultimate subparticles are called ultra-propertons.)\par\bigskip
What about the development of our universe? That is how it changes with 
respect to time. This force-like process does 
produce a  
``beginning'' for our universe and$,$ as mentioned$,$ 
in a remarkable step-by-step manner$,$  
it produces in the proper ``time'' ordered sequence  all of the material 
changes  
from the very  
beginning until the universe arrives at a stage such as that which 
we observe about us. It 
produces all the  Natural events that constitute the moment-to-moment 
changes that alter the appearance of a universe. 
This remarkable force-like process 
applied to an 
ultimate ultraword ((12/12/12) an ultra-logic-system) yields a solution to the general grand unification problem.  
\par\bigskip
 I say that this process is  ``remarkable'' 
but I haven't explained why. I give you one example. From human 
perception$,$ we often characterize certain Natural changes in a Natural-system 
as chaotic or random. This means that there seems to be no pattern for such 
changes -- that is no pattern that can be comprehended by the human mind. 
Indeed$,$ 
no human predictions can be 
made as to how individual objects will behave from one moment to another.  
From  our viewpoint$,$ 
there are  no harmonious or 
regular laws that can produce such individual changes. 
It can be shown that from the 
NSP-world viewpoint the opposite is true. This seemingly irregular behavior 
within our universe is actually only what {\bf we} can perceive of what is$,$ in 
reality$,$ an extremely regular process. How is this possible?\parm
Well$,$ it turns out that as a universe develops$,$ as it changes$,$ there are 
millions of other events taking place within the NSP-world that we can't 
perceive scientifically. These {\it ultranatural events} sustain and hold our 
universe together$,$ so to speak. It's the ultranatural events that are combined 
together with the Natural events we perceive that actually comprise  
a complete change. Thus it's simply a matter of perception. I wish there were 
words in any language that fully describe this wondrous  ``combining'' together 
process. There is a technical term$,$ ``ultrauniform$,$'' that can be used. But 
this gives no indication of what is actually occurring. The steps in this 
combining together process are so minuscule$,$ so small$,$ so refined that the 
human mind can't fully appreciate nor comprehend the $\Hyper {\b S}$-process. It doesn't 
correspond to anything that can at present be 
perceived or imagined by us.\parm 

Another startling aspect of the $\Hyper {\b S}$-process is that every 
Natural event is connected$,$ within the NSP-world$,$ with every other 
Natural event. No Natural changes$,$ from the NSP-world viewpoint$,$ are independent from 
one another. A Natural event taking place in one galaxy is related by the 
$\Hyper {\b S}$-process$,$ to events 
taking place in every other galaxy. But$,$ unfortunately$,$ we 
can't know$,$ except in general terms$,$ the actual composition of an ultimate 
ultraword nor describe in any human language the necessary ultranatural 
events that sustain this process. \par
Now$,$ what I've described$,$ as best as I can$,$ is the creation of an ``ideal'' 
universe. But what happens when there exist creatures within a universe 
that can alter its ideal development?  Well$,$ these creatures can  only alter 
the Natural events$,$ they can't alter the ultranatural events. No matter what 
these creatures do this ``glue'' that holds the universe together still 
remains $\lbrack\!\lbrack$8$\rbrack\!\rbrack$. I repeat$,$ we can have no knowledge as to what these ultranatural 
events are. They cannot be described in a human language$,$ ever. All we know is 
that they exist. \pars
From the human perspective$,$ there can be ``sudden changes'' in the behavior of 
a Natural-system at any time during its development 
$\lbrack\!\lbrack$7$\rbrack\!\rbrack$. But these changes are not 
truly ``sudden'' from the NSP-world viewpoint. Indeed$,$ we can magnify a point 
in time$,$ as we did with a point in space$,$ and investigate what happens at the 
moment of change. Again wondrous events occur that are difficult to describe 
in a human language. Technically$,$ the changes occur in an {\it ultrauniform} 
manner. This type of change$,$ from the NSP-world viewpoint$,$ can be compared with 
the mathematical concept of an  ``uniformly continuous'' change. 
But ultrauniform is ``infinitely'' more uniform$,$ so to speak$,$ than the 
usual standard concept of an uniformly continuous change.  
\pars  
\line{\leaderfill}
If we look at the GGU-model as a whole$,$ is there  a way to describe it in its 
entirety? Yes$,$ there is and this may be the most remarkable aspect of this 
research. The GGU-model can be characterized as behaving in its entirety like a 
super$,$ super$,$ super to an infinite degree$,$ {\it mind} or possibly a {\it computer}. 
The processes are 
similar 
to how an almost inconceivably powerful mind or maybe a computer would behave.  
\pars
{\quad}
\hrule
\smallskip
\hrule\bigskip\bigskip
\indent  In the next section titled ``The Mathematics$,$''  is reproduced$,$ with 
additional comments$,$ some of the actual mathematics that$,$ when interpreted$,$ 
describes 
the properties of the GGU-model that correspond to the nonmathematical 
descriptions. [The procedures in this section have been highly refined and added to.  The notion of ``instructions'' and ``instruction-information'' are now the fundamental entities that generate a universe. (See appendix.) ] \pars 
This portion of this discussion will be somewhat more technical than the 
previous portion.  
I will refer to 
section 3 as the motivation behind this mathematical model is discussed. 
The expression  ``Natural world'' refers to the collection of all entities 
that are 
categorized as Natural-systems. \pars
Relative to the behavior of a Natural-system$,$  a  general scientific 
approach is taken and it is assumed that scientists are interested in various types of 
descriptions for Natural-system behavior. It is not difficult to show that all 
forms of scientific description can be reduced to strings of symbols. 
Developmental paradigms are simply time related descriptions for the ideal 
behavior of any Natural-system viewed simply as strings of symbols. Although 
some of these collections of descriptions might be generated by a specific 
theory$,$ a general approach that a developmental paradigm describes a
sequence of Natural events is taken. The expression {\it Natural event} 
means an 
objectively real and physical occurrence that is categorized as ``natural'' by 
the physical scientist. This approach does not include 
any requirement that such a sequence be generated by some 
accepted and humanly comprehensible theory. On the other hand$,$ theory 
generated sequences are not excluded.\pars
On the first six pages of the next section are reproductions of pages taken 
from the paper ``Nonstandard consequence operators'' 
$\lbrack\!\lbrack$Herrmann$,$ R. A.  
{\it Kobe J. Math.} 4(1987): 1--14$\rbrack\!\rbrack$. They detail how the 
basic somewhat unusual model $\cal E$ is constructed. The terminology used is 
that of the abstract mathematical structure. For example$,$ certain subtle 
consequence operators are interpreted as ultralogics$,$ while an ultraword is an 
unreadable sentence. The theorems on these pages show some of the behavior of 
nonstandard consequence operators. The most significant discussion on these 
first six pages is the construction and embedding of the set $\cal E.$ 
Although the results on these first six 
pages show some of the interesting  
behavior 
of nonstandard 
consequence operators they
 are not specifically needed to comprehend what 
follows. Further$,$ the  ``time ordering'' concept
 considered throughout what 
follows can be
 replaced with the notion of the  ``universal event
 number.''\pars 

Since an ultralogic is based upon the selection of some nontrivial logical 
process$,$ there is a need to select a logical system that is common to all known 
logical systems used throughout scientific discourse. After some difficulty$,$  
a system$,$ denoted by $\b S$ and discussed at the beginning of section 
7.3  titled ultrawords$,$ was selected. 
I point out that the ordering of the sections of the 
included mathematics is not the same ordering in which the processes were 
originally discovered and used. Intuitively$,$ throughout the modeling 
of these linguistic concepts$,$ a specific ``frozen segment'' is considered as 
a description for a particular Natural event that occurs at a particular 
moment of time. From a scientific communication point of view$,$ 
this description is all that 
can be scientifically known about such an event and is substituted for it. 
\pars
 Developmental paradigms$,$ the deductive logic $\bf S$ and the like
 are embedded within a special but well-know 
mathematical structure called a superstructure by means of a fixed encoding and 
are  further embedded into a nonstandard structure.  The next step in the 
modeling process is to show that for a developmental paradigm written in a 
standard language there exists a new object$,$ called an ultraword$,$ that 
when the ultralogic (an intrinsic ultranatural process (i.e. an IUN-process)) 
$\Hyper {\b S}$ is applied to this ultraword the entire developmental 
paradigm (or the corresponding Natural event sequence) is logically produced. Theorem 
7.3.1  establishes that for each developmental paradigm such an 
ultraword exists.  Defining the {\it Natural world} as the collection of all 
Natural event sequences that correspond to the behavior of all systems that are 
categorized as Natural-systems$,$ it follows that ultrawords cannot be 
entities within the Natural world 
under the GGU-model interpretation. Also the force-like operator 
$\Hyper {\b S}$ cannot 
be applied within the Natural world. Rather then simply accepting that 
ultrawords are ``things'' that cannot be further described$,$ 
elsewhere  additional intuitive meaning for 
this concept is discussed. \parm
Relative to the collection of all events generated 
by the force-like process $\Hyper {\b  S},$ a question that arises.  
Are only the developmental paradigm 
events obtained as a result of the process $\Hyper {\b S}$ applied to an ultraword $w$?
A conjecture was no. Theorem 7.3.2  shows that other  ``descriptions'' for 
other types of events also occur when $\Hyper {\b S}$ is applied to a specific 
$w.$ A later  investigation$,$ Theorem 10.1.1$,$ shows the general composition of 
these new descriptions or events. But more importantly$,$ Theorems 7.3.2 and 
10.1.1 show 
that these new events must occur. Further$,$ they cannot occur within the Natural 
world and cannot be  described by any natural language. These events have been 
interpreted as events (called ultranatural events (i.e. UN-events)) that are needed to uphold and 
sustain a Natural-system's development. \par
\bigskip
The above discussed results do not answered Wheeler's basic question. The 
basic question is answered by Theorem 7.3.4 in an slightly more general mode.   
Every subnatural-system within the Natural-system called the universe can be 
associated with its own special system generating ultraword $w_i'.$ Corollary
7.3.4.1 says that if we select a set of ultrawords that generate each and every 
subnatural-system that is within our Natural universe$,$ then there (logically) exists an ultimate 
ultraword 
$w^\prime$ such that when the force-like process $\Hyper {\b S}$ is applied to 
$w^\prime$ each of the subnatural-system ultrawords are produced. Hence$,$  
all of the original subnatural-systems (the Natural event sequences) are produced by 
application of $\Hyper {\b S}$ to $w^\prime.$ Thus within the nonstandard physical 
world$,$ not within the Natural world$,$ there (logically) exists a force-like 
process the 
combines all of the Natural-systems together in a consistent manner and yields 
our Natural universe and all 
of the moment-to-moment alterations that comprise its 
development. The mathematical existence of the ultimate ultraword $w^\prime$
yields a solution to the ``general grand unification problem'' 
as described by Wheeler.\pars
Since the Natural event sequences are not necessarily predicted by a theory$,$ then is 
it possible that  the ideal developmental paradigms and
 the corresponding Natural 
event sequences are somehow ``preselected''? Theorem 7.2.1 coupled with 
the 
discussion in 10.4 yields the necessary NSP-world IUN-selection processes.  
It is at this point 
that one of the most basic and significant properties of 
the GGU-model becomes apparent. \pars
Suppose that you consider a partial denumerable developmental paradigm $d_i$
where one of its members $F_i$ is a frozen segment that 
represents a specific configuration 
for a Natural-system as it appears at a standard time interval $[t_i, t_{i+1})$    
Notice that $\b F_i \in 
\b T_i .$ The (cosmic or standard NSP-world) time $t_i$ 
is conceived of as the moment of time in the past when a time fracture 
occurs and all other members of $d_i$ represent behavior ``after'' such a time.  
There are infinitely many sequences (finite or denumerable) 
of standard frozen segments or *-frozen segments that can be adjoined to 
$d_i,$ and that yield 
other developmental paradigms that  describe Natural or ultranatural behavior 
for the same Natural-system but 
 for cosmic times ``prior to'' $t_i.$ This yields type $d$ 
or type $d^\prime$ development paradigm.\pars
Theorem 7.2.1 states that there is an external IUN-selection process 
that yields each of these developmental paradigms. A developmental paradigm 
that contains no additional frozen segments of any type
prior to $t_i$ along with its associated ultraword  
represents the maximum scenario. Note that a maximum scenario can yield by 
application of an ultralogic$,$ *-frozen segments as well as frozen segments. 
A developmental paradigm that contains only 
time ordered standard frozen segments before and after $t_i$ is a 
minimum scenario. An intermediate scenario contains some or all  
*-frozen segments representing ultranatural events prior to $t_i.$  
``Sudden alterations'' are modeled by the minimum and intermediate scenarios 
and can be 
used for various purposes such as the Patton and Wheeler concept of  
``mutability'' of Natural law or behavior. 
For the minimum and intermediate scenarios$,$ 
can we 
investigate what happens during the
NSP-world time when Natural law or Natural constants are altered? Can we  
``open up the time fracture$,$'' so to speak$,$ and look inside?\pars
As discussed in section 7.5$,$ applications of Theorem 7.5.1 yield a 
startling view of how these alterations are being made within the NSP-world. 
Furthermore$,$ the force-like process $ \Hyper {\b S}$ that produces all of the 
event sequences does so in a remarkable manner as described by Corollary 
7.4.1.2.\pars
As to various NSP-world objects that can mediate any GGU-model alterations$,$  
produce informational transmissions$,$ and be considered as the composition of  
the vacuum$,$ these are 
automatically generated by this mathematical model as shown by Theorem
9.3.1 where the descriptions are interpreted as descriptions for 
objects. \parm
As to how the assumed ``Laws of Nature'' that seem to exist today came into 
being$,$ the discussion in section 10.2 shows that these can also be assumed to 
have been produced by $\Hyper {\b S}$ applied to an ultraword. This answers 
another basic Wheeler question. The GGU-model cosmogony yields the various 
Natural laws that exist today. Further$,$ an ultraword such as $w"$ gives an 
external unification to the collection of all written physical theories. \parm
The last section in this paper deals with the ``\underbar{substance} out of 
which the (Natural) universe is made.'' It shows that using the method outlined in 
Theorem 9.3.1 the basic properties of propertons can be obtained.  
{\it Ultra-propertons} are  obtained with the aid of           
Theorem 11.1.1. Other properties of the
 infinitesimal and infinite hyperreal numbers
coupled with a very simple hyperfinite translation (affine 
operation) lead to intermediate propertons that can be finitely combined 
together to produce every basic entity within our Natural universe. \pars
[Note (1): Relative to cyclic$,$ multi-universe$,$ plasma or any cosmology that claims 
that our universe has no Natural time beginning or no Natural time ending$,$ 
such universes still have  a ``beginning'' from the
GGU-model viewpoint. No result in this paper is dependent upon a universe existing for 
only a finite 
period of time. Each Natural-system still has a beginning and ending with respect to an identified cycle. There is for each Natural-system within an $i$th cycle an ultraword $w_j^i.$ Then there is a cycle ultimate ultraword $(w^i)'.$  The multi-cycle-universe generating ultimate ultraword $w'$ still exists and the force-like
operator $\Hyper {\b S}$ still applies. That is, that $w'$ generates each cycle and the contents of each cycle. Also note that if a Natural-system $j$ is open-ended in that it continues to alter its appearance for all of time and has either no beginning or no ending in the intuitive sense$,$ then this also can be modeled by considering a denumerable sequence of basic time intervals $[a_i,b_i).$ 
Again there is an ultraword $w_j^i$ for each $[a_i,b_i)$ and$,$ hence$,$ an ultraword $(w_j)'$ that generates each of the $w_j^i.$ Thus $(w_j)'$ generates the entire Natural-system's behavior. Finally$,$ there is an ultimate ultraword $w'$ that generates each $(w_j)'$ and$,$ hence$,$ each Natural-system.  3 MAR 1996]\pars
[Note (2): For an ultimate GGU-model conclusion that shows how an GGU-model 
generated universe can be made to vary due to moment-to-moment perturbations$,$ see section 11.2 that starts on page 55. This ultimate GGU-model conclusion also specifically models the ``quantum'' participatory requirements. 24 DEC 1996]\pars
[Note (3): As mentioned previously$,$ the set of all ultranatural initial conditions that would lead to universe creation can be entirely compose of entities that behave like``initial conditions'' but these conditions cannot be described in any language associated with any entity within our universe. This fact does not mean that this set does not contain some ultranatural initial conditions that can be properly described by means of a comprehensible language. One such description follows the pattern of ``field fluctuations.'' The collection of all ultimate ideal ultrawords $w'$ and its collection of perturbative ultimate ultrawords $\{w_t'\}$ within the NSP-world substratum can be supposed to form a dense collection W'. These collections fluctuate fortuitously with respect to NSP-world spacetime. Whenever a fluctuation exceeds a specified spacetime parameter$,$ the ultralogic $\Hyper {\b S}$ acts upon the $w'$
and a universe is generated. All of the universes that are generated by this process are disjoint. 2 JAN 1999]\pars
[Note (4): Relative to the deBroglie statement and to all the areas that appear to be humanly comprehensible$,$ with the exception of an additional postulate relative to the continuous energy spectrum for electromagnetic radiation that is required for properton generation$,$ the GGU-model is based upon one and only one postulate. The postulate is that how nature combines together natural laws and processes to bring into being a natural event or a change in a natural system is modeled or mirrored by the mental processes human beings use to describe the produced events or changes. 
This hypothesis is verified by the largest amount of empirical evidence that could ever exist since whenever a scientist discusses or predicts natural system behavior human mental processes are applied. This includes all of the methods used to gather and analyze evidence that would tend to verify any scientific hypothesis. 3 JAN 1999]\parm
In the next section$,$ the references$,$ when they have been reproduced$,$ appear at 
the end of each  ``chapter.''\parm
\centerline{\bf Important Note}\pars
\indent In the mathematics section 3$,$ objects are chosen that seem 
to yield the simplest possible entities. This is done to minimize controversy 
and to 
allow most conclusions to be established in a convincing and 
straightforward manner.
\noindent The philosophy of science employed is the exact same 
philosophy of science used in theoretical cosmology and quantum logic 
investigations. In almost all cases$,$ the physical-like interpretations
correlate directly to the mathematical structure. The first section of part 3 
is a reproduction of a Kobe Math. J. paper. The second section of part 3 is 
a reproduction of portions of the book {\it Ultralogics and More} (see 
reference [2] of part 1) where all the details for the actual construction
of the nonstandard model can be found.  It should be self-evident that the 
results contained within this monograph are only the must basic and that 
further in-depth investigations should be pursued. \pars
Intuitively$,$ I am 
confident that all of the 
 theorems are correct. If any  ``proof'' is not 
convincingly established$,$ then this  should be easy to rectify. Finally$,$ I do 
not contend that this is THE solution to this problem and questions. 
Although these results are speculative in character$,$ they are no more 
hypothetical than the
Everett-Wheeler-Graham many-worlds 
interpretation or 
Hartle-Hawking quantum gravity model.  
I do contend that since the method was devised in 1979 that these results 
are the FIRST such
solutions obtained scientifically. The term ``scientifically'' refers to the 
sciences of concrete mathematical modeling$,$ mathematical logic and 
interpreting mathematical structures physically. \pars

\vfil\eject                          
{\leftskip 0.5in \rightskip 0.5in\noindent 3. {\bf The Mathematics}\parm
The following is a direct copy$,$ with numerously many printers errors corrected$,$ of the indicated portions of the indicated 
published paper.\pars

\font\ninerm=cmr9
\hrule\smallskip
\hrule\smallskip

The following is a direct copy of this published paper$,$ but with numerously many printer's errors corrected and with a few additional remarks (18 NOV 2010).\par\medskip
\hrule\smallskip
\hrule\smallskip
\noindent {\bf Herrmann$,$ R. A.\par
\noindent Kobe J. Math.}\par
\noindent \b 4(1987) 1-14\par
\bigskip
\centerline{\bf NONSTANDARD CONSEQUENCE OPERATORS}\bigskip
\centerline{By Robert A. Herrmann}\medskip
\centerline{\ninerm (Dedicated to Professor K. Is\'eki)}
\centerline{\ninerm (Receiver October 22, 1984)}\bigskip
\indent {\bf 1. Introduction}\parm
In 1963, Abraham Robinson applied his newly discovered nonstandard analysis to formal first-order languages and developed a nonstandard logic [11] relative to the ``truth'' concept and structures. Since that time not a great deal of fundamental research has been attempted in this specific area with one notable exception [3]. However, when results from this discipline are utilized they have yielded some highly significant and important developments such as those obtained by Henson [4]. \pars
The major purpose for this present investigation is to institute formally a more general study than previously pursued. In particular, we study nonstandard logics relative to consequence operators [2] [6] [12] [13] defined on a nonstandard language. Since the languages considered are not obtained by the usual constructive methods, then this will necessitate the construction of an entirely new foundation distinctly different from Robinson's basic embedding techniques. Some very basic results of this research were very briefly announced in a previous report [6].\pars
In order to remove ambiguity from the definition of the ``finite'' consequence operator, the definition of ``finite'' is the ordinary definition in that the empty set is finite and any nonempty set $\r A$ is finite if and only if there exists a bijection $\rm f\colon A \to [1,n],$ where $\rm [1,n] = \{x\mid n\in \nat,\ 1\leq x \leq n\}$  ($\nat$ is the set of natural numbers with zero). Unless otherwise stated, all sets $\r B$ that are infinite will also be assumed to be Dedekind-infinite. This occurs when a set $\r B$ is denumerable, since $\r B$ inherits a well-ordering from $\nat$, or $\r B$ is well-ordered [2, p. 248], or the Axiom of Choice is assumed. We note that within mathematics one is always allowed to make a finite choice from finitely many nonempty sets, among others [9, p. 1].
  
In 2, we give the basic definitions, notations and certain standard results are obtained that indicate the unusual behavior of the algebra of all consequence operators defined on a set. In 4, some standard properties relative to subalgebras and chains in the set of all consequence operators are investigated. Finally, the entire last section is devoted to the foundations of the theory of nonstandard consequence operators defined on a nonstandard language.\par\bigskip

\indent {\bf 2. Basic concepts}\parm
Our notations and definitions for the standard theory of consequence operators are taken from references [2][6][12][13]$,$ and we now recall the most pertinent of these. Let L be any  nonempty set that is often called a {\it language}$,$ $\power {\r L}$ denote the power set of L and for any set X let $F(\r X)$ denote the finite power set of X (i.e. the set of all finite subsets of X.)\parm
DEFINITION 2.1 A mapping $\r C\colon \power{\r L} \to \power {\r L}$ is a consequence operator (or closure operator) if for each $\r X,\ \r Y \in 
\power {\r L}$\pars
\indent\indent (i) $\rm X \subset C(X) = C(C(X)) \subset L$ and if\pars
\indent\indent (ii) $\rm X \subset Y$$,$ then $\rm C(X) \subset C(Y).$\pars
\noindent A consequence operator C defined on L is said to be {\it finite} ({\it finitary}$,$ or {\it algebraic}) if it satisfies\pars
\indent\indent (iii) $\rm C(X) = \cup\{C(A)\mid A \in F(\r X)\}.$\par\medskip
REMARK 2.2 The above axioms (i) (ii) (iii) are not independent. Indeed$,$ 
(i)(iii) imply (ii).\pars
Throughout this entire article the symbol ``C'' with or without subscripts 
or with or without symbols juxtapositioned to the right will always denote a consequence operator. The only other symbols that will denote  consequence operators are ``I'' and ``U''. The symbol $\cal C$ [resp. ${\cal C}_f$] denotes the set of all consequence operators [resp. finite consequence operators] defined on $\power {\r L}.$\parm
DEFINITION 2.3. (i) Let\ \r I \ denote the identity map defined
on $\power {\r L}.$ \par
\noindent (ii) Let $\r U\colon \power {\r L} \to \power {\r L}$ be defined as follows: for each $\r X \in \power {\r L},\ \r U(\r X) = \r L.$\par
\noindent (iii) For each $\r C_1,\ \r C_2 \in {\cal C},$ define $\r C_1 \leq
\r C_2$ iff $\r C_1(\r X) \subset \r C_2(\r X)$ for each $\r X \in 
\power {\r L}.$
(Note that $\leq$ is obviously a partial order defined on $\cal C$.)\par  
\noindent (iv) For each $\r C_1,\ \r C_2 \in {\cal C}$$,$ define $\r C_1 \lor \r C_2
\colon \power {\r L} \to \power {\r L}$ as follows: for each $\r X \in 
\power {\r L},\ (\r C_1 \lor \r C_2)(\r X)= \r C_2(\r X)\cup \r C_2(\r X).$\par
\noindent (v) For each $\r C_1,\ \r C_2 \in {\cal C},$ define
$\r C_1 \land \r C_2 \colon \power {\r L} \to \power {\r L}$ as follows: for each $\r X \in \power {\r L},\ (\r C_2 \land \r C_2)(\r X) = \r C_1(\r X) \cap \r C_2 (\r X).$ \par
\noindent (vi) For each $\r C_1,\ \r C_2 \in {\cal C}$ define $\r C_2 \lor_w \r C_2\colon \power {\r L} \to \power {\r L}$ as follows: for each 
$\r X \in \power {\r L},\ (\r C_1 \lor_w \r C_2)(\r X) = \cap\{\r Y\mid
\r X \subset \r Y\subset \r L$ and $\r Y = \r C_1(\r Y)= \r C_2(\r Y)\}.$\pars 
Prior to defining certain special consequence operators notice that $\r I,\  \r U \in {\cal C}_f$ and that \ I [resp. U] \ is a lower [resp. upper] unit for the algebras $\langle {\cal C}, \leq \rangle$ and $\langle {\cal C}_f,\leq \rangle.$\pars \bigskip

\line{\leaderfill}\medskip
\line{\leaderfill}\bigskip

{\bf 4. Nonstandard Consequence Operators}\parm
Let ${\cal A}$ be a nonempty finite set of symbols. It is often convenient to assume that $\cal A$ contains a symbol that represents a blank space. As usual any nonempty finite string of symbols from $\cal A,$ with repetitions$,$ is called a {\it word} [10$,$ p.222]. A word is also said to be an (intuitive) {\it readable sentence} [5$,$ p. 1]. We let \ W\ be the intuitive set of all words created from the {\it alphabet} $\cal A$. Note that in distinction to the usual approach$,$ \ W \ does not 
contain a symbol for the empty word. \pars
We accept the concept delineated by Markov [4]$,$ the so-called  ``abstraction 
of identity$,$'' and say that $\r w_1,\ \r w_2 \in \r W$ are ``equal'' if they 
are composed of the same symbols written in the same intuitive order (left to 
right). The {\it join} or juxtaposition operation between $\r w_1,\ \r w_2 \in 
\r W$ is the concept that yields the string $\r w_1 \r w_2$ or $\r w_2 \r 
w_1$. Thus \ W\ is closed under join. Notice that we may consider a 
denumerable formal language as a subset of \ W. (By adjoining a new symbol not in $\cal A$ and defining it as the unit, $\r W$ becomes a free monoid generated by the set ${\cal A} \cup \{\rm new\ symbol\}.$)\pars
Since \ W\ is denumerable$,$ then there exists an injection
$i\colon \r W \to \nat.$ Obviously$,$ if we are working with a formal language that is a subset of \ W$,$ 
then we may require $i$ restricted to a formal language to be a G\"odel 
numbering. Due to the join operation$,$ a fixed member of \ W\ that contains two 
or more 
distinct symbols can be represented by various {\it subwords} that are joined together to yield the given fixed word. The word ``mathematics'' is generated by the join of $\r w_1 = {\rm math},\ \r w_2 = \r e,\ \r w_4 = {\rm mat},\ \r w_4 = {\rm 
ics}.$ This word can also be formed by joining together 11 not necessarily distinct members of \ W. \pars
Let $i[\r W] = T$ and for each $n \in \nat,$ let $T^n = T^{[0,n]}$ denote the set of all mappings from $[0,n]$ into $T.$ Each element of $T^n$ is called a {\it partial sequence}$,$ even though this definition is a slight restriction of the usual one that appears in the literature.
Let $f \in T^n, n>0.$ Then the {\it order induced by f} is the simple inverse order determined by $f$ applied to the simple order on $[0,n].$ Formally$,$ for each $f(j),\ f(k)\in f[[0,n]],$ define $f(k)\leq_f f(j)$ iff $j\leq k,$ where $\leq$ is the simple order for $\nat$ restricted to $[0,n].$ In general$,$ we will not use this notation $\leq_f$ but rather we will indicate this (finite) order in the usual acceptable manner by 
writing the symbols
$f(n),f(n-1),\ldots,f(0)$ from left to right . Thus we symbolically let $f(n)\leq_f f(n-1)\leq_f \cdots \leq_f f(0)= f(n)f(n-1)\ldots f(0).$\pars
Let $f\in t^n.$ Define $\r w_f \in \r W$ as follows: $\r w_f = 
(i^{-1}(f(n)))(i^{-1}(f(n-1)))\cdots (i^{-1}(f(0))),$ where the operation indicated by juxtaposition is the join. We now define a relation on 
$P =\cup \{T^n\mid n \in \nat \}$ as follows: let $f,\ g\in P.$ Then for 
$f\in T^n$ and $g \in T^m,$ define $f \sim g$ iff $(i^{-1}(f(n)))\cdots (i^{-1}f(0)))=(i^{-1}(g(m)))\cdots (i^{-1}(g(0))).$ It is obvious that 
$\sim$ is an equivalence relation on $P.$ For  each $f \in P,$ $[f]$ denotes 
the equivalence class under $\sim$ that contains $f.$ Finally$,$ let
${\cal E}=\{[f]\mid f\in P\}.$ Observe that for each $[f] \in \cal E$ there exist $f_0,\ f_m \in [f]$ such that 
$f_0 \in T^0,\ f_m \in T^m$ and if there exists some $k\in \nat$ such that $0<k<m,$ then there exists some
$g_k \in [f]$ such that $g_k \in T^k$ and if $j\in \nat$ and $j >m,$ then there does not exist $g_j \in T^j$ such that $g_j \in [f].$ If we define the {\it size} of a word $\r w \in \r W$ (size(w)) to be the number of not necessarily distinct symbols counting left to right that appear in  \ W$,$ then the size(w) = $m+1.$ For each $\r w \in \r W,$ there is $f_0 \in T^0$ such that
$\r w = i^{-1}(f_0(0))$ and such an $f_m \in [f_0]$ such that size (w) = $m+1.$
On the other hand$,$ given $f \in P,$ then there is a $g_0 \in [f]$ such that 
$(i^{-1}(g_0(0))) \in \r W.$ Of course$,$ each $g \in [f]$ is interpreted to be 
the word $(i^{-1}(g(k)))\ldots (i^{-1}(g(0))).$\pars
Each $[f] \in \cal E$ is said to be a (formal) word or (formal) {\it readable
sentence.} All the intuitive concepts$,$ definitions and results relative to consequence operators defined for $\r A  \in \power {\r W}$ are now passed
to $\power {{\cal E}}$ by means the map $\theta(i(\r w))= [f_0]$.
In the usual manner$,$ the map $\theta$
is extended to subsets of each $\r A \in \power {\r W},$ n-ary relations and the like. For example$,$ let $\r w \in \r A \in \power {\r w}.$ Then there exists $f_w \in P$ such that $f_w \in T^0$ and $f_w(0)= i(\r w).$ Then 
$\theta(i(\r w)) = [f_w].$ In order to simplify notation$,$ the images of the extended 
  $(\theta\, i)$ composition will often be indicated by bold notation with the exception  
of customary relation symbols which will be understood relative to the context. 
For example$,$ if \ S \ is a subset of \ W$,$ then we write $\theta (i[\r S]) = \b S.$ [(12/27/12) Notice that $\b W = {\cal E}.$\pars

Let $\cal N$ be a superstructure constructed from the set $\r W\cup \nat$ as its set of atoms. (12/16/12 the ground set has now been so expanded so any symbols in W that are used for natural numbers are different from those in $\nat$.) Our standard structure is ${\cal M} = ({\cal N}, \in, = ).$ Let 
$\Hyper {\cal M}= (\Hyper {\cal N},\in,=)$ be a nonstandard and elementary extension of $\cal M$$.$ Further, $\Hyper {\cal M}$ is an enlargement.\pars

For an alphabet $\cal A$$,$ there exists $[g] \in \Hyper {\cal E}-{\cal E}$ such that there are only finitely many standard members of $\nat$ in the range of $g$ and these 
standard members injectively correspond to alphabet symbols in $\cal A$ [5$,$ p. 24]. On the other hand$,$ there exist $[g'] \in \Hyper {\cal E}-{\cal E}$ such that the range of $[g']$ does not correspond in this manner to elements in $\cal A$ [5$,$ p. 90]. \pars
Let $C\in {\cal H}$ map a family of sets $\cal B$ into ${\cal B}_0$. If $C$ satisfies either the Tarski axioms (i)$,$ (ii) or (i)$,$ (iii)$,$ or the *-transfer 
*(i)$,$ *(ii)$,$ or *(i)$,$ *(iii) of these axioms$,$ then $C$ is called a {\it subtle consequence operator}.  For example$,$ if $\r C \in {\cal C},$ then  it is immediate that $\Hyper {\b C}\colon \Hyper {(\power {\theta(\r A)}} \to \Hyper {\power {\theta (\r A))}}$ satisfies *(i) and *(ii) for the family of all internal 
subsets of $\Hyper {(\theta (\r A))}.$ This $\Hyper {\b C}$ is a subtle consequence operator. For any set $A \in {\cal N},$ let $\sig A =\{\hyper a\mid a \in A\}.$ (In general, this definition does not correspond to that used by other authors.) If for a subtle consequence operator $C$ there does not exist some similarly defined $D \in {\cal N}$ such that $C = \sig 
{\b D}$ or $C = \Hyper {\b D}$$,$ then $C$ is called a {\it purely}
subtle consequence operator. Let infinite $A \subset {\cal E}$ and $B = \hyper A - \sig A.$ Then the identity $I\colon \power B - \power B$ is a purely subtle consequence operator.\pars
There are certain technical procedures associated with the $\sigma$ map that take on a specific significance for consequence operators. Recall that $\cal N$ is closed under finitely many power set or finite power set 
iterations. Let $X,\ Y \in {\cal N}.$ It is not difficult to show that if ${\cal P}\colon \power X \to Y,$ then for each $A \in \power X,\ \Hyper (\power A) = \Hyper {\cal P}(\hyper A).$ Moreover$,$ if $F\colon \power X \to Y,$ where $F$ is the finite power set operator$,$ then for each $A \in \power A,\ \Hyper {(F(A))}=
\hyper F (\hyper A).$ If $\r C \in \cal C$ and $X \subset \r W$$,$ then $\b C\colon \power X \to \power X$ has the property that for each $A \in \power X,\ \Hyper {(\b C(A))} = \Hyper {\b C}(\hyper A).$\pars
Recall that we identify 
each $\Hyper n \in \hypernat$ with $n \in \nat$ since 
$\Hyper n$ is but a constant sequence with the value $n.$ Utilizing this fact$,$ we have the following
straightforward lemma the proof of which is omitted.\parm
\indent {\bf LEMMA 4.1}.

 (i) {\it Let $A \in \cal N.$ Then 
$^\sigma (F(A)) = F(^\sigma A)$. If also $A \subset ({\cal W} \cup {\cal E}),$ then $^\sigma 
(F(A)) = F(A).$}\pars

(ii) {\it Let ${\rm C} \in {\cal C}^\prime,\ \r B \subset \r X 
\subset \cal W.$} \pars 

(a) {\it $^\sigma (\b C(\b B)) = \b C(\b B).$}\pars 

(b) {\it $\Hyper {\b C}\bigm| \{\hyper {\b A}\mid {\b A} \in \power X \}=
 \{ (\hyper {\b A},\hyper {\b B}) 
\mid (\b A,\b B) \in \b C \} = {^\sigma \b C}.$} \pars 

(c) {\it If $\r F \in \r F(\r B),$ 
then $^\sigma (\b C(\b F)) \subset (^\sigma \b C )(^\sigma {\b F}) = 
(^\sigma \b C)(\b F).$ Also 
$^\sigma (\b C(\b B)) \subset {(^\sigma \b C) (\hyper {\b B})}$ and$,$ in general$,$ 
$^\sigma (\b C(\b B)) \not= (^\sigma \b C)(\hyper {\b B}),\ ^\sigma (\b C(\b F)) \not= 
(^\sigma \b C)(F).$} \pars 

(d) {\it If $C \in {\cal C}_f^\prime,$ then 
$^\sigma (\b C(\b B))= \bigcup \{^\sigma (\b C(\b F)) \mid \b F \in {^\sigma (F(\b B))}\} 
= \bigcup \{ ^\sigma (\b C(\b F)) \mid \b F \in F(^\sigma {\b B})\} = \bigcup \{\b C(\b F) 
\mid \b F \in F(\b B) \}.$}\parm

(A duplicate lemma holds, where $A \in \cal N$ and $C \in \cal C$, where $\cal C$ set of consequence operators defined on subsets of W if W is included as a subset of the ground set. The difference is that the ``bold'' notion does not appear.)\parm
Throughout the remainder of this paper$,$ we remove from $\cal C$ the one and only one inconsistent consequence operator $U.$ Thus notationally we let $\cal C$ denote the set of all consequence operators defined on infinite $ \r L = \r X \subset \r W$ with the exception of $U.$ Two types of chains will be investigated. Let \ T\ be any chain in $\langle {\cal C},\leq \rangle$ and
\  $\r T'$ \ be any chain with the additional property that for each $\r C \in \r T'$ there exists some $\r C' \in \r T'$ such that $\r C < \r C'.$\parm

{\bf THEOREM 4.2} {\it There exists some $C_0 \in \Hyper {\b T}$ such that for each $\r C \in \r T,\ \Hyper {\b C}\leq C_0.$ There exists some $C'_0 \in \Hyper {\b T'}$ such that $C'_0$ is a purely subtle consequence operator and for each $\r C \in \r T',\ \Hyper {\b C} < C_0'.$ Each member of $\Hyper {\b T}$ and $\Hyper {\b T'}$ are subtle consequence operators.}\pars
PROOF. Let $R = \{(x,y)\mid x,\ y \in \b T\}\ {\rm and}\ x\leq y \}$ and 
$R'\{(x,y)\mid x,\ y  \in \b T' \ {\rm and}\ x < y\}.$ In the usual manner$,$ it follows that $R$ and $R'$ are concurrent on the set $\b T$ and $\b T'$ respectively. Thus there is some $C_0 \in \Hyper {\b T}$ and $C'_0\in \Hyper {\b T'}$ such that for each $\r C \in \r T$ and $\r C' \in \r T,\ \Hyper {\b C} \leq C_0$ and $\Hyper {\b C'} < C_0'$ since $\Hyper {\cal M}$ is an enlargement. Note that the members of $\Hyper {\b T}$ and $\Hyper {\b T'}$ are defined on the set of all internal subsets of $\Hyper {\b L}.$
However$,$ if there is some similarly defined $D \in \cal N$ such that 
$C_0$ or $C_0' = \sig D,$ then since $\sig D$ is only defined for *-extensions of the (standard) members of $\power {\r L}$ and each 
$E \in \Hyper {\b T}\ {\rm or}\ \Hyper {\b T'}$ is defined on the internal subsets of $\Hyper {\b L}$ and there are internal subsets of $\Hyper {\b L}$ that are not *-extensions of standard sets we would have a contradiction. Of course$,$ each member of $\Hyper {\b T}$ or $\Hyper {\b T'}$ is a subtle consequence operator. Hence each $E \in \Hyper {\b T}$ or $\Hyper {\b T'}$ is either equal to some $\Hyper {\b C},$ where $\r C \in \r T$ or $\r C\in \r T'$ or it is a purely subtle consequence operator. Now there does not exist a $D \in \cal N$ such that $C_0' = \Hyper {\b D}$ since $C_0' \in \Hyper {\b T'}$ and $\Hyper {\b C} \not= C_0'$ for each $\Hyper {\b C} \in \sig {\b T'}$ would yield the contradiction that $\Hyper D \in \Hyper {\b T'}- \sig {\b T'}$ but $\Hyper {\b D} \in \sig {\bf {\cal C}}.$ Hence $C_0'$ is a purely subtle consequence operator. This completes the proof.\parm
Let $\r C \in \r T'$. Since $\Hyper {\b C} < C_0'$$,$ then $C'_0$ is  ``more
powerful'' than any $\r C\in \r T'$ in the following sense. If $\r B \in \power {\r L},$ then for each $\r C \in \r T'$ it follows that $\b C({\b B}) \subset \Hyper(\b C(\b B)) = \Hyper {\b C}(\Hyper {\b B}) \subset C_0'(\Hyper {\b B}).$ Recall that$,$ for $\r C \in \cal C,$ a set $\r B \in \power {\r L}$ is called a {\it C-deductive system} if $\r C(\r B) = \r B.$ From this point on$,$ all results are restricted to chains in $\langle {\cal C}_f, \leq \rangle.$\pars
{\bf THEOREM 4.3.} {\it Let $\r C \in \r T \cup \r T'$ and $\r B \in \power {\r L}.$ Then there exists a 
*-finite $F_0 \in \Hyper (F(\b B))$ such that $\b C(\b B)\subset \Hyper {\b C}(F_0) \subset \Hyper {\b C}(\Hyper {\b B}) = \Hyper (\b C({\b B}))$ and $\Hyper {\b C}(F_0) \cap \b L = \b C(\b B) = \Hyper {\b C}(F_0) \cap \b C(\b B).$}\pars
PROOF. Consider the binary relation $Q = \{(x,y)\mid x \in \b C(\b B),\ y\in F(\b B)\ {\r and}\ x\in \b C(y)\}.$ By axiom (iii)$,$ the domain of $Q$ is 
$\b C(\b B).$ Let $(x_1,y_1),\ldots,(x_n,y_n) \in Q.$ By theorem 1 in [6$,$ p. 64]$,$ (the monotone theorem) we have that $\b C(y_1)\cup \cdots\cup \b C(y_n) \subset \b C(y_1 \cup \cdots\cup y_n).$ Since $F = y_1 \cup \cdots \cup y_n \in F(\b B),$ then $(x_1,F),\ldots,(x_n,F) \in Q.$ Thus $Q$ is concurrent on ${\b C}(\b B).$ Hence there exists some $F_0 \in \Hyper {(F(\b B))}$ such that $\sig(\b C(\b B)) = \b C(\b B)\subset \Hyper {\b C}(F_0) \subset \Hyper {\b C}(\Hyper {\b B}) = \Hyper (\b C(\b B)).$ Since $\sig {\b L} = \b L,$ then $\Hyper {\b C}(F_0) \cap \b L = 
\b C(\b B) = \Hyper {\b C}(F_0) \cap \b C(\b B).$\parm
{\bf COROLLARY 4.3.1} {\it If $\r C\in {\cal C}_f$ and $\r B \in \power {\r 
L}$ is a C-deductive system$,$ then there exists a *-finite $F_0 \subset \Hyper 
{\b B}$ such  that  $\Hyper {\b C}(F_0) \cap \b L = \b B.$}\pars 
PROOF. Simply consider the one element chain $\r T = \{ \r C\}.$\parm 
{\bf COROLLARY 4.3.2.} {\it Let $\r C \in {\cal C}_f.$ There there exists a *-finite $F_1 \subset \Hyper {\b L}$ such that for each C-deductive system $\r B 
\subset \r L,\ \Hyper {\b C}(F_1)\cap \b B = \b B.$}\pars 
PROOF. Let $\r T = \{\r C\}$ and the  ``B'' in theorem 4.3 equal \ L. The 
result now follows in a straightforward manner.\parm 
{\bf THEOREM 4.4.} {\it Let $\r B \in \power {\r L}.$\pars 
{\rm (i)} There exists a *-finite $F_B \in \Hyper {(F(\b B))}$ and a subtle 
consequence operator $C_B  \in \Hyper {\b T}$ such that for all $\r C \in \r 
T,\ \sig(\b C(\b B)) = \b C(\b B) \subset C_B(F_B).$\pars 
{\rm (ii)} There exists a *-finite $F_B \in \Hyper {(F(\b B))}$ and a purely subtle 
consequence operator $C_B' \in \Hyper {\b T'}$ such that for all $\r C\in \r 
T',\ \sig(\b C(\b B)) = \b C(\b B) \subset C_B'(F_B).$}\pars 
PROOF.  Consider the ``binary'' relation $Q=\{((x,z),(y,w))\mid x \in \b T,\ 
y\in \b T,\ w \in F(\b B),\ z\in x(w),\ z\in \power {\b L}, z \in 
x(w),\ {\rm and}\ x(w) \subset y(w)\}.$ Let $\{((x_1,z_1),(y_1,w_1)),\ldots, 
((x_n,z_n),(y_n,w_n))\} \subset Q.$ Notice that $F = w_1\cup\cdots\cup w_n \in 
F(\b B)$ and for the set
$K =\{x_1,\ldots, x_n\},$ let $D$ be the largest member of $K$ with respect 
to $\leq.$ It follows that $z_i \in x_i(w_i) \subset x_i(F) \subset D(F)$ for 
each $ i = 1,\ldots, n.$ Hence $\{((x_1,z_1),(D,F)),\ldots,((x_n,z_n),(D,F))\} 
\subset Q$ implies that $Q$ is concurrent on its domain. Consequently$,$ there 
exists some $(C_B,F_B) \in \Hyper {\b T} \times \Hyper (F(\b B))$ such that 
for each $(x,z) \in \ {\rm domain\ of}\ Q,\ (\Hyper (x,z), (C_B,F_B))\in 
\Hyper Q.$  Or$,$ for each $(u,v) \in \ \sig ({\rm domain \ of}\ Q),\ ((u,v),(C_B,F_B)) 
\in \Hyper Q.$ Let arbitrary $\r C \in \r T$ and $b \in C(B).$ Then 
there exists some $F' \in F(\b B)$ such that $\b b \in \b C(F').$ Thus $(\Hyper 
{\b C},\Hyper {\b b}) \in \ \sig ({\rm domain\ of}\ Q).$ Consequently$,$ for each 
$\r C \in \r T$ and $\b b \in \b C(\b B),\  \b b =\Hyper {\b b} \in (\Hyper {\b C})(F_B) 
\subset C_B(F_B).$ This all implies that for each $\r C \in \r T,$ 
$\sig (\b C(\b B)) = \b C(\b B) \subset C_B(F_B).$\pars 
(ii) Change the relation $Q$ to $Q'$ by requiring that $x \not= y.$ Replace 
$D$ in the proof of (i) above with $D'$ is greater than and not equal to the 
largest member of $K.$ Such a $D'$ exists in $\b T'$ from the definition of 
$\b T'.$ Continue the proof in the same manner in order to obtain $C_B'$ and 
$F_B'.$ The fact that $C_B'$ is a purely subtle consequence operator follows 
in the same manner as in the proof of theorem 4.2.\parm
{\bf COROLLARY 4.4.1} {\it There exists a \r [resp. purely\r ] subtle consequence 
operator $C_L \in \Hyper {\b T}$ \r [resp. $\Hyper {\b T'}$\r ] and a *-finite $F_L 
\in \Hyper (F(\b L))$ such that for all $\r C \in \r T$ \r [resp. $\r T'$\r ]
and each $\r B \in \power {\r L},\ \b B\subset \b C(\b B) \subset C_L(F_L).$}
\pars
PROOF. Simply let  ``B'' in theorem 4.4 be equal to \ L. Then there exists a 
[resp. purely] subtle $C_L \in \Hyper {\b T}$ [resp. $\Hyper {\b T'}$] and 
$F_L \in \Hyper (F(\b L))$ such that for all $\r C\in \r T$ [resp. $\r T'$] 
$\b C(\b L) \subset C_L(F_L).$ If $\r B \in \power {\r L}$ and $\r C \in \r T$ 
[resp. $\r T'$]$,$ then $\r B \subset \r C(\r B)\subset \r C(\r L).$ Thus for 
each $\r B \in \power {\r L}$ and $\r C\in \r T$ [resp. $\r T'$] 
$\b B \subset \b C(\b B) \subset C_L(F_L)$ and the theorem is 
established.\parm
The nonstandard results in this section have important applications to 
mathematical philosophy. We present two such applications. Let $\cal F$ be the 
symbolic alphabet for any formal language \ L \ with the usual assortment of 
primitive symbols [10$,$ p. 59]. We note that it is possible to mimic the 
construction of \ L \ within $\cal E$ itself. If this is done$,$ then it is not 
necessary to consider the map $\
$ and we may restrict our 
attention entirely to the sets ${\cal E}$ and $\Hyper {\cal E}.$ \pars
Let \ S \ denote the predicate consequence operator by the standard rules for 
predicate (proof-theory) deduction as they appear on pages 59 and 60 of 
reference [10]. Hence $\r A \in \power {\r L},\ \r S(\r A) =\{x\mid x \in \r L\ 
{\rm and}\ \r A \vdash x\}.$ It is not difficult to restrict the modus ponens 
rule of inference in such a manner that a denumerable set $\r T' = \{\r 
C_n\mid n \in \nat \}$ of consequence operators defined on $\power {\r L}$ is 
generated with the following properties.\parm
(i) For each $ \r A \in \power {\r L},\ \r S(\r A)= \bigcup \{ \r C_n(\r 
A)\mid n\in \nat \}$ and $\r C_n \not= \ \r S$ for any $n \in \nat.$\pars
(ii) For each $\r C \in \r T'$ there is a $\r C'$ such that $\r C < \r C'$ [5$,$ 
p.57]. Let $\r A \in \power {\r L}$ be any S-deductive system. The $\r A =\r 
S(\r A)= \bigcup \{\r C_n(\r A)\mid n\in \nat\}$ yields that \ A \ is
a $\r C_n$-deductive system for each $n \in \nat.$ Thus  \ S and \ $\r C_n$  
$n \in \nat$ are consequence operators defined on $\power {\r A}$ as well as
on $\power {\r L}.$\parm 
{\bf THEOREM 4.5.} {\it Let \ $\r L$ \ be a first-order language and $\r A \in 
\power {\r L}.$ Then there exists a purely subtle $C_1 \in \Hyper {\r T'}$ and 
a *-finite $F_1 \in \Hyper (F(\b A))$ such that for each $\r B \in \power {\r 
A}$ and each $\r C \in \r T'$\pars
{\rm (i)} $\b C(\b B) \subset C_1(F_1),$\pars
{\rm (ii)} $\b S(\b B) \subset C_1(F_1) \subset \Hyper {\b S(F_1)} \subset \Hyper 
{(\b S(\b A))}.$\pars
{\rm (iii)} $\Hyper {\b S(F_1)} \cap \b L = \b S(\b A) = C_1(F_1)\cap \b L.$}\pars
PROOF. The same proof as for corollary 4.4.1 yields that there is some purely 
subtle $C_1 \in\Hyper {\b T'}$ and $F_1 \in \Hyper {(F(\b A))}$ such that for 
each $\r B \in \power {\r A}$ and each $\r C \in \r T',\ \b C(\b B) \subset 
C_1(F_1)$ and (i) follows. From (i)$,$ it follows that $\bigcup \{ \sig(\b C
(\b B))\mid \r C\in \r T'\}= \bigcup \{\b C(\b B)\mid \r C  \in \r T'\} 
= \b S(\b B) = \sig(\b S(\b B))\subset C_1(F_1)$ and the first part of (ii) 
holds. By 
*-transfer $C_1 < \Hyper {\b S}$ and $C_1$ and $\Hyper {\b S}$ are defined on 
internal subsets of $\Hyper {\b A}.$ Thus $C_1(F_1) \subset \Hyper {\b S(F_1)}
\subset \Hyper {\b S}(\Hyper {\b A}) = \Hyper (\b S(\b A))$ by the *-monotone 
property. This completes (ii). Since $\b S(\b A) \subset C_1(F_1) \subset 
\Hyper {\b S}(F_1) \subset \Hyper (\b S(\b A))$ from (ii)$,$ then (iii) follows
and the theorem is proved.\parm
REMARK 4.6. Of course$,$ it is well known that there exists some $F \in \Hyper 
(F(\b A))$ such that $\b S(\b A) \supset \b A \subset F \subset \Hyper {\b 
A}$ and *-transfer of axiom (i) yields that $\Hyper {\b S(F)} \subset \Hyper 
{\b S}(\Hyper {\b A}) = \Hyper {(\b S(\b A))}.$ However$,$ $F_1$ of theorem 4.5 
is of a special nature in that the purely subtle $C_1$ applied to $F_1$ yields 
the indicated properties. Also theorem 4.5 holds for many other infinite 
languages and deductive processes.\pars
Let \ L \ be a language and let $M$ be a structure in which \ L \ can be 
interpreted in the usual manner. A consequence operator \ C \ is  {\it sound} 
for $M$ if whenever $\r A \in \power {\r L}$ has the property that 
$M\models \r A,$ then $M\models \r C(\r A).$ As usual$,$ $T(M) =\{ x\mid
x \in \r L \ {\rm and}\ M \models x\}.$ Obviously$,$ if \ C \ is sound for $M$$,$ 
then $T(M)$ is a C-deductive system. \pars
Corollary 4.3.1 implies that there exists *-finite $F_0 \subset \Hyper 
{(T(M))}$ such that $\Hyper {\b C}(F_0)\cap \b L = T(M).$ Notice that the fact 
that $F_0$ is *-finite implies that $F_0$ is *-recursive. Moreover$,$ trivially$,$ $F_0$ is a 
*-axiom system for $\Hyper {\b C(F_0)},$ and we do not lack knowledge about 
the behavior of $F_0$ since any formal property about $\b C$ or recursive sets$,$ 
among others$,$ must hold for $\Hyper {\b C}$ or $F_0$ when property 
interpreted.
If \ L \ is a first-order language$,$ then \ S \  is sound for first-order 
structures. Theorem 4.5 not only yields a *-finite $F_1$ but a purely subtle 
consequence operator $C_1$ such that$,$ trivially$,$ $F_1$ is a *-axiom for 
$C_1(F_1)$ and for $\Hyper {\b S(F_1)}.$ In this case$,$ we have that 
$\Hyper {\b S}(F_1)\cap \b L = T(M) = C_1(F_1)\cap \b L.$ By the use of 
internal and external objects$,$ the nonstandard logics $\{\Hyper {\b C},\Hyper 
{\b L}\},\ \{C_1,\Hyper {\b L}\}$ and $\{\Hyper {\b S}, \Hyper {\b L}\}$ 
technically by-pass a portion of G\"odel's first incompleteness theorem.\pars
By definition $\r b \in \r S(\r B),\ \r B \in \power {\r L}$ iff there is a 
finite length ``proof'' of \ b \ from the premises \ B. It follows$,$ that for 
each $b \in \Hyper {(T(M))}$ there exists a *-finite length proof of $b$ from 
a *-finite set of premises $F_1.$ \bigskip
\centerline{\bf References}\parm
\noindent 1. \ A. Abian: The theory of sets and transfinite arithmetic, W. B. Saunders Co., Philadelphia and London, 1965.\pars
\noindent 2. \ W. Dziobiak: The lattice of strengthenings of a strongly finite 
consequence operator$,$ Studia Logica$,$ {\bf 40}(2) (1981)$,$ 177-193.\pars
\noindent 3. \ J. R. Geiser: Nonstandard logics$,$ J. Symbolic Logic$,$ {\bf 
33}(1968)$,$ 236-250.\pars
\noindent 4. \ C. W. Henson: The isomorphism property in nonstandard analysis 
and its use in the theory of Banach spaces$,$ J. Symbolic Logic$,$ {\bf 39}(1974)$,$ 
717-731.\pars
\noindent 5.\ R. A. Herrmann: The mathematics for mathematical philosophy$,$ 
Monograph \#130$,$ Institute for Mathematical Philosophy Press$,$ Annapolis$,$ 
1983. (Incorporated into ``The Theory of Ultralogics$,$''  http://xxx.arxiv.org/abs/math.GM/9903081 and 9903082.)\pars
\noindent 6. \ R. A. Herrmann: Mathematical Philosophy$,$ Abstracts A. M. S. 
{\bf 2}(6)(1981)$,$ 527.\pars
\noindent 7. \ T. Tech: The axiom of choice, North-Holland and American Elsevier Publishing Co., 1973.\pars
\noindent 8. \ J. Los and R. Suszko: Remarks on sentential logics$,$ Indag. 
Math.$,$ {\bf 20}(1958)$,$ 177-183.\pars
\noindent 9. \ A. A. Markov: Theory of algorithms$,$ Amer. Math. Soc. 
Translations$,$ Ser. 2$,$ {\bf 15}(1960)$,$ 1-14.\pars
\noindent 10. \ E. Mendelson: Introduction to mathematical logic$,$ D. Van 
Nostrand$,$ New York$,$ 1979.\pars
\noindent 11. \ A. Robinson: On languages which are based on non-standard 
arithmetic$,$ Nagoya Math. J.$,$ {\bf 22}(1963)$,$ 83-118.\pars
\noindent 12. A. Tarski: Logic$,$ semantics and metamathematics$,$ Oxford 
University Press$,$ Oxford 1956.\pars
\noindent 13. \  R. W\'ojcicki: Some remarks on the consequence operation in 
sentential logics$,$ Fund. Math. {\bf 68}(1970)$,$ 269-279.\pars
\bigskip\bigskip
\noindent{Robert A. Herrmann}\par
\noindent{Mathematics Department$,$}\par
\noindent{U. S. Naval Academy$,$}\par
\noindent{ANNAPOLIS$,$ MD 21402}\par
\noindent{U. S. A.}\par
\noindent The material that begins on the next page (23) is taken from  
``The Theory of Ultralogics'' (Ultralogics and More) with a minor addition or two.  Although much here continues to be applicable to the model, a new approach using the notion of instruction-information and logic-systems has been introduced in the appendix. In what follows, symbols have been changed. $\r W = {\cal W}$ and $\r T = {\cal E}$.\par}\par
\vfil\eject                              
\leftline{\bf 7.1 Introduction.}
\medskip
Consider the real line. If you believe that time is the ordinary continuum$,$ then the 
entire real line can be your {time line}. Otherwise$,$ you may consider only a 
subset of the real line as a time line. In the original version of this section$,$ the time concept for the
GGU-model was presented in a unnecessarily complex form. As shown in [3]$,$ one can assume an
absolute substratum time within the NSP-world. It is the infinitesimal light-clock time measures
that may be altered by physical processes. In my view$,$ the theory of quantum electrodynamics
would not exist without such a NSP-world time concept.\pars 
Consider a small interval $[a,b),\ a<b$ as our basic time interval where as the real numbers
increase the time is intuitively considered to be increasing. In the following approach$,$ one may
apply the concept of the persistence of mental version relative to descriptions for the behavior of a
Natural-system at a moment of time within this interval. An exceptionally small subinterval can be
chosen within $[a,b)$ as a maximum subinterval length $=M.$ 
``Time'' and the size of a ``time'' interval as they are used in this and the following sections refer to an intuitive concept used to aid in comprehending the notation of an event sequence - an event ordering concept.
First$,$ let $a = t_0.$ Then choose $t_1$ such
that $a< t_1 <b.$
There is a partition $t_1, \ldots, t_m$ of $[a,b)$ such that $t_0<t_1< \cdots < t_m < b$ and
$t_{j+1}-t_j \leq M.$ The final subinterval $[t_m, b)$ is now separated$,$ by induction$,$ say be
taking midpoints$,$ into an increasing sequence of times $\{t_q\}$ such that $t_m <t_q <b$
for each $q$ and 
$\lim_{q \to \infty}\ t_q = b.$\pars  
Assume the prototype $[a,b)$ with the time subintervals as defined above. Let $[t_j, t_{j+1})$ be
any of the time subintervals in 
$[a,b).$  For each such subinterval$,$ let ${\r W}_i$  denote the readable 
sentence\parm
\line{\hfil
This$\sp$frozen$\sp$segment$\sp$gives$\sp$a$\sp$description$\sp$for$\sp$the$\sp$\hfil}\pars
\line{\hfil
time$\sp$interval$\sp$that$\sp$has$\sp$as$\sp$its$\sp$leftmost$\sp$endpoint$\sp$the \hfil}\pars
\line{\hfil
$\sp$time$\sp\lceil t_i\rceil\sp$that$\sp$corresponds$\sp$to$\sp$the$\sp$natural$\sp$number$\sp$i.\hfil}\parm 
\noindent Let ${\r T}_i = \{ \r x\r W_i\mid  \r x \in {\cal W}\}.$ 
 The set ${\r T}_i$ is called 
a {{\it totality}} and each member of any such ${\r T}_i$ is called a
{{\it frozen segment.}} Notice that since
 the empty word is not a member of ${\cal W},$ then the cardinality of each 
member of ${\r T}_i$ is greater than that of ${\r W}_i.$ Each ${\rm T}_i$ is 
a (Dedekind) denumerable set$,$ and if $i \not=j,$ then ${\rm T}_i \cap {\rm T}_j 
= \emptyset.$ \pars

\hrule\smallskip\hrule \smallskip
$\r W_i$ is only an identifier and may be altered. ``Time$,$'' either its measure or otherwise$,$ is not the 
actual underlying interpretation for these intervals. Time refers to an 
external event ordering concept$,$ an intuitive {\it event sequence}.  
For most purposes$,$ simply 
call these intervals  ``event intervals.'' [These event intervals can correspond to the universal
event numbers concept.] In the above descriptions for  
$\r W_i,$ simply replace ``time$\sp$interval'' with ``event$\sp$interval'' 
and replace the second instance of the word ``time'' with the word  
``event.'' If this event sequence interpretation is made$,$ then other compatible  
interpretations would be necessary when applying some of the following 
results. ((Added 1/27/2015) The refined developmental paradigm approach in Herrmann (2013b) should be consulted. The notion of a ``continuous'' development is a matter of philosophic choice. As here presented in order to have a meaningful description, this approach can be considered as an exceptionally refined approximation for a continuous development. Or, a continuous development can be considered as an approximation for this development paradigm approach.) \pars \hrule\smallskip\hrule\smallskip 

I point out two minor aspects of the above constructions. First$,$
 within 
certain descriptions there 
are often  ``symbols'' used for real$,$ complex$,$ natural numbers etc. These 
objects also exist as abstract objects within the structure ${\cal M}.$ No 
inconsistent interpretations should occur when these objects are 
specifically modeled within ${\cal M}$ since to my knowledge all of 
the usual mathematical objects used within physical analysis are disjoint 
from ${\cal E}$ as well as disjoint from any finite Cartesian product of
${\cal E}$ with itself. If for future research within physical applications 
finite partial sequences of natural numbers and the finite equivalence classes 
that appear in ${\cal E}$ are needed and are combined into one model for 
different purposes than the study of 
descriptions$,$ then certain modifications would need to be made so that 
interpretations would remain consistent. Secondly$,$ I have tried whenever 
intuitive strings are used or sets of such strings are 
defined to 
use Roman letter notation for such objects. This  
only applies for the intuitive model. Also $\r W_i$ is only an identifier and may be altered.\par\bigskip
\leftline{\bf 7.2 Developmental Paradigms}
\medskip
It is clear that if one considers a time interval of the type $(-
\infty,+\infty),$ $(-\infty,b)$ or $[a, +\infty),$  
then  each of these may be considered as the union of a denumerable collection of time 
intervals of the type $[a,b)$ with common endpoint names displayed. Further$,$ although 
$[a,b)$ is to be considered as subdivided into denumerably 
many subintervals$,$ it is not necessary that each of the time intervals $[t_j,t_{j+1})\subset [a,b)$ 
be accorded a corresponding description for the appearance of a specific 
Natural-system that is distinct from all others that occur throughout the time subinterval.
Repeated descriptions only containing a different last natural number i  in the next to last position
will suffice. Each basic developmental paradigm will be restricted$,$ at present$,$ to such a time
interval $[a,b).$\pars
 Where human 
perception and descriptive ability is concerned$,$ the least controversial 
approach would be to consider only finitely many descriptive choices as 
appropriate. A finite set is recursive and such a choice$,$ since the result is 
such a set$,$ would be considered to be the simplest type of algorithm. You ``simply'' 
check to see if an expression is a member of such a finite set.   
If we limited ourselves to finitely many human choices for Natural 
system descriptions 
from the set of all totalities and did not allow a denumerable or a continuum 
set to be chosen$,$ then the next result establishes that within the 
{Nonstandard Physical world (i.e. NSP-world)} 
such a finite-type of choice can be applied and a continuum of descriptions 
obtained.\pars
The following theorem is not insignificant even if we are willing to accept a 
denumerable set of distinct descriptions --- descriptions that are not only 
distinct in the next to the last symbol$,$ but are also distinctly different in 
other aspects as well. For$,$ if this is the case$,$ the results of Theorem 
7.2.1 still apply. The same finite-type of process in the NSP-world yields 
such a denumerable set as well. \pars
The  term  ``NSP-world'' will signify a certain second type of interpretation 
for nonstandard entities. In particular$,$ the subtle logics$,$ unreadable 
sentences$,$ etc. This 
interpretation will be developed throughout the remainder of this book. One 
important aspect of how descriptions are to be interpreted is that a 
description 
correlates directly to an assumed or observed real Natural phenomenon$,$ and 
conversely. In these investigations$,$ the phenomenon is called an {{\it 
event.}}\par
In order to simplify matters a bit$,$ the following notation is employed.  Let ${\cal T}= \{{\r T}_i\mid i \in
\nat \}.$ Let $F({\cal T})$ be the set of all {\bf nonempty} and {\it finite}
subsets of ${\cal T}.$ This symbol has been used previously to include the 
empty set$,$ this set is now excluded. Now let ${\r A} \in F({\cal T}).$ 
Then there exists a finite choice set \ s \ such that $\r x \in \r s$ iff
there exists a unique ${\r T}_i \in {\r A}$ and $ \r x \in {\r T}_i.$ Now let the 
set $\cal C$ denote the set of all such finite choice sets. As to interpreting these results within 
the NSP-world$,$ the following is essential. Within nonstandard analysis the 
term  ``hyper'' is often used for the result of the * map. For example$,$ you 
have $\hyperreal$ as the hyperreals since $\real$ is termed the real numbers. 
For certain$,$ but not all concepts$,$ the term ``hyper'' or the corresponding 
* notation will be universally replaced by the term ``ultra.'' Thus$,$ certain 
purely  subtle 
words or *-words become ``ultrawords'' within the developmental paradigm 
interpretation. [Note: such a word was previously called a superword.] Of 
course$,$ for other scientific or philosophical systems$,$ such abstract 
mathematical objects can be reinterpreted by an appropriate technical term 
taken from those disciplines. \pars
As usual$,$ we are working within any enlargement and all of the 
above intuitive objects are embedded into the G-structure. Recall$,$ that to 
simplify 
expressions$,$ we often suppress within our first-order statements a specific 
superstructure element that bounds a specific quantifier.
[Note 2 MAY 1998: The material between the [[ and the ]] has been altered from the original that appeared in the 1993 revision.] [[Although theorem 7.2.1 may not be insignificant$,$ it is also not necessary for the other portions of this research. The general axiom of choice can be applied to generate formally developmental paradigms. \pars

\hrule\smallskip\hrule
\smallskip
The $\cal S$ in Theorem 7.2.1 determines a special intrinsic ultranatural selection 
process 
(i.e. {\bf  IUN-selection process}). 
This process is obtained by application of  hyperfinite choice.
\par
\smallskip
\hrule\smallskip\hrule
\medskip

{\bf Theorem 7.2.1} 
{\sl Let $\emptyset \not= \gamma \subset \nat$ and $\widetilde{\cal T} = \{\b T_i 
\mid i \in \gamma \}.$  There exists a set of 
sets $\cal S$ determined by hyperfinite set $Q$ and hyper
finite choice defined on $Q$ such that:\pars {\rm 
(i )} $s^\prime  \in \cal S$ iff for each $ \b T \in \widetilde{\cal T}$ there 
is one and only one $[g] \in \Hyper {\b T}$ such that $[g] \in s^\prime,$  
 and if $x \in s',$ then there is some $ \b T \in \widetilde{\cal T}$ and some $[g] \in \Hyper {\b T}$ such that $x = [g].$ 
{\rm (}If 
$\Hyper [g] \in {^\sigma {\b T}},$ then $[g] = [f] \in \b T.${\rm)}} \pars 
Proof. (i) Let $A \in F(\widetilde{\cal T}).$ Then from the definition of 
$\widetilde{\cal T},$ there exists some $ n \in \nat$ such that $A = \{
\b T_{j_i} \mid i = 0, \ldots, n \land j_i \in \nat \}.$ From the definition of 
$\b T_k,$ each $\b T_k$ is denumerable. Notice that any $[f] \in \b T_k$ is 
associated with a unique member of $i[{\cal W}].$ Simply consider the unique $f_0 \in 
[f].$ The unique member of $i[{\cal W}]$ is by definition $f_0(0).$ Thus each member of 
$\b T_k$ can be specifically identified. Hence$,$ for each $\b T_i$ there is a 
denumerable $M_i \subset \nat$ and a bijection $h_i\colon M_i \to \b T_i$ such 
that $a_i \in \b T_i$ iff there is a $k_i \in M_i$ and $h_i(k_i) = a_i.$ 
Consequently$,$ for each $i = 0,\ldots,n$ and $a_{j_i} \in \b T_{j_i},$ we have 
that $h_{j_i}(k_{j_i}) = a_{j_i},$ and conversely for each $i =0,\dots,n$ and 
$k_{j_i} \in M_{j_i},\ h_{j_i}(k_{j_i}) \in \b T_{j_i}.$ Obviously$,$ 
$\{h_{j_i}(k_{j_i}) \mid i = 0, \ldots, n \}$ is a finite choice set.
All of the above may be translated into the following sentence that holds in 
${\cal M}.$ (Note: Choice sets are usually considered as the range of choice functions. Further$,$ ``bounded formula simplification'' has been used.)\pars
\line{\hfil $ \forall y(y \in F(\widetilde{\cal T}) \to \exists s((s \in \power {{\cal E}})\land \forall 
x((x \in y) \to \exists z((z \in x) \land (z \in s) \land $\hfil}\pars
\line{\hfil $\forall w(w \in {\cal E} \to ((w  \in s) \land(w \in x) \iff (w = z)))))\land$\hfil}\pars
\line{(7.2.1)\hfil$\forall u(u \in {\cal E} \to ((u \in s) \iff \exists x_1((x_1 \in y)\land (u \in x_1))))))$\qquad\hfil }\parm
\noindent For each $A \in F(\widetilde{\cal T}),$ let $S_A$ be the set of all 
such choice sets generated by the predicate that follows the first $\to$ formed from (7.2.1) by deleting 
the $\exists s$ and letting $y = A.$ Of course$,$ this set exists within our set 
theory. Now let ${\cal C} = \{S_A \mid A \in F(\widetilde{\cal T})\}.$ \pars
Consider $\Hyper {\cal C}$ and $\Hyper (S_A).$ Then $s \in \Hyper (S_A)$ iff 
$s$ satisfies (7.2.1) as interpreted in $\Hyper {\cal M}.$ Since we are 
working in an enlargement$,$ there exists an internal $Q \in \Hyper 
(F(\widetilde{\cal T}))$ such that $^\sigma{\widetilde{\cal T}} \subset Q 
\subset \Hyper {\widetilde{\cal T}}.$ Recall that $^\sigma{\widetilde{\cal T}} 
=\{\Hyper {\b T} \mid \b T \in \widetilde{\cal T}\}.$ Also $^\sigma{\b T} 
\subset \Hyper {\b T}$ for each $\b T \in \widetilde{\cal T}.$  From the 
definition of $\Hyper {\cal C},$ there is an internal set $S_Q$ and  $s \in  
S_Q$ iff $s$ satisfies the internal defining predicate for members of $S_Q$
and this set is the set of all such $s.$ $(\Rightarrow)$ Consequently$,$ 
since for each $\b T\in \widetilde{\cal T},\ \Hyper {\b T} \in Q$$,$ then the 
generally external $s^\prime = \{ s\cap \Hyper {\b T}\mid \b T \in 
\widetilde{\cal T}\}$ satisfies the $\Rightarrow$ for (i). Note$,$ however$,$ that for 
$\Hyper {\b T},\ \b T \in \widetilde{\cal T},$ it is possible that 
$s \cap \Hyper {\b T} = \{\Hyper {[f]}\}$ and $\Hyper {[f]}\in {^\sigma{\b 
T}}.$ In this case$,$ by the finiteness of $[f]$ it follows that $[f] = \Hyper 
{[f]}$ implies that $s \cap\Hyper {\b T} = \{[f]\}.$ Now let
${\cal S} = \{s^\prime \mid s \in S_Q\}.$ In general$,$ $\cal S$ is an external 
object. \pars
($\Leftarrow$) Consider the internal set $S_Q.$  Let $s^\prime$ be the set as defined by the right-hand side of (i). For each internal $x \in s^\prime$
and applying$,$ if necessary$,$ the *-axiom of choice for *-finite sets$,$ we have the internal set $A_x = \{y\mid (y \in S_Q)\land (x \in y)\}$ is nonempty. The set $\{A_x\mid x \in s^\prime\}$ has the finite intersection property. For$,$ let nonempty internal $B=\{x_1,\ldots, x_n\}.$ Then the set $A_B =\{y\mid (y \in S_Q)\land (x_1 \in y)\cdots \land (x_n \in y)\}$ is internal and nonempty by the *-axiom of choice for *-finite sets. Since we are in an enlargement and $s^\prime$ is countable$,$ then $D=\bigcap\{A_x\mid x \in s^\prime\} \not= \emptyset.$ Now take any $s\in D.$ Then $s \in S_Q$ and from the definition of ${\cal S},\ s^\prime \in {\cal S}$. This completes the proof.\qed 
[Note: Theorem 7.2.1 may be used to model physical developmental paradigms associated with event sequences.]
\par
Although it is not necessary$,$ for this particular investigation$,$ the set $\cal S$ may  be considered               
{{\it a set of all developmental paradigms.}}  Apparently$,$  $\cal S$ 
contains every possible developmental paradigm for all possible frozen segments
and $\cal S$ contains paradigms for any *-totality $\Hyper {\b T}.$ There are 
*-frozen segments contained in various $s^\prime$ that can be assumed to be 
unreadable sentences since $^\sigma{\b T} \not= \Hyper {\b T}.$]]\pars
Let $A \in F(\widetilde{\cal T})$ and $M(A)$ be a subset of $S_A$ for which 
there exists a written set of rules that selects some specific member of 
$S_A.$ Obviously$,$ this may be modeled by means of functional relations. 
First$,$ $M(A) \subset S_A$ and it follows$,$ from the difference in 
cardinalities$,$ that 
there are infinitely many members of $\Hyper (S_A)$ for which there does not 
exist a readable rule that will select such members. However$,$ this does not 
preclude the possibility that there is a set of purely unreadable sentences 
that do determine a specific member of 
$\hyper {S_A} - {^\sigma M(A)}.$ This might come about in the following 
manner. Suppose that $H$ is an infinite set of formal sentences that is 
interpreted to be a set of rules for the selection of distinct members of $M(A).$ 
Suppose we have a bijection $h\colon M(A) \to \b H$ that represents this 
selection process. Let $\Hyper {\cal M}$ be at least a polysaturated 
enlargement of $\cal M,$ and consider $^\sigma f: {^\sigma(M(A))} \to {^\sigma 
{\b H}}.$ The map $^\sigma f$ is also a bijection and $^\sigma f: 
{^\sigma(M(A))} \to \Hyper {\b H}.$ Since $\vert {^\sigma(M(A))} \vert <
\vert {\cal M} \vert,$ it is  well-known that there exists an internal map 
$h\colon A^\prime \to \Hyper {\b H}$ such that $h\mid {^\sigma (M(A))} = {^\sigma 
f},$
 and $A^\prime,\ h[A^\prime]$ are internal. Further$,$ for internal $A^\prime \cap 
\Hyper {(S_A)} =B,\ {^\sigma (M(A))} \subset B.$ However$,$ $^\sigma (M(A))$ is 
external.  This yields that $h$ is defined on $B$ and 
$B\cap(\Hyper {S_A} - {^\sigma (M(A))}) \not= \emptyset.$ Also$,$ $^\sigma {\b 
H} \subset h[B] \subset \Hyper {\b H}$ implies$,$ since $h[B]$ is 
internal$,$ that $^\sigma H \not= h[B].$ Consequently$,$ in this case$,$ $h[B]$ may 
be interpreted as a set of *-rules
that determine the selection of members of 
$B.$ That is to say that there is some $[g] \in h[B] - {^\sigma H}$ and a $[k] 
\in \Hyper {S_A} - {^\sigma (M(A))}$ such that $([k],[g]) \in h.$  As it will 
be shown in the next section$,$ the set $H$ can be so constructed that if $[g] 
\in h[B] - {^\sigma H},$ then $[g]$ is unreadable. \parm
\leftline{\bf 7.3 Ultrawords}
\medskip
Ordinary propositional logic is not compatible with deductive quantum logic$,$ 
intuitionistic logic$,$ among others. In this section$,$ a subsystem of 
propositional logic is investigated which rectifies this incompatibility. I 
remark that when a standard propositional language L or an informal language 
P isomorphic to L is considered$,$ it will always be the case that the L or P is  
minimal relative to its applications. This signifies that if L or P is 
employed  
in our investigation for a developmental paradigm$,$ then L or P is constructed 
only from those distinct propositional atoms that correspond to distinct 
members of d$,$ etc. The same minimizing process is always assumed for the 
following constructions. \pars
Let B be a formal or$,$ informal nonempty set of propositions. Construct the 
language $\r P_0$  in the usual manner from B (with superfluous parentheses 
removed) so that $\r P_0$ forms the smallest set of formulas that contains B
and such that $\r P_0$ is closed under the two binary operations $\land$ and 
$\to$ as they are formally or informally expressed. Of course$,$ this language 
may be constructed inductively or by letting $\r P_0$ be the intersection of 
all collections of such formula closed under $\land$ and $\to.$ \pars
We now define the deductive system $S.$  Assume substitutivity$,$
parenthesis 
reduction and the like.  Let $\r d 
= \{\r F_i \mid i \in \nat \}= \r B$ be a 
development paradigm$,$ where each $\r F_i$ is a 
readable frozen segment and describes the behavior of a Natural-system over a 
time subinterval. Let the set of axioms be the schemata\parm
\line{(1) \hfil $({\cal A} \land {\cal B}) \to {\cal A},\ {\cal A} \in \r B$\hfil}\pars
\line{(2)\hfil $({\cal A} \land {\cal B}) \to {\cal B}$ \hfil }\pars
\line{(3)\hfil ${\cal A} \land ({\cal B} \land {\cal C}) \to ({\cal A}\land 
{\cal B}) \land {\cal C},$ \hfil }\pars
\line{(4)\hfil $ ({\cal A} \land {\cal B}) \land {\cal C} \to {\cal A} \land ({\cal 
B} \land {\cal C}).$ \hfil}\parm
If $\r P_0$ is considered as informal$,$ which appears to be necessary for some 
applications$,$ where the parentheses are replaced by the concept of symbol 
strings being to the   ``left'' or   ``right'' of other symbol strings and the 
concept of strengths of connectives is used (i.e. $\r A \land \r B \to \r C$ 
means $((\r A \land \r B) \to \r C$)$,$ then axioms 3 -- 4 and the parentheses 
in (1) and (2) may be omitted. The one rule of inference is Modus Ponens (MP). 
Proofs or demonstrations from hypotheses $\Gamma$ contain finitely many steps$,$ 
hypotheses may be inserted as steps and the last step in the proof is either a 
theorem if $\Gamma = \emptyset$ or if $ \Gamma \not= \emptyset,$ then the last 
step is a consequence of ( a deduction from ) $\Gamma.$ Notice that repeated application of (4)
along with (MP) will allow all left parentheses to be shifted to the right with the exception of the
(suppressed) outermost left one. Thus this leads to the concept of left to right ordering of a
formula. This allows for the suppression of such parentheses. In all the following$,$ this
suppression will be done and replaced with formula left to right ordering.  \pars
For each $\Gamma \subset \r P_0,$  let $S(\Gamma)$ denote the set of all 
formal theorems and consequences obtained from the above defined system $S.$ 
Since hypotheses may be inserted$,$ for each $\Gamma \subset \r P_0,\ 
\Gamma \subset S(\Gamma) \subset \r P_0.$ This implies that $S(\Gamma) \subset 
S(S(\Gamma)).$ So$,$ let $\r A  \in S(S(\Gamma)).$ The general concept of 
combining together finitely many steps from various proofs to yield another 
formal proof leads to the result that $\r A \in S(\Gamma).$ Therefore$,$ 
$S(\Gamma) = S(S(\Gamma)).$ Finally$,$ the finite step requirement also yields 
the result that if $\r A \in S(\Gamma),$ then there exists a finite $\r F 
\subset \Gamma$ such that $\r A \in S(\r F).$ Consequently$,$ $S$ is a finitary 
consequence operator and observe that if $C$ is the propositional consequence 
operator$,$ then $S(\Gamma) {\subset\atop \not=} C(\Gamma).$ Of course$,$ we may 
now apply the nonstandard theory of consequence operators to $S.$ \pars
It is well-known that the axiom schemata chosen for $S$ are theorems in 
intuitionistic logic. Now consider quantum logic with the {Mittelstaedt 
conditional} $i_1(\r A,\r B) = \r A^\perp \lor (\r A \land \r B).$ [1]  Notice 
that $i_1(\r A \land  \r B,\r B)= (\r A \land \r B)^\perp \lor ((\r A \land \r B) 
\land \r B)= (\r A \land \r B)^\perp \lor (\r A \land  \r B)= I$ ( the upper 
unit.) Then $i_1((\r A \land \r B),\r A)= (\r A \land \r B)^\perp \lor
((\r A \land \r B) \land  \r A) = (\r A \land \r B)^\perp \lor (\r A \land \r 
B) = I;\ i_1((\r A \land \r B) \land \r C,\r A \land (\r B \land \r C)) = 
((\r A \land \r B) \land \r C)^\perp \lor (\r A \land (\r B \land \r C)) = I=
i_1(\r A \land (\r B \land \r C), (\r A \land \r B) \land \r C).$ Thus with 
respect to the interpretation of ${\cal A} \to {\cal B}$ as  conditional  
$i_1$ the axiom schemata  for the system $S$ are theorems and the system $S$ 
is compatible with deductive quantum logic under the Mittelstaedt 
conditional.\pars
In what follows$,$ assume $\land$ is interpreted as $\sp {\rm and}\sp.$ Recall that $\r d 
= \{\r F_i \mid i \in \nat \}= \r B$  is a 
development paradigm$,$ where each $\r F_i$ is a 
readable frozen segment and describes the behavior of a Natural-system over a 
time subinterval.  Although it is sufficient in most cases to consider a 
formal language$,$ it is more convenient to employ an isomorphic informal 
language. This 
is especially necessary if one wishes to more closely analyze certain special 
subtle objects. Let $\r M_0 = \r d.$ Define $\r M_1 = \{\r F_0\sp{\rm and}\sp\r 
F_1\}.$ Assume that $\r M_n$ is defined. Define $\r M_{n+1} = \{\r x\sp{\rm 
and}\sp\r F_{n+1}\mid \r x \in \r M_n\}.$ From the fact that $\r d$ is a 
developmental paradigm$,$ where the last two symbols in each member of $\r d$ is 
the time indicator  ``i.''$,$ it follows that no member of $\r d$ is a member 
of $\r M_n$ for $n > 0.$ Now let ${\rm M_d} = \bigcup \{\r M_n \mid n 
\in \nat\}.$ Intuitively$,$ $\sp{\rm and}\sp$ behaves as a conjunction and each 
$\r F_i$ as an atom within our language. Notice the important formal demonstration fact that for
an hypothesis consisting of any member of $\r M_n,\ n>0,$ repeated applications of
(1)$,$ (MP)$,$ (2)$,$ (MP) will lead to the members of $\r d$ appearing in the proper time ordering at
increasing (formal) demonstration step numbers.\pars
\hrule\smallskip\hrule
\smallskip
The next theorem shows the existence of the most basic ultraword $w.$ The 
$\Hyper {\b S}$ is the basic compatible ultralogic. This ultraword and the 
corresponding ultralogic generate a selected Natural event sequence along with 
numerous ultranatural events.
\pars
\hrule\smallskip\hrule
\medskip  

{\bf Theorem 7.3.1} {\sl For $\b d = \{\b F_i \mid i\in \nat \},$ there exists 
an ultraword $w \in \Hyper {\bf M_d} - \Hyper {\b d}$ such that $\b F_i \in 
\Hyper {\b S}(\{w\})$ {\rm (}i.e. $w \Hyper {\vdash_S} \b F_i{\rm )}$ for each $i \in 
\nat.$}\pars 
Proof. Consider the binary relation $G = \{(x,y)\mid (x \in \b d) \land (y \in 
{\bf M_d - d}) \land ( x \in \b S(\{y\}) \}.$ Suppose that $\{(x_1,y_1),\ldots 
(x_n,y_n)\}\subset G.$ For each $i=1,\ldots,n$ there is a unique $k_i \in \nat$ such 
that $x_i = \b F_{k_i}.$ Let $m = \max \{k_i\mid (x_i = \b F_{k_i})\land
(i = 1,\ldots,n) \}.$ Let $b \in \b M_{m+1}.$ It follows immediately that 
$x_i \in \b S(\{b\})$ for each $i = 1,\ldots, n$ and$,$ from the construction of
$\r d,\ b \notin \b d.$ Thus $\{(x_1,b),\ldots,(x_n,b)\} \subset G.$  
Consequently$,$ $G$ is a concurrent relation. Hence$,$ there exists some $w \in 
\Hyper {\bf M_d}- \Hyper {\b d}$ such that $^\sigma \b F_i = \b F_i \in \Hyper {\b 
S}(\{w\})$ for each $i \in \nat.$ This completes the proof. [See note 3.]\qed
Observe that $w$ in Theorem 7.3.1 has all of the formally expressible 
properties of a readable word. For example$,$ $w$ has a hyperfinite 
length$,$ among other properties. However$,$ since $\r d$ is a denumerable set$,$ 
each ultraword has a very special property. \pars
Recall that for each $[g] \in \cal E$ there exists a unique $m \in \nat$ and 
$f^\prime \in  T^m$ such that $[f^\prime] = [g]$ and for each $k$ such that $m 
< k \in \nat,$ there does not exist $g^\prime \in T^k$ such that $[g^\prime] = 
[g].$ The function $f^\prime \in T^m$ determines all of the alphabet symbols$,$ 
the symbol used for the blank space$,$ and the like$,$ and determines there 
position within the intuitive word being represented by $[g].$ Also for each 
$j$ such that
$0\leq j\leq m,\ f^\prime(j) = i(\r a) \in i[{\cal W}] = T,$ where 
$i(\r a)$ is the ``encoding'' in $i[{\cal W}]$ of the symbol ``a''. For each $m \in 
\nat,$ let $P_m = \{f \mid ( f \in T^m) \land (\exists z ((z \in {\cal E}) 
\land (f \in z)\land \forall x((x \in \nat)\land (x > m)\to \neg \exists y((y \in T^x)\land 
(y \in z)))))\}.$  An element $n \in \Hyper{\cal W}$ is a {{\it subtle 
alphabet symbol}} if there exists $ m \in \nat$  and $f \in \Hyper (P_m),$ or
if $m \in \nat_\infty$ and $f \in P_m,$ and some $j \in \hypernat$ such that 
$f(j) = n.$ A symbol is a {{\it pure subtle alphabet symbol}} if 
$f(j) = n \notin i[{\cal W}].$ Subtle {alphabet symbols can be characterized}
in $\hyper {\cal E}$ for they are singleton objects. A $[g] \in \hyper {\cal 
E}$ represents a subtle alphabet symbol iff there exists some $f\in (\Hyper 
T)^0$ such that $[f] = [g]= [(0,f(0))],\ f = \{(0,f(0))\}.$ \parm

{\bf Theorem 7.3.2} {\sl Let $\r d = \{\r F_i \mid i\in \nat\}$ be a denumerable 
developmental paradigm. For an ultraword $ w = [g] \in \hyper {\cal E}$ that 
exists by Theorem 7.3.1$,$ there exists $\delta \in \nat_\infty$ and infinitely 
many $\nu \in \nat_\infty$ such that $[f] = [g],\  f\in P_\delta$ and each 
$f(\nu) \in \Hyper {({i[\r P_0]})} - ({i[\r P_0]})$ is a pure subtle alphabet symbol. }\pars
Proof. Since $\b d$ is denumerable$,$ consider a bijection $h\colon \nat \to \b 
d$ such that $h(n) = \b F_n = [(0,q_n)],\ q_n \in i[\r d]\subset i[\r P_0].$ 
From the 
definition of ${\bf M_d - d},$ if $[g] \in {\bf M_d - d},$ there exists a unique 
$m,\ n \in \nat \ (n\geq 1) $ and $f^\prime \in (i[\r P_0])^m$ such that 
$h[[0,n]] \subset \b S(\{[g]\})$ and (\b 0) $[f^\prime] = [g],$ and (\b 1) $m 
= 2n,$ (\b 2) for each even $2k\leq m, \ (k \geq 0),\ f^\prime(2k) = q_k \in 
i([\r P_0]) \subset i[{\cal W}],\ [(0,q_k)] \in \b d.$ All such $q_k$ are distinct 
and $[(0,q_k)] \in \b S(\{f^\prime]\}).$ For each odd $2k+1 \leq m,\ f^\prime 
(2k+1) = i(\sp{\rm and}\sp).$ (\b 3) For each $h(k) \in h[[0,n]]$ there exist 
distinct even $2k \leq m$ such that $f^\prime(2k) = 
q_k \in i[\r P_0] \subset i{\cal W}]$ and conversely$,$ and $h(k) = [(0,q_k)] \in \b d$ 
all such $q_k$ being distinct. Also note that for each $i \in \nat,\  h(i) \in 
\b S(\{[g]\})$ iff $h(i) \in h[[0,n]].$ (\b 4) There exists a unique $ r \in 
\nat$ such that $r > m$ and an $f" \in P_r$ such that $[f"] = [g].$  (\b 5) For 
each of the $n+1$ distinct $k$'s that exist from the first part of (\b 2)$,$ 
there  exists at least $n+1$ distinct $r_k \in \nat$ such that $0\leq r_n \leq 
r$ and $n+1$ distinct $f"(r_k) = i(\r b_k),$ where $\r b_k$ is a symbol 
representing the natural number that appears as the next to the last symbol in 
a member of $\r d.$ \pars
Let $w = [g] \in \Hyper {\bf M} - \Hyper {\b d} $ be the ultraword that exists 
by Theorem 7.3.1. Then there exists some $\delta \in \hypernat$ and a unique 
$\nu \in \hypernat$ as well as 
$f^\prime \in (\Hyper {\bf i[P_0]})^\delta   $ such that $\hyper h[[0,\nu]] 
\subset \Hyper {\b S}(\{[f^\prime]\}) = \Hyper {\b S}(\{[g]\})$ with the 
*-transfer of properties (\b 1) --- (\b 5). From Theorem 7.3.1$,$ 
$^\sigma(h[\nat]) \subset \Hyper {\b S}(\{[f^\prime]\})$ yields$,$ by 
application$,$ of (\b 3) that $h[\nat] =  {^\sigma(h[\nat])} \subset \hyper 
h[[0,\nu]].$ This implies that $\hyper h[[0,\nu]]$ is infinite and internal. 
From (\b 1)$,$ we obtain that $\delta \in \nat_\infty.$ From (\b 4)$,$ there 
exists a unique $\rho > \delta$$,$ hence $\rho \in \nat_\infty$$,$ and a unique
$f \in P_\delta$ such that $[f] = [g] = [f^\prime].$ From (\b 5)$,$ there exist 
at least $\nu$ *-distinct$,$ and hence distinct$,$ $\rho_k \in \hypernat$ such 
that the $\nu$ distinct $f(\rho_k) \in \Hyper (i[\r P_0]).$ Since $\Hyper 
{\cal M}$ is an ultrapower or ultralimit enlargement based upon $\nat$$,$ it 
follows that $\vert[0,\nu]\vert \geq 2^{\aleph_0}.$ Consequently$,$ the cardinality 
of the $\nu$  distinct $f(\rho_k)$ is greater than or equal to $2^{\aleph_0}.$ 
Since $i[{\cal W}]$ is denumerable$,$ the cardinality of the set of purely 
subtle alphabet symbols contained in the set of $\nu$ distinct $f(\rho_k) \geq 
2^{\aleph_0}.$ This complete the proof. \qed
With respect to the proof of Theorem 7.3.2$,$ the function $f$ determines the 
alphabet composition of the ultraword $w.$ The word $w$ is unreadable not only 
due to its infinite length but also due to the fact that it is composed of 
infinitely many purely subtle alphabet symbols.\pars
The developmental paradigm d utilized for the two previous theorems is 
composed entirely of readable sentences. We now investigate what happens if a 
developmental paradigm contains countably many unreadable sentences. Let the nonempty developmental paradigm $d^\prime$ be composed of at most 
countably many members of $\hyper {\cal E} - \cal E$ and let $d^\prime \subset 
\Hyper {\b B} \subset \Hyper {\b P_0}.$ Construct$,$ as previously$,$ the set 
$\rm M_B$ from $\r B,$ rather than from d and suppose that $\r B \cap 
\r M_i = \emptyset,\ i \not= 0.$ [This last requirement for $\r B$ can be achieved 
as 
follows: construct a special symbol not originally in $\cal A.$ Then this 
symbol along with $\cal A$ is considered the alphabet.  
Next only consider a $\r B$ that 
does not contain this special symbol within any of its members. Now use this 
special symbol consistently with or without the spacing symbol to construct 
$\r M_i\ i \not= 0.$ Of course $\land$ is interpreted as this special symbol 
with or without the spacing symbol in the axiom system $S$.] \par 

\smallskip
\hrule\smallskip\hrule
\smallskip
The $w$ that exists by the next theorem generates$,$ upon applying 
the ultralogic $\Hyper {\b S},$ any selected event sequence that contains not 
only previously selected Natural events but also selected UN-events.\par
\smallskip\hrule\smallskip\hrule
\medskip

{\bf Theorem 7.3.3} {\sl Let $d^\prime = \{[g_i]\mid i\in \nat  \}.$ Then 
there exists an ultraword $w\in \Hyper {\bf M_B} - \Hyper {\b B}$ such that for 
each $ i\in \nat,\ [g_i] \in \Hyper {\b S}(\{w\}).$}\pars
Proof. Consider the internal binary  relation $G = \{(x,y)\mid (x \in \Hyper 
{\b B})\land (y \in \Hyper {\bf M_B} - \Hyper {\b B})\land (x \in \Hyper {\b 
S}(\{y\})\}.$ Note that members of $d^\prime$ are members of $^\sigma {\cal E}$ 
or$,$ at the most$,$ denumerably many members of $\hyper {\cal E} -{^\sigma{\cal 
E}}.$ From the analysis in the proof of Theorem 7.3.1$,$ for a finite $\r F \subset \r B,$ 
there exists some $\r y \in {\rm M_B}- \r B$ such that $\r F 
\subset S(\{\r y\}).$ It follows by *-transfer
that if $F$ is a finite or *-finite subset of $\Hyper {\b B},$ then there exists 
some $y \in \Hyper {\bf M_B} - \Hyper {\b B}$ such that $F \subset \Hyper {\b 
S}(\{y\}).$ As in the proof of Theorem 7.3.1$,$ this yields that $G$ is at least 
concurrent on $\Hyper {\b B}.$ However$,$ $d^\prime \subset \Hyper {\b B}$ and  
$\vert d^\prime \vert \leq \aleph_0.$ From $\aleph_1$-saturation$,$ there exists 
some $w \in\Hyper {\bf M_B} -\Hyper {\b B}$ such that for each $[g_i] \in 
d^\prime,\ [g_i] \in \Hyper {\b S}(\{w\}).$ This completes the proof. \qed

Let $\emptyset \not= \lambda \subset \nat$ and ${\cal D}_j = \{d_{ij} \mid i \in \lambda \}$ and each $d_i \subset \Hyper {\b B}$ 
is considered to be a developmental paradigm either of type $d$ or type 
$d^\prime$ and $\r B \cap \r M_i = \emptyset,\ i \not=0.$ Notice that ${\cal D}_j$ may be either a finite or denumerable set and Theorem 7.3.1 holds for the case that $\r d \subset \r B,$ where $w \in \Hyper {\bf M_B} -\Hyper {\b B}.$ 
For each $d_{ij} \in {\cal D}_j,$ use the Axiom of Choice 
to select an ultraword $w_{ij} \in \Hyper {\bf M_B} - \Hyper {\b B}$ that exists 
by Theorems 7.3.1 (extended) and 7.3.3. Let $\{w_{ij} \mid i \in \lambda\}$ be such 
a set of ultrawords. \pars

\smallskip
\hrule\smallskip\hrule
\smallskip
This next theorem shows the existence of an ultimate ultraword that by 
application of the ultralogic yields all of the other needed ultrawords. Thus 
$w_i'$ generates$,$ after application of the ultralogic $\Hyper {\b S},$ 
a Natural-system's event sequence$,$ whether it contains only Natural events
or selected UN-events. Then $w'$ and $\Hyper {\b S(\{w'\})}$ solve Wheeler's  ``general grand unification problem.''\par
\smallskip
\hrule\smallskip\hrule
\medskip

{\bf Theorem 7.3.4} {\sl There exists an ultimate ultraword $w_j' \in \Hyper {\bf  
M_B} - \Hyper {\b B}$ such that for each $i \in \lambda,\ w_{ij} \in \Hyper {\b 
S}(\{w_j' \})$ and$,$ hence$,$ for each $d_{ij} \in {\cal D}_j,\ d_{ij} \subset \Hyper {\b S}(\{w_{ij}\})\subset \Hyper {\b S}(\{w_j'\}).$ }\pars
Proof. For each finite $\{\r F_1,\ldots,\r F_n\} \subset {\rm M_B} - {\r B}$ 
there is a natural number$,$ say $m,$ such that for $i = 1,\ldots, n,
\ \r F_i \in \r M_j$ for some $j \leq m.$ Hence$,$ taking $\r b \in \r M_{m+1},$ we 
obtain that each $\r F_i \in S(\{\r b\}).$ Observe that $\r b \notin {\r B.}$ 
By *-transfer$,$ it follows that the 
internal relation $G =\{(x,y)\mid (x \in \Hyper {\bf M_B} -\Hyper 
{\b B})\land (y \in \Hyper {\bf M_B} - \Hyper {\b B})\land (x \in \Hyper {\b 
S}(\{y\})\}$ is concurrent on internal $\Hyper {\bf M_B} -\Hyper 
{\b B}$ and $\{w_{ij}\mid i \in \lambda \} \subset \Hyper {\bf M_B} -\Hyper 
{\b B}.$ Again $\aleph_1$-saturation yields that there is some $w_j' \in 
\Hyper {\bf M_B} -\Hyper {\b B}$ such that for each $i \in \lambda,\ w_{ij} \in 
\Hyper {\b S}(\{w_j'\}).$ The last property is obtained from $d_{ij} \subset 
\Hyper {\b S}(\{w_{ij}\}) \subset \Hyper {\b S}(\Hyper {\b S}(\{w_j^\prime\}))= 
\Hyper {\b S}(\{w_j^\prime\})$ since $\{w_i\}$ is an internal subset of $\Hyper
{\b P_0}.$ This completes the proof.\qed 
{\bf Corollary 7.3.4.1} {\sl Let $\emptyset\not= \gamma \subset \nat.$ There exists an ultimate ultraword $w' \in \Hyper {\bf M_B} - \Hyper {\b B}$ such that for each $j \in \gamma,\ w_j' \in \Hyper {\b 
S}(\{w' \})$ and$,$ hence$,$ for each $d_{ij} \in \bigcup{\cal D}_j,\ d_{ij} \subset \Hyper {\b S}(\{w_j'\})\subset \Hyper {\b S}(\{w'\}).$ }\parm     
The same analysis used to obtain Theorem 7.3.2  
can be applied to the 
ultrawords of Theorems 7.3.3 and 7.3.4. \par
\bigskip
\leftline{\bf 7.4 Ultracontinuous Deduction}
\medskip
In 1968$,$ a special {topology on the set of all nonempty  subsets} of a given set 
$X$ was constructed and investigated by your author. We apply a similar 
topology to subsets of $\cal E.$ \pars
Suppose that nonempty $X \subset \cal E.$ Let $\tau$ be the discrete topology 
on $X.$ In order to topologize $\power X,$ proceed as follows: for each $G \in 
\tau,$ let $N(G) = \{A\mid (A \subset X)\land (A \subset G)\}=\power G.$ 
Consider ${\cal B} = \{N(G)\mid G \in \tau \}$  to be a base for a topology $\tau_1$ on $\power X.$ Let $A \in N(G_1) 
\cap N(G_1).$ The discrete topology implies that $N(A)$ is a base element and 
that $N(A) \subset N(G_1) \cap N(G_2).$ 
There is only one 
member of $\cal B$ that contains $X$ and this is $\power X.$ Thus if $\power 
X$ is covered by members of $\cal B,$  then $N(X)=\power X$ is one of these covering 
objects. Thus $(\power X, \tau_1)$ is a compact space. Further$,$ since 
$N(\emptyset) \subset N(G)$ for each $G \in \tau,$ the space  $(\power X, 
\tau_1)$ is connected. The topology $\tau_1$ is a special case of a more 
general topology with the same properties. [2] Suppose that $D \subset X.$ Let 
$D \in N(G)= \power G, G \in \tau.$ Then $D \in N(D) \subset N(G).$ This 
yields that the nonstandard monad is $\monad D = \bigcap \{\hyper N(G) \mid N(G) \in {\cal B}\}= \Hyper 
{(\power D)} = \Hyper {\power {\hyper D}}.$ \par

\smallskip
\hrule\smallskip\hrule
\smallskip
This next theorem shows that the ultralogic $\Hyper {\b S}$ produces Natural and 
ultranatural events in an ultracontinuous manner.\par
\smallskip
\hrule\smallskip\hrule
\medskip

{\bf Theorem 7.4.1} {\sl Any consequence operator $C \colon (\power X, 
\tau_1) \to (\power X,\tau_1)$ is continuous.}\pars
Proof. Let $A \in \power X$ and $H \in \Hyper {\b C}[\monad A].$ Then there 
exists some $B \in \monad A$ such that $\Hyper {\b C}(B) = H.$ Hence$,$ $B \in 
\Hyper {\power {\hyper A}}.$ By *-transfer of a basic property of our 
consequence operators$,$ $\Hyper {\b C}(B) \subset \Hyper {\b C}(\hyper A) = 
\Hyper {(\b C(A))}.$ Thus $\Hyper {(\b C(B))} \in \Hyper {(\power {\b 
C(A))}}$ implies that $\Hyper {\b C}(B) \in \monad A.$ Therefore$,$ $\Hyper {\b 
C}[\monad A] \subset \monad {[\b C(A)]}.$ Consequently$,$ $\b C$ is continuous. \qed
{\bf Corollary 7.4.1.1} {\sl For any $X \subset \cal E$$,$ and any consequence 
operator $\b C \colon \power X \to \power X,$ the map $\Hyper {\b C}\colon 
\Hyper {(\power X)} \to  \Hyper {(\power X)}$ is ultracontinuous.}\parm
{\bf Corollary 7.4.1.2} {\sl Let $\r d$ {\r [}resp. $d^\prime$$,$ $\r d$ or 
$d^\prime${\r ]} be a developmental 
paradigm as defined for Theorem 7.3.1 {\r [}resp. Theorem 7.3.3$,$ 7.3.4{\r ]}. Let 
$w$ be a 
ultraword that exists by Theorem 7.3.1 {\r [}resp Theorem 7.3.3$,$ 7.3.4{\r ]}. Then 
$\r d$ {\r [}resp. $d^\prime$$,$ $\r d$ or $d^\prime${\r ]} is obtained by means of a 
ultracontinuous subtle deductive process applied to $\{w\}.$ }\parm
Recall that in the real valued case$,$ a function $f \colon [a,b] \to \real$ is 
uniformly continuous on $[a,b]$ iff for each $p,\ q \in \Hyper {[a,b]}$ such 
that $ p -q \in \monad 0,$ then $f(p) - f(q) \in \monad 0.$ If $D \subset [a,b]$ 
is compact$,$ then $p,\ q \in \Hyper D$ and $p-q \in \monad 0$ imply that there is 
a standard $ r \in D$ such that $p, \ q \in \monad r.$ Also$,$ for each $ r\in D$ 
and any $p,\ q \in \monad r,$ it follows that $p-q \in  \monad r.$ Thus$,$ if 
compact $D \subset [a,b],$ then $f:D \to \real$ is uniformly continuous iff 
for every $r \in D$ and each $p,\ q \in \monad r,\ \hyper f(p),\ \hyper f(q) 
\in \monad {f(r)}.$ With this characterization in mind$,$ it is clear that any 
consequence operator $\b C \colon \power X \to \power X$ satisfies the 
following statement. For each $A \in \power X$ and each $ p,\ q \in \monad A,\ 
\Hyper {\b C(p)},\ \Hyper {\b C(q)} \in \monad {\b C(A)}.$\pars
From the above discussion$,$ one can think of ultracontinuity as being a type 
of {ultrauniform continuity}.\par
\bigskip
\leftline{\bf 7.5 Hypercontinuous Gluing}
\medskip 
There are various methods that can be used to investigate the behavior of 
adjacent frozen segments. All of these methods depend upon a significant 
result relative to discrete real or vector valued functions. The major goal in 
this section is to present a complete proof of this major result and to 
indicate how it is applied. \pars
First$,$ as our standard structure$,$ consider either the intuitive real numbers 
as atoms or axiomatically a standard structure with atoms ${\bf ZFR} = {\bf ZF} 
+ {\bf AC}  + {\cal W}({\rm atoms}) +A{\rm (atoms)} + \vert  A \vert = \r c,$ where $A$ 
is isomorphic to the real 
numbers and ${\cal W} \cap A = \emptyset.$  
 Then$,$ as done previously$,$ there is a model 
$\langle C,\in,= \rangle $ within our ${\bf ZF} + {\bf AC}$ 
model for  ${\bf ZFR},$ where $A$ has all of the 
ordered field properties as the real numbers. A superstructure $\langle {\cal 
R}, \in, = \rangle $ is constructed in the usual manner$,$ where the  superstructure 
$\langle {\cal N}, \in, = \rangle $ is   a substructure. Proceeding as in 
Chapter 2$,$ construct $\hyper {\cal M}_1 = \langle \Hyper {\cal R}, \in = 
\rangle$ and ${\cal Y}_1.$ The structure ${\cal Y}_1$ is called\break
 the {{\it
Extended Grundlegend Structure}} --- the EGS.  The Grundlegend
Structure is a substructure of ${\cal Y}_1.$\pars
It is important to realized in  what follows that the objects utilized for the 
G-structure {\it interpretations} are nonempty finite equivalence classes of 
partial sequences. Due to this fact$,$ the following results should not lead to 
ambiguous interpretations. \pars
As a preliminary to the technical aspects of this final section$,$ we 
introduce the following 
definition. A function $f\colon [a,b] \to \realp m$ is {{\it 
differentiable-C}} on $[a,b] $ if it is continuously differentiable on $ 
(a,b)$ except at finitely many removable discontinuities. This definition is 
extended to the end points $\{a,b\}$ by application of one-sided 
derivatives.  For any $[a,b],$ consider a partition $ P = 
\{a_0,a_1,\cdots,a_n,a_{n+1}\},\ n\geq 1,\ a = a_0,\ b = a_{n+1} $ and
$ a_{j-1} < a_j,\ 1\leq j\leq n+1.$ For any such
partition $ P,$ let the real valued function $ g $ be
defined on the set $ D = [a_0,a_1)\cup  (a_1,a_2) \cup \cdots \cup 
(a_n,a_{n+1}] $ as follows: for each  $x\in [a_0,a_1),$ let $ g(x) = 
r_1 \in \real$; for each $ x \in (a_{j-1},a_j),$ let $ g(x) = r_j \in 
\real ,\ 1 < i \leq n$; for each $ x \in (a_n,b],$ let $ g(x) = r_{n+1} 
\in \real .$ 
It is obvious that $ g $ is a type of simple step 
function. Notationally$,$ let
$ {\cal F}(A,B) $  denote the set of all functions with domain $ A $
and codomain $ B.$  \par
\smallskip
\hrule\smallskip\hrule\smallskip
Theorem 7.5.1. shows that if a time fracture occurs of the minimum or intermediate type, then there exists an ultracontinuous, ultrauniform and ultrasmooth alteration process within the NSP-world that yields all of the alterations.\par
\smallskip
\hrule\smallskip\hrule
\medskip
{\bf Theorem 7.5.1} {\sl There exists a function 
$ G\in \Hyper ({\cal F}([a,b],\real))  $ with the following properties.\pars
\indent\indent (i) The function $ G $ is *-continuously *-differentiable and *-uniformly
*-continuous on $ \hyper {\eskip [a,b]},$\hfil\break
\indent\indent (ii) for each odd $ n\in \hypernat ,\ (n \geq 3),\ G $ is *-
differentiable-C of order $ n $ on $ \hyper {\eskip [a,b]},$\hfil\break
\indent\indent (iii) for each even $ n\in \hypernat ,\  G $ is *-continuously 
*-differentiable in $\hyper {[a,b]} $ except at finitely many 
points,\hfil\break
\indent\indent (iv) if $ c = \min \{r_1,\cdots,r_{n+1}\},\ d=\max 
\{r_1,\cdots,r_{n+1}\} $$,$ then the range of $ G = \hyper {\eskip [c,d]},\ \st 
G $
at least maps $ D $ into $ [c,d] $ and $ (\st G)\vert D = g.$}\pars
   Proof. First$,$ for any real $ c,d,$ where $ d \not= 0,$ consider the 
finite set of functions $$h_j(x,c,d) = (1/2)(r_{j+1}-r_j)\Bigl(\sin \bigl((x 
-c)\pi/(2d)\bigr)+ 1\Bigr) + r_j,\eqno (7.5.1)$$
$1\leq j \leq n.$ Each $ h_j $ is continuously differentiable for any order 
at each $ x \in \real .$ Observe that for each odd $ m\in \nat,$ each
m'th derivative $ h_j^{(m)} $ is continuous at $ (c+d) $ and $ (c-d) $ 
and $ h_j^{(m)}(c+d) = h_j^{(m)}(c-d) = 0 $ for each $ j.$ \pars
Let positive $ \delta \in \monad 0.$ Consider the finite set of internal 
intervals $ \{[a_0,a_1 - \delta),(a_1 + \delta , a_2 - \delta),\cdots,
(a_n + \delta,b]\} $ obtained from the partition $ P.$ Denote these intervals 
in the expressed order by $ I_j,\ 1\leq j\leq n+1.$ Define the internal 
function $$G_1 =\{(x,r_1)\vert x\in I_1 \} \cup \cdots \cup \{(x,r_{n+1})\vert 
x\in I_{n+1} \}.\eqno (7.5.2)$$
Let internal $ I_j^\dagger = [a_j - \delta, a_j + \delta],\ 1 \leq j \leq n,$ 
and for each $ x\in I_j^\dagger,$ let internal      
$$G_j(x) = (1/2)(r_{j+1}-r_j)\Bigl(\hyper {\sin}
 \bigl((x-a_j)\pi/(2\delta)\bigr)+1 \Bigr) + r_j.\eqno (7.5.3)$$
Define the internal function
$$G_2 =\{(x,G_1(x))\vert x\in I_1^\dagger \}\cup \cdots \cup \{(x,G_n(x))\vert 
x\in I_n^\dagger \}.\eqno (7.5.4)$$
The final step is to define $ G = G_1 \cup G_2$. Then $ G\in \Hyper ({\cal 
F}([a,b],\real)).$\pars
By *-transfer$,$ the function $ G_1 $ has an internal *-continuous *-
derivative $ G_1^{(1)} $ such that $ G_1^{(1)}(x) = 0 $ for each 
$ x\in I_1 \cup\cdots\cup I_{n+1}.$ Applying *-transfer to the properties of 
the functions $ h_j(x,c,d),$ it follows that $ G_2 $ has a unique internal 
*-continuous *-derivative $$G_2^{(1)} =(1/(4\delta))(r_{j+1} - r_j)\pi
\Bigl(\hyper \cos \bigl( (x - a_j)\pi/(2\delta)\bigr)\Bigr)\eqno (7.5.5)$$
for each $ x\in I_1^\dagger \cup\cdots\cup I_n^\dagger.$  The results that 
the *-left limit for the internal $ G_1^{(1)} $ and the *-right limit for 
internal $ G_2^{(1)} $ at each $ a_j - \delta $ as well as the *-left 
limit of $ G_2^{(1)} $ and *-right limit of $ G_1^{(1)} $ at each $ a_j 
+\delta $ are equal to $ 0 $ and $ 0 = G_2^{(1)}(a_j - \delta) = 
G_2^{(1)}(a_j+\delta) $ imply that internal $ G $ has a *-continuous *-
derivative $ G^{(1)} = G_1^{(1)} \cup G_2^{(1)} $ defined on $ \hyper 
{\eskip [a,b]}.$\pars
A similar analysis and *-transfer yield that for each $ m \in \hypernat,\ m 
\geq 2,\ G $ has an internal *-continuous *-derivative $ G^{(m)} $ defined 
at each $ x\in \hyper {\eskip [a,b]} $ except at the points $ a_j\pm \delta $ 
whenever $ r_{j+1} \not= r_j.$ However$,$ it is obvious from the definition of 
the functions $ h_j $ that for each odd $ m\in \hypernat, \ m\geq 3,$ each 
internal $ G^{(m)} $ can be made *-continuous at each $ a_j\pm \delta $ by 
simply defining $ G^{(m)}(a_j\pm \delta) = 0 $ and with this parts (i)$,$ 
(ii)$,$ and (iii) are established.\pars
For part (iv)$,$ assume that $ r_j \leq r_{j+1}.$ From the definition of the 
functions $ h_j,$ it follows that for each $ x \in I_j\cup I_j^\dagger\cup
I_{j+1},\ r_j \leq G(x) \leq r_{j+1}.$ The nonstandard intermediate value 
theorem implies that $ G\bigl[ \hyper {\eskip [a_j,a_{j+1}]}\bigr] = \hyper 
{\eskip [r_j,r_{j+1}]} $ and in like manner for the case that $ r_{j+1}< r_j.$
Hence$,$ $ G\bigl[\hyper {\eskip [a,b]}\bigr] = \hyper {\eskip [c,d]}.$ Clearly$,$ $\st {\Hyper {D}} =
[a,b].$ If $ p\in D $ and $ x\in \monad p \cap {\Hyper {D}} ,$ then $ G(x) = r_j=g(p) $
for some $ j $ such that $ 1 \leq j \leq n+1.$ This completes the 
proof. \qed
The nonstandard approximation theorem 7.5.1 can be extended easily to functions 
that map $ D $ into $\realp m.$ For example$,$ assume that $ F\colon D \to 
\realp 3,$  the component functions $ F_1,\  F_2 $ are continuously 
differentiable on $ [a,b];$ but that $ F_3 $ is a $ g $ type step 
function on $ D.$ Then letting $ H = (\hyper F_1,\hyper F_2,G),$
on $ \hyper {\eskip [a,b]},$ where $ G $ is defined in Theorem 4.1$,$ we 
have an internal *-continuously *-differentiable function $ H\colon \hyper {\eskip [a,b]} 
\to \hyperrealp 3,$ with the property that $ \st H\vert D = F.$\pars
With respect to Theorem 7.5.1$,$ it is interesting to note that if $h_j$ is 
defined on $\real,$ then for even orders $ n \in \nat,$ 
$$\vert\  h_j^{(n)}(c \pm d)\ \vert = \bigg|{{(r_{j+1} -r_j)\pi^n}\over {2^{n+1}d^n}}
\biggl| = 0 \leqno (7.5.6)$$
for $ r_{j+1} = r_j$ but not 0 otherwise. If $r_{j+1} - r_j \not= 0,$ then 
$G_2^{(n)}(a_j \pm \delta)$ is an infinite nonstandard real number. Indeed$,$ if 
$m_i$ is an increasing sequence of even numbers in $\hypernat$  and $r_{j+1} 
\not= r_j,$ then $\vert G_2^{(m_i)}(a_j \pm \delta) \vert$ forms a decreasing 
sequence of nonstandard infinite numbers. The next result is obvious from the 
previous result. \parm
{\bf Corollary 7.5.1.1} {\sl For each $ n \in \hypernat,$ then internal 
$G^{(n)} = G_1^{(n)} \cup G_2^{(n)}$ is *-bounded on $\Hyper [a,b].$}\parm
Let $D(a,b)$ be the set of all bounded and piecewise continuously 
differentiable functions defined on $[a,b].$  By considering 
all of the possible (finitely many) subintervals$,$ where $f \in D(a,b),$ it 
follows from the Riemann sum approach that for each real $\nu > 0,$ there 
exists a real $\nu_1 > 0$ such that for each real $\nu_i,\ 0<\nu_i < \nu_1,$ 
a sequence of partitions $P_i = \{a = b_0^i < \cdots < b_{k_i}^i = b \}$ can 
be selected such that the ${\rm mesh}(P_i) \leq \nu_i$ and 
$$ \vert (f(b) - f(a)) - \sum_{n=1}^{k_i} f^\prime(t_n)(b_n^i - b_{n-1}^i) 
\vert < \nu \leqno (7.5.7)$$ for any $t_n \in (b_{n-1}^i,b_n^i).$ \pars
Moreover$,$ for any given number $M,$ the sequence of partitions can be so 
constructed such that there exists a $j$ such that for each $i > j,\ k_i >M,$ 
where $P_i$ and $P_j$ are partitions within the sequence of partitions. By 
*-transfer of these facts and by application of Theorem 7.5.1 and its 
corollary we have the next result.\parm
{\bf Corollary 7.5.1.2} {\sl For each $n \in \nat$ and each internal 
$G^{(n)},$ 
the difference $G^{(n)}(b) - G^{(n)}(a)$ is infinitesimally close to an 
(externally) infinity *-finite sum of infinitesimals.}\parm
A developmental paradigm is a very general object and$,$ therefore$,$  can be 
used for numerous applications. At present$,$ developmental paradigms are still 
being viewed from the {{\it substratum}} or external world. For what follows$,$ 
it is assumed that a developmental paradigm d traces the evolutionary history 
of a specifically named natural system or systems. In this first application$,$ 
let each $\r F_i \in \r d$ have the following property ({\b P}).\pars
 {\leftskip=0.5in \rightskip= 0.5in \noindent  $\r F_i$ describes 
``the general behavior and characteristics of the named natural system $S_1$ 
as well as the behavior and characteristics of named constituents contained 
within $S_1$ at time $t_i.$''\par}\pars
Recall that for $\r F_i,\ \r F_{i+1} \in \r d,$ there exist unique functions 
$f_0 \in \b F_i = [f],\ g_0 \in \b F_{i+1} = [g]$ such that $f_0,\ g_0 \in 
T^0$ and $\{(0,f_0(0))\} \in [f],\ \{(0,g_0(0))\} \in [g].$ Thus$,$ to each 
$\r F_j \in \r d,$ correspond the unique natural number $f_0(0).$ Let $D = 
[t_{i-1}, t_i) \cup (t_i,t_{i+1}]$ and define $f_1\colon D \to \nat$ as 
follows: for each $x \in  [t_{i-1}, t_i),$ let $f_1(x) = f_0(0);$ for each $ x 
\in (t_i,t_{i+1}],$ let $f_1(x) = g_0(0).$ Application of theorem 7.5.1 yields 
the internal function $G$ such that $G\vert D = f_1.$ For these physical 
applications$,$ utilize the term ``substratum'' in the place of the technical 
terms  ``pure nonstandard.'' [Note: Of course$,$ elsewhere$,$ the term  
``pure NSP-world'' or simply the ``{NSP-world}'' is used as a specific name for 
what has here been declared as the substratum.] This yields the following 
statements$,$ where the symbols $\r F_i$ and $\r F_{i+1}$ are defined and 
characterized by the expression inside the quotation marks in property (\b 
P).\parm
{\leftskip=0.5in \rightskip=0.5in \noindent (\b A): There exists a substratum 
hypercontinuous$,$ hypersmooth$,$ hyperuniform process $G$ that 
binds together $\r F_i$ and $\r F_{i+1}.$\par}\pars
{\leftskip=0.5in \rightskip=0.5in \noindent (\b B): There exists a substratum 
hypercontinuous$,$ hypersmooth$,$ hyperuniform alteration process $G$ that 
transforms $\r F_i$ into $\r F_{i+1}.$\par}\pars
{\leftskip=0.5in \rightskip=0.5in \noindent (\b C): There exists an 
ultracontinuous  subtle force-like (i.e. deductive) process that yields 
$\r F_i$ for each time $t_i$ within the development of the natural 
system.\par}\parm
In order to justify (\b A) and (\b B)$,$ specific measures of physical properties 
associated with constituents may be coupled together. Assume that for a subword 
$\r r_i \in F_i \in \r d,$ the symbols $\r r_i$ denote a numerical quantity 
that aids in characterizing the behavior of an object in a system $S_1$ or the 
system itself.  Let $(\b M_1)$ be the statement:\parm
{\leftskip=0.5in \rightskip=0.5in \noindent  ``There exists a substratum  
hypercontinuous$,$ hypersmooth$,$ hyperuniform functional process $G_i$ such that 
$G_i$ when restricted to the standard mathematical domain it is $f_i$ and such 
that $G_i$ hypercontinuously changes $r_i$ for system $S_1$ at time $t_i$ into 
$r_{i+1}$ for system $S_1$ at time $t_{i+1}.$''\par}\parm
\noindent This modeling procedure yields the following interpretation:\parm
{\leftskip=0.5in \rightskip=0.5in \noindent (\b D) If there exists a continuous or 
uniform [resp. discrete] functional process $f_i$ that changes $r_i$ for $S_1$
at time $t_i$ into $r_{i+1}$ for $S_1$ at time $t_{i+1},$ then $(\b 
M_1).$\par}\parm
At a particular moment $t_i,$ two natural systems $S_1$ and $S_2$ may 
interface. More generally$,$ two very distinct developmental paradigms may  
exist one $\r d_1$ at times prior to $t_i$ (in the $t_i$ past) and one $\r d_2$ 
at time after $t_i$ (in the $t_i$ future). We might refer to the time $t_i$ as 
a {{\it standard time fracture.}} Consider the developmental paradigm 
$\r d_3 = \r d_1 \cup \r d_2.$ In this case$,$ the paradigms may be either of type
$\r d$ or $d^\prime.$ For the type $d^\prime,$ the corresponding system need 
not be considered a natural system but could be a pure substratum system.\pars
At $t_i$ an $\r F_i \in  \r d_3$ can be characterized by statement (\b P) 
(with the term natural removed if $\r F_i$ is a member of a $d^\prime$).  
In like manner$,$ $\r F_{i+1}$ at time $t_{i+1}$ can be characterized by (\b 
P). Statements (\b A)$,$ (\b B)$,$ (\b C) can now be applied to $\r d_3$ and a 
modified statement (\b D)$,$ where the second symbol string $S_1$ is changed to 
$S_2.$ Notice that this modeling applies to the actual human ability that only 
allows for two discrete descriptions to be given$,$ one for the interval 
$[t_{i-1},t_i)$ and one for the interval $(t_i,t_{i+1}].$ From the modeling 
viewpoint$,$ this is often sufficient since the length of the time intervals can 
be made smaller than Planck time. \pars
Recall that an analysis of the scientific method used in the investigation of 
natural system should take place exterior to the language used to describe the 
specific system development. Suppose that $\cal D$ is the language accepted 
for a scientific discipline and that within $\cal D$ various expressions from 
mathematical theories are used. Further$,$ suppose that enough of the modern 
theory of sets is employed so that the EGS can be constructed. The following 
statement would hold true for $\cal D.$ \parm
{\leftskip=0.5in \rightskip=0.5in \noindent {\it
If by application of first-order logic to a set of non-mathematical premises 
taken from $\cal D$ it is claimed that it is not logically possible for 
statements such as \r (\b A\r )$,$ \r (\b B\r )$,$ \r (\b C\r ) and \r (\b D\r ) to hold$,$ then the set of 
premises is inconsistent.}\par}\parm
\centerline{\bf CHAPTER 7 REFERENCES}
\medskip
\noindent {\bf  1} Beltrametti$,$ E. G.and G. Cassinelli$,$ The logic of quantum 
mechanics$,$ in {\it Encyclopedia of Mathematics and its Application}$,$ 
Vol. 15$,$ Addison-Wesley$,$ Reading$,$ 1981.\pars
\noindent {\bf 2} Herrmann$,$ R. A.$,$ {\it Some Characteristics of Topologies on 
Subsets of a Power Set}$,$ University Microfilm$,$ M-1469$,$ 1968.\pars 
\noindent {\bf 3} Herrmann$,$ R. A.$,$ {\it Einstein Corrected,} (1993) I. M. P. http://www.arXiv.org/abs/math/0312189 
\parm
\centerline{NOTES}\pars
\noindent [1] The actual members, $\rm F_i,$  of a developmental paradigm $\rm d$ need not be unique. However, the specific information contained in each readable word used for a specific $i \in \nat$  is unique. \pars
 \noindent [2] Depending upon the application, a single standard word may also be termed as an ultraword. \pars
\noindent [3] For a new more detailed method to obtain ultrawords for a refined developmental paradigm, see pages 4 -- 7 of http://arxiv.org/abs/math/0605120 

\vskip 0.25in
\line{\leaderfill}
\line{\leaderfill}
\vskip 0.25in

\leftline{\bf 9.1 An Extension.}\par 
\medskip
Although it is often not necessary$,$ we assume when its useful that we are 
working 
within the {EGS}. Further$,$ this structure is assumed to be $2^{\vert {\cal 
M}_1\vert}$-saturated$,$ where ${\cal M}_1 =\langle {\cal N}, \in,= \rangle$ and the ground set is ${\cal W}' \cup \real$, 
  even though such a degree of saturation can usually be 
reduced. Referring to the paragraph prior to Theorem 7.3.3$,$ it can be assumed 
that the developmental paradigm $d^\prime \subset \Hyper {\b B} \subset \Hyper
{\b P_0.}$ It is not assumed that such a developmental paradigm is obtained 
from the process discussed in Theorem 7.2.1$,$ although a modification of the 
proof of Theorem 7.2.1 appears possible in order to allow this method of 
selection. \parm
{\bf Theorem 9.1.1} {\sl Let $d^\prime = \{[g_i] \mid i \in \lambda \},\ \vert 
\lambda \vert < 2^{\vert {\cal M}_1 \vert}.$ There exists an ultraword $w \in 
\Hyper {\bf M_B} - \Hyper {\b B}$ such that for each $ i \in \lambda,\ 
[g_i] \in \Hyper {\b S}(\{w\}).$}\pars
Proof. The same as Theorem 7.3.3 with the change in saturation.\qed
Let ${\cal D} = \{d_i \mid i \in \lambda\}, \ \vert \lambda \vert< 2^{\vert 
{\cal M}_1\vert},\ \vert d_i \vert < 2^{\vert {\cal M}_1\vert}$ and each $d_i 
\subset \Hyper {\b B}$ 
is considered to be a developmental paradigm either of type $d$ or type 
$d^\prime.$ 
For each $d_i \in \cal D,$ use the Axiom of Choice 
to select an ultraword $w_i \in \Hyper {\bf M_B} - \Hyper {\b B}$ that exists 
by Theorems 9.1.1. Let $\{w_i \mid i \in \lambda\}$ be such 
a set of ultrawords. \parm
{\bf Theorem 9.1.2} {\sl There exists an ultraword $w^\prime \in \Hyper {\bf  
M_B} - \Hyper {\b B}$ such that for each $i \in \lambda,\ w_i \in \Hyper {\b 
S}(\{w^\prime \})$ and$,$ hence$,$ for each $d_i \in {\cal D},\ d_i \subset \Hyper {\b 
S}(\{w^\prime\}).$ }\pars
Proof. The same as Theorem 7.3.4 with the change in saturation.\qed
\bigskip\vfil\eject
\leftline{\bf 9.2 NSP-World Alphabets.}
\medskip
First$,$ recall the following definition.
$P_m = \{f \mid ( f \in T^m) \land (\exists z ((z \in {\cal E}) 
\land (f \in z)\land \forall x((x \in \nat)\land (x > m)\to \neg \exists y((y \in T^x)\land 
(y \in z)))))\}.$ The set $T=i[{\cal W}].$ The set $P_m$ determines the 
unique partial sequence $f \in [g] \in {\cal E}$ that yields$,$ for 
each $ j \in \nat$ such that $0\leq j \leq m,\ f(j) = i(\r a),$ where 
$i(\r a)$ is an ``encoding'' in $i[{\cal W}]$ of the alphabet symbol  ``a'' used to 
construct our intuitive language $\cal W.$ The set $[g]$ represents an 
intuitive word constructed from such an {alphabet of symbols.}pars
Within the discipline of Mathematical Logic$,$ it is assumed
that there exists 
symbols --- a sequence of variables --- each one of which corresponds$,$ in a one-to-one 
manner$,$ to a natural number. Further$,$ under the subject matter of {generalized 
first-order theories} [2]$,$ it is 
assumed that the cardinality of the set of 
constants is greater than $\aleph_0.$ In the forthcoming investigation$,$ it may 
be useful to consider an alphabet that injectively corresponds to the real 
numbers $\real.$ This yields a new alphabet ${\cal A}^\prime$ containing 
our original 
alphabet. A new collection of words ${\cal W}^\prime$ composed of 
nonempty finite strings of such alphabet symbols may be constructed. It may also be 
useful to well-order $\real.$ The set $\cal E$ also exists with respect to the 
set of words ${\cal W}^\prime.$ Using the {ESG}$,$ many  
 previous results in this book now hold with respect to 
${\cal W}^\prime$ and for the case that we are working in a $2^{\vert{\cal 
M}_1\vert}$-saturated enlargement.\pars

 With respect to this {extended language}$,$ if you 
wish to except the possibility$,$ a definition as to what constitutes a {purely 
subtle alphabet symbol} would need to be altered in the obvious fashion. 
Indeed$,$ for $T$ in the definition of $P_m,$ we need to substitute $T^\prime = i[{\cal 
W}^\prime].$ Then the altered definition would read that $r \in \Hyper i[{\cal W}]$ is 
a {{\it pure subtle alphabet symbol}} if there exists an $m \in \nat$ and $f 
\in \Hyper {(P_m)},$ or if $m \in \hypernat - \nat$ an $f \in P_m,$ and some 
$j \in \hyperreal$ such that $f(j) = r \notin i[{\cal W}^\prime].$   Notice that if one chooses, then $r$ corresponds at an $r' \in \Hyper {\cal W}'.$   Further$,$ 
some of the previous theorems also hold when the proofs are modified.\pars

Although these extended languages are of interest to the mathematician$,$ most 
of science is content with approximating a real number by means of a rational 
number. In all that follows$,$ the cardinality of our language$,$ if not 
denumerable$,$ will  be 
specified.  All theorems from this book that are used to establish a result 
relative to a denumerable language will be stated without qualification. If 
a theorem has not been reestablished for a higher language but can be so 
reestablished$,$ then the theorem will be termed an {{\it extended}} theorem.
 \par
\bigskip
\leftline{\bf 9.3 General Paradigms.}
\medskip
There is the developmental paradigm$,$ and for nondetailed descriptions the 
{{\it general developmental paradigm.}} But now we have something totally new 
--- the {{\it general paradigm}}. It is important to note that the general 
paradigm is considered to be distinct from developmental paradigms$,$ although 
certain results that hold for general paradigms will hold for developmental 
paradigms and conversely. For example$,$ associated with each general paradigm 
${\rm G_A}$ 
is an {ultraword} $w_g$ such that the set ${\bf G_A} \subset  
\Hyper {\b S}(\{w_g\})$ and all other theorems relative to such 
ultrawords hold for general paradigms. The {general paradigm} is a collection 
of words that discuss$,$ in general$,$ the behavior of entities and other 
constituents of a natural system. They$,$ usually$,$ do not contain a time 
statement $\r W_i$ as it appears in section 7.1 for developmental paradigm 
descriptions. Our interest in this section is relative to only two such general 
paradigms. The reader can easily generate many other general paradigms.  
\pars
Let $\r c^\prime$ be a symbol that denotes some fixed real number and $ 
n^\prime $ a symbol that denotes a natural number. [Note: what follows is 
easily extended to an {extended language}.] Suppose that you have a theory 
which includes each member of the following set ($i$ suppressed.\parm
\line{\hfil ${\rm G_A} = \{{\rm An\sp elementary\sp particle\sp 
\alpha(\r n^\prime)\sp with\sp}$\hfil}\pars
\line{(9.3.1)\hfil  ${\rm kinetic\sp energy\sp \r c^\prime}{+}1/(\r n^\prime).\mid 
n \in G \land n \not= 0\},$\hfil}\parm
\noindent where $G$ is a denumerable subset of the real numbers. The set ${\rm G_A}$ is 
of particular interest when $G = \nat.$ 
Theories that 
include such sentences consider such particles as {{\it free in 
space.}}\pars
\hrule
\smallskip
\hrule
\smallskip
Theorem 9.3.1 shows that it is necessary for propertons to exist.\par
\smallskip
\hrule
\smallskip
\hrule
\smallskip
Of particular interest is the composition of members of $\Hyper {\bf G_A - 
G_A}.$  Notice that $\vert {\bf G_A} \vert = \vert G\vert$ 
since  $z_1,\ z_2 \in {\bf G_A}$  and $z_1 \not= z_2$ iff $[x_1] = z_1,\ 
[x_2] = z_2,\ x_1(30)= x_1(2) \not= x_2(30)= x_2(2),\ x_1(2),\ x_2(2) \in 
G.$ Now consider the bijection 
$K\colon {\bf G_A} \to G.$ \parm
{\bf Theorem 9.3.1} {\sl The set $[g] \in \Hyper {\bf G_A} - {\bf G_A}$ iff
there exists a unique  $f\in \Hyper (P_{55})$ and $\nu \in \Hyper G-G$ such that $[g] = [f],$ and 
$f(55) = i(\r A),\ f(54) = i(\r n),\ f(53) = i(\sp), \cdots, f(30) = f(2), 
\cdots, f(3) = i((), f(2)=\nu \in \Hyper G - G \subset \hyperreal - \real,\ 
f(1) 
= i()),\ f(0) = (.).$ }\pars
Proof. From the definition of $\rm G_A$ the sentence \pars
\line{\hfil $ \forall z(z \in {\cal E}\to (( z \in {\bf G_A}) \iff \exists ! x\exists ! w(
(w \in G) \land (x \in P_{55}) 
\land (x \in z) \land $\hfil}\pars
\line{\hfil $((55,i(\r A)) \in x) \land ((54,i(\r n)) \in x) \land \cdots 
\land (x(30) = x(2)) \land 
\cdots \land $ \hfil}\pars
\line{\hfil $((3,i(()) \in x)\land (x(2)=w)\land(K(z) = w) \land$\hfil}\pars 
\line{(9.3.2)\hfil$((1,i())) \in x) \land ((0,i(.)) \in x)))).$\qquad\hfil}\parm
\noindent holds in $\cal M,$ hence in $\Hyper {\cal M}.$ From the fact that 
$K$ is a bijection$,$ it follows that $\hyper K[\Hyper {\bf G_A} - {\bf G_A}] = 
\Hyper G -G \subset \hyperreal - \real.$ The result now follows from 
*-transfer. \qed
Using Theorem 9.3.1$,$
each member of $\Hyper 
{\bf G_A} - {\bf G_A},$ when interpreted by considering $i^{-1},$ has only two 
positions with a single missing object since positions 30 and 2 do not 
correspond to any symbol string in our language $\cal W.$ 
 This 
interpretation still retains a {vast amount of content}$,$ however. For a specific 
member$,$ you could substitute a {new constructed symbol}$,$ not in $\cal W,$ 
into these 
two missing positions. Depending upon what type of pure nonstandard number 
this inserted symbol represents$,$ the content of such a sentence could be 
startling. Let $\Gamma^\prime$ be a nonempty set of new symbols disjoint from $\cal 
W$ and assume that $\Gamma^\prime$ is injectively mapped by $H$ into 
$\Hyper G - G.$ \pars
Although human ability may preclude the actual construction of more 
than denumerably many new symbols$,$ you might consider this mapping to be 
onto if you accept the ideas of {extended languages} with a greater cardinality.  
As previously$,$ denote these new symbols by $\gamma^\prime.$  Now let \par
\line{\hfil ${\rm G_A^\prime} = \{{\rm An\sp elementary\sp particle\sp 
\alpha(\gamma\prime)\sp with\sp}$\hfil}\pars
\line{(9.3.3)\hfil ${\rm kinetic\sp energy\sp \r c^\prime}{+}1/(\gamma^\prime).\mid 
H(\gamma^\prime) \in \Hyper G - G\},$\qquad \hfil}\parm
\noindent This leads to the following interpretation stated in terms of describing 
sets.\parm
{\leftskip=0.5in \rightskip=0.5in \noindent (1) {\it The describing set $\rm G_A$ (mathematically) exists 
iff the describing set $\rm G_A^\prime$ (mathematically) exists.}\par}\par
\bigskip
\leftline{\bf 9.4 Interpretations}
\medskip
Recall that the Natural world portion of the NSP-world model may contain 
{{\it undetectable}} objects$,$ where  ``undetectable'' means that there does 
not appear to exist  human$,$ or humanly constructible machine sensors that 
directly detect the objects or directly measure any of the objects physical 
properties. The rules of the scientific method utilized within the 
micro-world of subatomic physics allow all such 
undetectable Natural objects to be accepted as existing reality.[1] The 
properties of such objects are indirectly deduced from the observed properties 
of gross matter. In order to have indirect evidence of the objectively real 
existence of such objects$,$ such indirectly obtained behavior will usually satisfy a 
specifically accepted model.\par
Although the numerical quantities associated with these undetectable Natural 
(i.e. standard) world objects$,$ if they really do exist$,$ cannot be directly 
measured$,$ these quantities are still represented by standard mathematical 
entities. By the rules of correspondence for interpreting pure NSP-world 
entities$,$ the members of ${\rm G_A^\prime}$ must be considered as {undetectable
pure NSP-world objects$,$} assuming any of them exist in this background 
world. On the other hand$,$ when viewed within the EGS$,$ any finite as well as 
many infinite subsets of $\rm 
G_A^\prime$ are internal sets. 
Consequently$,$ some finite collects of such objects {\it may be}  assumed to {indirectly 
effect behavior in the Natural world.} \par
The concept of {{\it realism}} often dictates that all interpreted members of 
a mathematical model be considered as existing in reality. The philosophy of 
science that accepts only {{\it partial realism}} allows for the following technique. 
One can stop at any point within a mathematically generated physical 
interpretation. 
Then proceed from that point to deduce an intuitive physical theory$,$ but only 
using other not interpreted mathematical formalism as {auxiliary constructs} or 
as {catalysts}.  With respect to the 
NSP-world$,$ another aspect of interpretation enters the picture. Assuming 
realism$,$ then the question remains which$,$ if any$,$ of these NSP-entities 
actually indirectly influence Natural world processes?  This interpretation 
process allows for the possibility that none of these pure NSP-world 
entities has 
any effect upon the standard world. These ideas should always be kept in 
mind.\parm
If you accept that such particles as described by $\rm G_A$ can exist in 
reality$,$ then the {philosophy of realism} leads to the next interpretation. 
\parm
{\leftskip 0.5in \rightskip 0.5in \noindent (2) {\it If there exist elementary 
particles with Natural-system behavior described by $\rm G_A$$,$  then there 
exist pure NSP-world objects that display within the NSP-world behavior 
described by members of $\rm G_A^\prime.$}\par}\parm
\noindent The concept of {absolute realism} would require that the 
acceptance of the elementary 
particles described by $\rm G_A$ is indirect evidence for the existence 
of the $\rm G_A^\prime$ described objects. I caution the reader that the 
interpretation we apply to such sets of sentences as $\rm G_A$ are only to be 
applied to such sets of sentences.\pars
The EGS may$,$ of course$,$ be interpreted in infinitely many different ways. 
Indeed$,$ the {NSP-world model with its physical-type language} can also be 
applied in infinitely many ways to infinitely many scenarios. I have applied it 
to such models as the {GGU-model}$,$ among others. In this section$,$ I consider 
another possible interpretation relative to those {Big Bang cosmologies} that 
postulate real 
objects at or near infinite temperature$,$ energy or pressure. These theories 
incorporate the concept of the {{\it initial singularity(ies).}} \pars
One of the great difficulties with many Big Bang cosmologies is that no 
meaningful physical interpretation for formation of the initial singularity is 
forthcoming from the theory itself. The fact that a proper and acceptable 
theory for {creation of the universe} requires that consideration not only be 
given to the moment of {zero cosmic time} but to what might have occurred 
``prior'' to that moment in the nontime period is what partially influenced 
 {Wheeler} to consider 
the concept of a  {{\it pregeometry.}}[3]$,$ [4] It is totally unsatisfactory 
to dismiss such questions as ``unmeaningful'' simply because they cannot be 
discussed in your favorite theory. Scientists must search for a {broader 
theory} 
to include not only the question but a possible answer.\pars

Although the {initial singularity} for a Big Bang type of state of affairs 
apparently cannot be discussed in a meaningful manner by many standard 
physical theories$,$ it can be discussed by application of our NSP-world 
language. Let $\r c^\prime$  be a symbol that represents any fixed real 
number. Define \parm
\line{\hfil ${\rm G_B} = \{{\rm An\sp elementary\sp particle\sp 
\alpha(\r n^\prime)\sp with\sp}$\hfil}\pars
\line{(9.4.1)\hfil ${\rm total\sp energy\sp \r c^\prime}{+}\r n^\prime.\mid 
n \in \nat\},$ \hfil}\parm
Application of Theorem 9.3.1 to $\rm G_B$ yields the set \parm
\line{\hfil ${\rm G_B^\prime} = \{{\rm An\sp elementary\sp particle\sp 
\alpha(\gamma^\prime)\sp with\sp}$\hfil}\pars
\line{(9.4.2)\hfil ${\rm total\sp energy\sp \r c^\prime}{+}\gamma^\prime.\mid 
\gamma \in \hypernat-\nat\},$ \hfil}\parm
{\leftskip 0.5in \rightskip 0.5in \noindent (3) {\it If there exist elementary 
particles with Natural-system behavior described by $\rm G_B$$,$  then there 
exist pure NSP-world objects that display within the NSP-world behavior 
described by members of $\rm G_B^\prime.$}\par}\parm
The particles being described by $\rm G_B^\prime$ have various {infinite 
energies}. These infinite energies {\bf do not} behave in the same manner as 
would the real number energy measures discussed in ${\rm G_B}.$ As is usual 
when a metalanguage physical theory is generated from a formalism$,$ we can 
further extend and investigate the properties of the $\rm G_B^\prime$ objects 
by imposing upon them the corresponding behavior of the {positive infinite 
hyperreal numbers}. This produces some interesting propositions. Hence$,$ we are 
able to use a nonstandard physical world language in order to give further 
insight into the state of affairs at or near a cosmic initial singularity. 
This gives {\it one} solution to a portion of the pregeometry problem. I point 
out that there are other NSP-world models for the beginnings of our universe$,$ 
if there was such a beginning. Of course$,$ the statements in $\rm G_B^\prime$  
need not be related at all to any Natural world physical scenario$,$ but could 
refer only to the behavior of pure NSP-world objects. \pars
Notice that Theorems such as 7.3.1 and 7.3.4 relative to the generation of 
developmental paradigms by ultrawords$,$ also apply to general paradigms$,$ where 
$\rm M_d,\ M_B, P_0$ are defined appropriately.   
The 
following is a slight extension of Theorem 7.3.2 for general paradigms. 
Theorem 9.4.1 will also hold for developmental paradigms.\pars

\hrule
\smallskip
\hrule
\smallskip
Theorem 9.4.1 shows that there exists an ultraword $w$ such that the basic 
ultralogic $\Hyper {\b S}$ when applied to $w$ yields not only a 
specific collection
of elementary particles but also a collection of propertons. Further$,$ it follows 
that there exists an ultraword $w^{\prime\prime}$ such that when $\Hyper {\b 
S}$ is applied to $w^{\prime\prime}$ the result is the collection 
all of the elementary particles 
that are claimed to exist and comprise the material portion of our 
universe.\par
\smallskip
\hrule
\smallskip
\hrule
\smallskip
{\bf Theorem 9.4.1} {\sl Let $\rm G_C$ be any denumerable general paradigm. 
Then there exists an ultraword  $w \in \Hyper {\b P_0}$ such that for each $\b 
F \in {\bf G_C},\ \b F \in \Hyper {\b S}(\{w\})$ and there exist infinitely 
many $[g] \in \Hyper {\bf G_C} - {\bf G_C} $ such that $[g] \in \Hyper {\b 
S}(\{w\}).$} \pars
Proof. In the proof of Theorem 7.3.2$,$ it is shown that there exists some $\nu 
\in \hypernat - \nat $ such that $\hyper h[[0,\nu]] \subset \Hyper {\b 
S}(\{w\})$  and $\hyper h[[0,\nu]] \subset \Hyper {\bf G_C}.$ Since 
$\vert \hyper h[[0,\nu]] \vert \geq 2^{\aleph_0},$ then $\vert \hyper h[[0,\nu]] 
-h[\nat] \vert \geq 2^{\aleph_0}$ for $h$ is a bijection. This completes the 
proof. \qed
{\bf Corollary 9.4.1.1} {\sl Theorem 9.4.1 holds$,$ where $\rm G_C$ is replaced 
by a developmental paradigm.}\parm
{\leftskip 0.5in \rightskip 0.5in \noindent (4) {\it Let $\rm G_C$ be a 
denumerable general paradigm. There exists an intrinsic ultranatural 
process$,$ $\Hyper {\b S},$ such that objects described by members of 
$\rm G_C$ are produced by $\Hyper {\b S}.$ 
During this production$,$ numerously many pure NSP-objects as described by 
statements in $\Hyper {\bf G_C} - {\bf G_C}$ are produced.}\par}\par

\bigskip
\line{\leaderfill}\smallskip
\centerline{\bf CHAPTER 9 REFERENCES}
\medskip
\noindent {\bf 1} Evans$,$ R.$,$ {\it The Atomic Nucleus}$,$ McGraw-Hill$,$ New York$,$ 
1955.\pars
\noindent {\bf 2} Mendelson$,$ E.$,$ {\it Introduction to Mathematical Logic}$,$ Ed. 
2$,$ D. Van Nostrand Co.$,$ New York$,$ 1979.\pars
\noindent {\bf 3} Misner$,$ Thorne and Wheeler$,$ {\it Gravitation}$,$ W. H. 
Freeman$,$ San Francisco$,$ 1973.\pars                                      \noindent {\bf 4} Patton C. and A. Wheeler$,$ Is physics legislated by 
cosmogony? in {\it Quantum Gravity}$,$ ed. Isham$,$ Penrose and Sciama$,$ Oxford 
University Press$,$ Oxford$,$ 1977$,$ 538--605.\pars
\bigskip
\leftline{\bf 10.1 More About Ultrawords.}\par 
\medskip
Previously$,$ we slightly investigated the composition of an ultraword
$w \in  \Hyper {\bf M} - \b d.$ Using the idea of the minimum informal language 
$\r P_0\subset \r P,$  where denumerable $\r B = \r d$ and $\r P$ is a propositional 
language$,$ our interest now lies in completely determining 
the composition of $\Hyper {\b S}(\{w\}).$ [Note: since our language is 
informal axiom (3) and (4) are redundant in that superfluous parentheses have 
been removed.] First$,$ two defined sets.\parm
\line{\hfil A $=\{\r x \mid \r x \in \r P_0$ is an instance of an axiom for 
S$\}$\qquad\hfil(10.1.1)}\pars  
\line{\hfil C $= \{\r x \mid \r x \in \r P_0$ is a finite $(\geq 1)$ conjunction of 
members of d$\}$\qquad\hfil(10.1.2)}\parm 
Notice that it is also possible to refine the set C by considering C to be an 
ordered conjunction with respect to the ordering of the indexing set used to 
index members of d. Further$,$ as usual$,$ we have that ${\rm A,\ C,\ d}$ are mutually disjoint.\pars
\hrule
\smallskip
\hrule
\smallskip
Theorem 10.1.1 shows that the contents of the ultranatural events that are 
produced when 
the ultralogic $\Hyper {\b S}$ is applied to an ultraword $w$ cannot be described 
in any standard human language.  However$,$ certain general properties  
can be described.  
\par
\smallskip
\hrule
\smallskip
\hrule
\medskip 
{\bf Theorem 10.1.1} { \sl Let $w \in \Hyper {\bf M_d} - \Hyper {\b d}$ 
be an ultraword for infinite $\b d \subset \Hyper {\b S}(\{w\}).$ Then $\Hyper 
{\b S}(\{w\}) = \Hyper {\b A} \cup Q_1 \cup d_1^\prime,$ where for internal *-
finite $d_1^\prime,\ \b d \subset d_1^\prime \subset \Hyper {\b d}$ and 
internal $Q_1 \subset \Hyper {\b C}$ is composed of *-finite $(\geq 1)$ 
conjunctions (i.e. $i({\rm\sp and\sp})$) of distinct members of $d_1^\prime$ and $w \in Q_1.$ Further$,$ each member of 
$d_1^\prime$ and no other *-propositions  is used to form the *-finite 
conjunctions in $Q_1,$ the only *-propositions in $\Hyper {\b S(\{w\})}$ are those in 
$w,$ and $\Hyper {\b A},\ Q_1$ and $d_1^\prime$ are mutually 
disjoint.}\pars 
Proof. The intent is to show that if $\rm w \in M_d - d,$ then $\rm 
S(\{ w \})=A \cup Q \cup d^\prime,$ where $\rm Q \subset C,$ finite $ \rm 
d^\prime \subset d$ and $\rm Q$ is composed of finite $(\geq 1)$ conjunctions 
of members of $\rm d^\prime,$  each member of $\rm d^\prime$ is used to form 
these conjunctions and no other propositions.\pars
Let $\r J$ be the set of propositional atoms in the composite w. (\b 0) Then $\rm J 
\subset 
S(\{w\}).$ If K is the set of all propositional atoms in $\rm S(\{w\}),$ then 
$\rm J \subset K.$ Let $\rm b \in K - J.$ It is obvious that $ \rm b \notin 
\rm S(\{ w \})$ since otherwise $\rm \{w,b\} \subset S_0 (\{ w \})$ but
$\rm \not\models_{S_0} w \to b.$  
Thus$,$ $\rm J = K.$  Consequently$,$ $\rm J \subset 
S(\{w\}),\ J \subset d$ and there does not exists an $\rm F \in d - J,$ such 
that $\rm F \in S(\{w\}).$ (\b 1) {\it Let $\rm J = d^\prime.$ The only 
propositional atoms in
                                                            
 $\rm S(\{w\})$ are those in w.} Obviously $\rm A \subset S(\{\emptyset\}).$ 
\pars
Assume the language $\r P_0$ is inductively defined from the set of atoms 
d. Recall that for our axioms ${\cal X} = {\cal D} \to {\cal F}$$,$ the 
strongest connective in $\cal X$ is $\to.$ While in $\cal D,$ or $\cal F$ when 
applicable$,$ the strongest connective is $\land.$  
Since $\emptyset \subset 
\rm \{ w \},$ it follows that $\rm S(\{w\}) = S(\emptyset) \cup S(\{w\}).$ Let 
$\rm b \in S(\emptyset).$ The only steps in the formal proof for b contain 
axioms
 or follows from modus ponens. Suppose that step $\rm B_k =b$ is the 
first modus ponens step obtained from steps $\rm B_i,\ B_j,\ i,j < k,$ 
where $\rm B_i =A \to b,\ B_j = A.$ The strongest connective for each axiom is 
$\to.$ However$,$ since $A \to b$ is an axiom$,$ the strongest connective in 
$\rm A$ is $\land.$ This contradicts the requirement that A must also be an axiom 
with strongest connect $\to.$ Thus no modus ponens step can occur in a formal 
proof for b. Hence$,$ (\b 2) $\rm A = S(\emptyset).$ (No modus ponens step can occur 
using two axioms.)\pars
Let $\rm B_k= b_1 \in P_0$ and suppose (a) that $\rm b_1 = w,$ or (b) $\rm b_1 
\not= w$ and is 
the first nonaxiom step that appears 
in a formal 
demonstration from the hypothesis w. Assume (b). Then all steps $\rm B_i 
\in \{w\} \cup A,
\ 0\leq i < k.$  
Then the  only way that ${\rm b_1}$ can be 
obtained is by means of modus ponens.  However$,$ all other steps$,$ not including 
that which is w$,$ are axioms. No modus ponens step can occur using two axioms. 
Thus one of the steps used for modus ponens must not be an axiom. The only 
nonaxiom that occurs prior to the step $\rm B_k$ is the step $\rm B_m = w.$ 
Hence$,$ one of the steps required for $\rm B_k$  must be $\rm B_m = w.$ The 
other step must be an axiom of the form $\rm w \to b_1$ and $\rm b_1 \not= w.$
Thus$,$ from the definition of 
the axioms (\b 3) $\rm b_1$ is either a finite $(\geq 1)$ conjunction of 
atoms in $\r d^\prime,$ or a single member of $\r d^\prime.$           
 Assume strong 
induction. Hence$,$ for $\rm n > 1,$ statement ({\bf 3}) holds for all $\rm r,\ 
1\leq r\leq n.$  
A
similar argument shows that ({\bf 3}) holds for the $\rm b_{n+1}$ nonaxiom 
step. Thus by induction$,$ ({\b 3}) holds for all nonaxiom steps. \pars
Hence$,$ there exists a $\rm Q \subset C$ such that each member of $\r Q$ is 
composed of  finitely many $(\geq 1)$ distinct members of $\rm d^\prime$ and 
the set $\rm G(Q)$ of all the proposition atoms that appear in  any member of $\rm 
Q = d^\prime = J$ since $\rm w \in Q.$ Moreover$,$ (\b 4) $\rm S(\{w\}) = A \cup 
d^\prime \cup Q$ and (\b 5) $\rm A,\ d^\prime,\ Q$ are mutually 
disjoint.\pars
\line{\hfil$ \forall x(x \in {\bf M_d - d} \to \exists y\exists z((y \in 
F(\b d)) \land (z \subset \b C)\land (\b S(\{x\}) =$\hfil}\pars 
\line{\hfil $\b A \cup y\cup z)\land (\b A \cap y = \emptyset)\land 
(\b A \cap z = \emptyset)\land (x \in z)\land$\hfil}\pars
\line{(10.1.3)\hfil$(y\cap z= \emptyset)\land \b G(z) = y)).$\qquad\hfil }\parm
\noindent holds in $\cal M,$ hence also in $\Hyper {\cal M}.$ So$,$ let $w$ be 
an ultraword. Then there exists internal $Q_1 \subset \Hyper {\b C},\ w \in Q_1$ and *-
finite $d_1^\prime \subset \hyper {\b d}$ such that
 $\b d \subset \Hyper {\b S}(\{w\}) = 
\hyper {\b A} \cup d_1^\prime \cup Q_1;\ \hyper {\b A},\ d_1^\prime,\ Q_1$ are 
mutually disjoint and $\Hyper G(Q_1) =d_1^\prime = \hyper {\b J}.$ Hence$,$ $\b d \subset d_1^\prime.$\pars
Now to analyze the objects in $Q_1.$ Let $\rm d = \{F_i \mid i \in \nat\}.$ 
Consider a bijection $h\colon \nat \to \b d$ defined by $h(n) = {\bf F_n} = 
[f],$ where $f \in T^0$ is the special member of $\bf F_n$ such that 
$f = \{(0,f(0))\},\  f(0) =i({\rm F_n}) = q_n \in i[\r d].$ From the above 
analysis$,$ (A) $[g] \in \b S(\{w\}) - \b A - \b d,\ (w \in {\bf M_d - d}),$ iff there 
exist $k,\ j \in \nat$ such that $ k< j$ and $f_1^\prime \in i[\r P_0]^{2(j-
k)}$ such that $[f_1^\prime] = [g],$ and this leads to (B) that for each even 
$2p,\ 0\leq 2p \leq 2(j-k); \ f_1^\prime(2p) = q_{k+p} \in i[\r P_0] \subset 
i[{\cal W}],\ [(0,q_{k+p})] \in \b d^\prime,$ all such $q_{k+p}$ being distinct. For 
each odd $2p+1$ such that $0\leq 2p+1 \leq 2(j-k),\ f_1^\prime(2p+1) = i({\rm
\sp and\sp}).$ Also (C) $h(p) \in h[[k,j]]$ iff there exists an even $2p$ such 
that $0\leq 2p\leq 2(j-k)$ and $f_1^\prime (2p) = h(p) =q_{k+p} \in i[\r 
P_0].$ [Note that 0 is considered to be an even number.] \pars
By *-transfer of the above statements (A)$,$ (B) and (C)$,$  $[g] \in Q_1$ iff there 
exists some $j,\ k \in \hypernat,\ k < j,$ and $f^\prime \in \Hyper (i[\r P_0])^{2(j-
k)}$ such that $[f^\prime]= [g]$ and $\Hyper h[[k,j]] \subset  \Hyper {\b d}.$ 
Moreover$,$ each $\Hyper h(r), \ r \in [k,j]$ is a distinct member of $\Hyper {\b 
d}.$ The conjunction ``codes'' for $i({\rm \sp and\sp}) \in i[{\cal W}]$ that are generated 
by each odd $2p+1$ are all the same and there are *-finitely many of them. 
Hence$,$ $Q_1$ is the *-finite $(\geq 1)$ conjunctions of distinct members of 
$d_1^\prime,$ no other *-propositions are utilized and since $\Hyper {\b G}(Q_1) = 
d_1^\prime,$ all members of $d_1^\prime $ are employed for these  
conjunctions. This completes the proof. \qed 
{\bf Corollary 10.1.1.1} {\sl Let $ w \in \Hyper {\bf M_d} - \Hyper {\b d}$ be an 
ultraword for denumerable {\r d} such that $\b d \subset \Hyper {\b S}(\{w\}).$ 
Then $\Hyper {\b S}(\{w\}) \cap {\bf P_0} = \bf A \cup Q \cup d$ and $\rm  
A,\ Q,\ d$ are mutually disjoint. The set \r Q is composed of finite $\geq 1$ 
conjunctions of members of \r d and all of the members of \r d are employed to 
obtain these conjunctions.}\pars
Proof. Recall that due to the finitary character of our standard objects
$^\sigma {\b A} = \b A = \hyper {\b A} \cap {\bf P_0}$. In like manner$,$ since 
$\b d \subset d_1^\prime,$ $d_1^\prime \cap {\bf P_0} = \b d.$ Now ${\bf P_0} 
\cap Q_1$ are all of the standard members of $Q_1.$ For each $ k \in 
\hypernat,\ \Hyper h(k) = {\bf F_k} \in \Hyper {\b d}$ and conversely. 
Further$,$ $\bf F_k \in d$ iff $ k\in \nat.$ Restricting $k,\ j \in \nat$ in the 
above theorem yields standard finite $\geq 1$ conjunctions of standard members 
of $d_1^\prime;$ 
hence$,$ members of $\b d.$ Since ultraword $w \in Q_1,$ we know that there  
exists some $\eta \in \hypernat-\nat$ and $f_1^\prime \in \Hyper (i[\r P_
0])^{2\eta},$ where $f_1^\prime$ satisfies the *-transfer of the properties 
listed in the above theorem . Since finite conjunctions of standard members of 
$d_1^\prime$ are *-finite conjunctions of members of $d_1^\prime$ and $\b d = 
\b d \cap d_1^\prime,$ it follows that all possible finite conjunctions of 
members of \b d that are characterized by the function $f_1^\prime \in i[{\rm 
P_0}]^{2(j-k)}$ are members of $Q_1$ for each such $j,\ k < \eta.$ Also for 
such $j,\ k$ the values of $f_1^\prime$ are standard.                                                            
On the other hand$,$ any value of $f_1^\prime$ is nonstandard iff it 
corresponds to a 
member of $d_1^\prime - \b d.$ Thus $Q_1 \cap {\bf P_0} = \b Q$ and this 
completes the proof. \qed
If it is assumed that each member of d describes a Natural event (i.e. 
N-event) at times indicated by $\r X_i,$ dropping the $\r X_i$ may still yield a 
denumerable developmental paradigm without specifically generated symbols such 
as the ``$i.$'' Noting that $d_1^\prime$ is *-finite and internal leads to the 
conclusion that we can have little or no knowledge about the word-like 
construction of each member of $d_1^\prime-\b d.$ These pure nonstandard 
objects 
can be considered as describing pure NSP-world events$,$ as will soon be 
demonstrated.  
Therefore$,$ it is important to understand the following 
interpretation scheme$,$ where descriptions are corresponded to events. \parm
{\leftskip 0.5in \rightskip 0.5in \noindent {\it Standard or internal 
NSP-world events or sets of events are interpreted as directly or indirectly 
influencing N-world events. Certain external objects$,$ such as the standard part 
operator$,$ among others$,$ are also interpreted as directly or indirectly 
influencing N-world events.}\par}\parm 
Notice that standard events can directly or indirectly affect standard 
events. In the micro-world$,$ the term {{\it indirect evidence}} or verification 
is a different idea than indirect influences. You can have direct or indirect 
evidence of direct or indirect influences when considered within the N-world. 
An indirect influence occurs when there exists$,$ or there is assumed to 
exist$,$ a  mediating  ``something'' between two events. Of course$,$ indirect 
evidence refers to  behavior that can be observed by normally 
accepted human sensors as such  behavior is  assumed to be caused by unobserved 
events. 
However$,$ the evidence for pure NSP-world events that directly or indirectly 
influence N-world events must be indirect evidence under the above 
interpretation. \pars
In order to formally  consider NSP-world events for the formation of objective 
standard reality$,$ proceed as follows: let $\cal O$ be the subset of $\cal W$
that describes those Natural events that are used to obtain developmental or general 
paradigms and the like. Let ${\rm E_j} \in \cal O.$ Linguistically$,$ assume that 
each $\rm E_j$  has the spacing symbol $\sp$ 
immediately to the right. Thus 
within each $\r T_i,$ there is a finite symbol string ${\rm F_i = E_i} \in \cal O$ that can be 
joined by the justaposition (i.e. join) operation to other event descriptions. 
Assume that 
${\cal W}_1$ is the set of nonempty symbol strings (with repetitions) formed 
(i.e. any finite permutation) from members of $\cal O$ by the join operation. These finite strings of symbols 
generate the basic elements for our partial sequences. \pars
Obviously$,$ ${\cal W}_1 \subset {\cal W}.$ Consider $\r T_i^\prime = \{\r X\r 
W_i \mid \r X \in {\cal W}_1\}$ and note that in many applications the time 
indicator $\r W_i$ need not be of significance for a given $\rm E_j$
 in some of the strings. Obviously$,$ $\r T_i^\prime \subset \r T_i$ for each 
$i.$ For our isomorphism $i$  onto $i[{\cal W]},$ the following hold.\parm
\line{ \hfil $ \forall y(y\in {\cal E}\to  (y\in \b T_i^\prime \iff \exists x\exists 
f\exists w((\emptyset \not= w \in F(i[{\cal O}])) \land (x \in \nat)\land 
 $\hfil}\pars
\line{\hfil $(f(0)=i[W_i])\land (f \in P)\land\forall z((z\in \nat)\land (0<z\leq x) \to $\hfil}\pars
\line{(10.1.4)\hfil $f(z) \in w)\land (f \in y  
)))).$\qquad\qquad\hfil}\pars

\line{\hfil$\forall x( x \in \nat \to \exists f\exists w((\emptyset \not= w 
\in F(i[{\cal O}])) \land (f \in P) \land$\qquad\hfil}\pars
\line{(10.1.5)\hfil $\forall z(z \in \nat \to (0< z\leq x \iff f(z) \in w)))).$\qquad\qquad\hfil}\pars
\line{\hfil$\forall w(\emptyset \not= w \in F(i[{\cal O}])\to \exists x\exists 
f\exists y((f \in P)\land (x \in \nat) \land(y \in \b T_i')\land$\hfil }\pars
\line{(10.1.6)\hfil $(f \in y)\land\forall z(z \in \nat \to (0< z\leq x \iff f(z) \in w)))).$ \qquad\hfil}\parm  
Since each finite segment of a developmental pardigm corresponds to a member of 
${\b T_i'},$ each nonfinite hyperfinite segment should correspond to a member of $\Hyper (\b T_i^\prime) - \b T_i^\prime$ and it should be certain individual segments of such members of $\Hyper (\b T_i^\prime) - \b T_i^\prime$ that correspond to the ultranatural events produced by an ultraword; UN-events that cannot be 
eliminated from an NSP-world developmental paradigms. [Note: for a scientific language$,$ 10.1.4 - 10.1.6 and other such statements would correspond to an ${\cal W}'$ as generated by$,$ at the least$,$ a denumerable alphabet such as used in 9.2$,$ 9.3.]\par
\bigskip
\leftline{\bf 10.2 Laws and Rules.}
\medskip 
One of the basic requirements of human mental activity is the ability to 
{recognize the symbolic differences} between finitely long strings of symbols as 
necessitated by our reading ability and to apply linguistic rules finitely 
many times. G\"odel numberings specifically utilize such recognitions and the 
rules for the generation of recursive functions must be comprehended with 
respect to finitely many applications. Observe that G\"odel number recognition 
is an   ``ordered'' process while some fixed intuitive order is not necessary 
for the application of the rules that generate recursive functions.\pars
In general$,$ the simplest  ``rule'' for ordered or unordered {finite human
choice}$,$ a rule that is assumed to be humanly comprehensible by finite 
recognition$,$ is to simply  {{\it list}} the results of our choice (assuming  
that they are symbolically representible in some fashion) as a partial finite 
sequence for ordered choice or as a finite set of finitely long symbol strings 
for an unordered choice. Hence$,$ the end result for a finite choice can itself 
be considered as an algorithm ``for that choice only.'' The next application 
of such a {{\it finite choice rule}} would yield the exact same partial 
sequence or choice set. Another more general rule would be a statement which 
would say that you should  ``choose a specific number of objects'' from a fixed 
set (of statements). Yet$,$ a more general rule would be that you simply are 
required to  ``choose a finite set of all such objects,'' where the term  
``finite'' is intuitively known. Of course$,$ there are numerous specifically 
described algorithms that will also yield finite choice sets.\pars
From the symbol string viewpoint$,$ there are trivial machine programmable  
algorithms that allow for the comparison of finitely long symbols with each 
member of a finite set of symbol strings B that will determine whether or not 
a specific symbol string is a member of B. These programs duplicate the 
results of human symbol recognition. As is well-known$,$ there has not been an 
algorithm described that allows us to determine whether or not a given finite 
symbol string is a member of the set of all theorems of such theories as formal {Peano 
Arithmetic}. If one accepts {Church's Thesis}$,$ then no such algorithm will ever 
be described. \pars
Define the general finite human choice relation on a set $A$ as $H_0(A) = 
\{(A,x) \mid x \in F_0(A)\},$ where $F_0$ is the finite power set operator 
(including the empty set = no choice is made). 
Obviously$,$ the inverse $H_0^{-1}$ is a function from $F(A)$ onto $\{A\}.$ 
There are choice operators that produce sets with  a specific number of elements 
that can be easily defined. Let $F_1(A)$  be the set of all singleton subsets 
of $A.$  The axioms of set theory state that such a set of singleton sets 
exists. Define $H_1(A) = \{(A,x) \mid x \in F_1(A) \},$ etc.  Considering such functions as defined on sets $X$ that are members of 
a  superstructure$,$ then these relations are subsets of $\power X \times \power 
X$ and as such are also members of the superstructure.\par

\smallskip
\hrule
\smallskip
\hrule
\smallskip
The next discussion shows that  the IUN-selection processes $\Hyper {\bf C_i},
 \ i \geq 0$ 
exist as formal objects within the NSP-world.\par
\smallskip
 \hrule
\smallskip
\hrule
\smallskip

Let $\r A =\r P_0.$ Observe that $^\sigma {\bf H_0(A)} = \{ (\hyper {\b A}, x) \mid 
x \in F_0(\b A)\}$ and $\Hyper {\bf H_i(A)} = \{(\hyper {\b A}, x) \mid x \in \Hyper 
{(F_i(\b A))}\}\ (i\geq 0).$ Now $\Hyper {(F_0 (\b A))} = \Hyper {F_0}(\hyper 
{\b A})$ is the set of all *-finite subsets of $\hyper {\b A}.$ On the other 
hand$,$ for the $i > 0$ cardinal subsets$,$ $\Hyper {(F_i (\b A))} = F_i(\hyper {\b A})$ for each $ i \geq 1.$ 
With respect to an ultraword $w$ that generates the general and developmental 
paradigms$,$ we know that $ w \in \Hyper {\bf P_0} - {\bf P_0}$ and that 
$(\Hyper {\bf P_0}, \{x\}) \in \Hyper {\bf H_1}(\bf P_0).$ The actual finite 
choice operators are characterized by th set-theoretic second projector 
operator $P_2$ as it is defined on $H_i(\r A).$ This operator embedded by the 
$\theta$ is the same as $P_2$ as it is defined on $\bf H_i(A).$ Thus$,$ 
when $h = (\b A, x) \in \bf H_i(A),$ then we can define $ x = P_2(h) = C_i(h) 
= {\bf C_i}(h).$ The maps $C_i$ and $\bf C_i,$ formally defined below$,$ are the 
specific finite choice operators. For consistency$,$ we let $C_i$ and $\bf C_i$ 
denote the appropriate finite choice operators for $H_i(\r A)$ and $\bf H_i(A)$$,$ 
respectively. \pars
Since the $\Hyper P_2$ defined on say $\bf H_i(A)$ is the same as the 
set-theoretic second projection operator $P_2,$ it would be possible to denote 
$\Hyper {\bf C_i}$ as $\bf C_i$ on internal objects. For consistency$,$ the 
notation $\Hyper {\bf C_i}$ for these special finite choice operators is 
retained.  
Formally$,$ let ${\bf C_i\colon H_i(A)} \to F_i(\b A).$ Observe that $^\sigma {\bf 
C_i} = \{\Hyper (a,b) \mid (a,b) \in {\bf C_i}\} =\{((\hyper {\b A},b),b)\mid 
b\in F_i(\b A)\} \subset \Hyper {\bf C_i};$ and$,$ for $b \in F_i(\b A),\ {\bf 
C_i}((\b A,b)) = b$ implies that $^\sigma({\bf C_i}((\b A,b))) = \{\hyper 
a\mid a \in b\} = b$ from the construction of $\cal E.$ Thus in 
contradistinction to the consequence operator$,$ for each $(\hyper {\b A},b)\in 
{^\sigma {\bf H_i}},$ the image $(^\sigma {\bf C_i})((\hyper {\b A},b)) = 
{^\sigma({\bf C_i}((\b A, b)))} = (\Hyper {\bf C_i})((\hyper {\b A},b)) = b.$ 
Consequently$,$ the set map $^\sigma{\bf C_i}\colon {^\sigma{\bf H_i}} \to 
F_i(\b A) = {^\sigma(F_i(\b A))}$ and $\Hyper {\bf C_i} \mid {^\sigma {\bf 
H_i}} = {^\sigma {\bf C_i}}.$ Finally$,$ it is not difficult to extend these 
finite choice results to general internal sets.\pars
In the proofs of such theorems as 7.2.1$,$ finite and other choice sets are 
selected due to their set-theoretic existence. The finite choice operators 
$C_i$ are not specifically applied since these operators are only intended as 
a mathematical model for apparently effective human processes --- procedures 
that generate acceptable algorithms. As is well-known$,$  
there are other describable rules that also lead to finite or infinite 
collections of statements. Of course$,$ with respect to a G\"odel  encoding 
$i$ for the set of all words $\cal W$ the finite choice of readable sentences 
in $\cal E$ is one-to-one and effectively related to a finite and$,$ hence$,$ 
recursive subset of $\nat.$\pars
From this discussion$,$ the descriptions of the finite choice operators would 
determine a subset of the set of all algorithms (``rules'' written in the 
language $\cal W$) that allow for the selection of readable sentences. Notice 
that before algorithms are applied there may be yet another set of readable 
sentences that yields conditions that must exist prior to an application of 
such an algorithm and that these application rules can be modeled by members 
of $\cal E.$ \pars
In order to be as unbiased as possible$,$ it has been required for N-world 
applications that the set of all frozen segments be infinite. Thus$,$ within the 
proof of Theorem 7.2.1$,$ every N-world developmental$,$ as well as a general 
paradigm$,$ is a proper subset of a {*-finite NSP-world paradigm}$,$ and the *-finite 
paradigm is obtained by application of the *-finite choice operator $\Hyper 
{\bf C_0}.$ As has been shown$,$ such *-finite paradigms contain pure 
unreadable (subtle) sentences that may be interpreted for developmental 
paradigms as {pure refined NSP-world behavior} and for general paradigms as 
specific pure NSP-world ultranatural events or objects.\pars
Letting $\Gamma$ correspond to the formal theory of Peano Arithmetic$,$ then assuming 
Church's Thesis$,$ there would not exist a N-world algorithm (in any human 
language) that allows for the determination of whether or not a statement F in 
the formal language used to express $\Gamma$ is a member of $\Gamma.$ By 
application of the *-finite choice operator $\Hyper {\bf C_0},$ however$,$ there 
does exist a *-finite $\Gamma^\prime$ such that $^\sigma\Gamma = \Gamma 
\subset \Gamma^\prime$ and$,$ hence$,$ within the NSP-world a ``rule'' that allows 
the determination of whether or not $\r F \in \Gamma^\prime.$ If such internal 
processes mirror the only allowable procedures in the NSP-world for such a 
``rule,'' then it might be argued that we do not have an effective NSP-world 
process that determines whether or not F is a member of $\Gamma$ for $\Gamma$ 
is external. \pars 
As previously alluded to at the beginning of this section$,$ 
when a G\"odel encoding $i$ is utilized with the N-world$,$ the injection $i$ is 
not a surjection. When such G\"odel encodings are studied$,$ it is usually {\it 
assumed}$,$ without any further discussion$,$ that there is some human mental 
process that allows us to recognize that one natural number representation (whether in prime 
factored form or not) is or is not distinct from another such representation. 
It is not an unreasonable assumption to assume that the same {\it effective} 
(but external) process exists within the NSP-world. Thus within the {NSP-
world there is a  ``process'' that determines whether or not an object is a 
member of} $\hypernat -\nat= \nat_\infty$ or $\nat.$ Indeed$,$ from the ultraproduct 
construction of our nonstandard model$,$ a few differences can be detected by 
the human mathematician. Consequently$,$ this assumed NSP-world effective 
process would allow a determination of whether or not $\b F = [f_m]$ is a 
member of $\Gamma$ by recalling that $f_m \in P_m$ signifies that $[f_m] \in 
\Hyper {\Gamma} -\Gamma$ implies $m \in \nat_\infty\simeq
\Hyper {({i[{\cal W}]})} - {i[{\cal W}]}.$\pars 
The above NSP-world recognition process is 
equivalent$,$ as defined in Theorem 7.2.1$,$ to various applications of a single 
(external) set-theoretic intersection. Therefore$,$ there are internal 
processes$,$ such as $\Hyper {\bf C_0},$ that yield pure NSP-world developmental 
paradigms and a second (external) but acceptable NSP-world effective process 
that produces specific N-world objects. Relative to our modeling procedures$,$ 
it can be concluded that both of these processes are {intrinsic ultranatural 
processes.} \pars 
With respect to Theorem 10.1.1$,$ the NSP-world developmental or general 
paradigm generated by an ultraword is *-finite and$,$ hence$,$ specifically NSP-
world obtainable prior to application of $\Hyper {\b S}$ through application 
of $\Hyper {\bf C_0}$ to $\Hyper {\b d}.$ However$,$ this composition can be 
reversed. The NSP-world (IUN) process $\Hyper {\bf C_1}$ can be applied to the 
appropriate $\Hyper {\bf M_d}$ type set and an appropriate ultraword $w \in 
\Hyper {\bf M_d}$ obtained. Composing $\Hyper {\bf C_1}$ with $\Hyper {\b S}$  
would yield $d_1^\prime$ in a slightly less conspicuous manner. Obviously$,$ 
different ultrawords generate different standard and nonstandard developmental 
or general paradigms.\pars 
To complete the actual mental-type processes that lead to the proper ordered 
event sequences$,$ the above discussion for the finite choice operators is 
extended to the human mental ability of ordering a finite set in terms of 
rational number subscripts. New choice operators are defined that model not 
just the selection of a specific set of elements that is of a fixed finite
cardinality but also choosing the elements in the required rational number 
ordering. 
The ultrawords $w$ that exist are *-finite in 
length. By application of the inverses of the $f$ and $\tau$ functions of 
section 7.1$,$ where they may be considered as extended standard functions 
$\hyper f$ and $\hyper \tau$$,$ there would be from analysis of extended 
theorem 7.3.2 a hyperfinite set composed of standard or nonstandard frozen 
segments contained in an ultraword. Further$,$ in theorem 7.3.2$,$ the chosen 
function $f$ does not specifically differentiate 
each standard or nonstandard frozen 
segment with respect to its ``time'' stamp subscript. There does exist$,$ 
however$,$ another function in the *-equivalence class $[g]=w$ that will make 
this differentiation. It should not be difficult to establish that after 
application of the ultralogic $\Hyper {\b S},$ there is applied an 
appropriate 
mental-like hyperfinite ordered choice operator (an IUN-selection 
process) and that this would yield that various types of event sequences. 
Please note that each event sequence has a beginning point of observation. 
This point of observation need not indicate the actual moment when a specific 
Natural-system began its development.\par

\smallskip
\hrule
\smallskip
\hrule
\smallskip
In the following discussion$,$ we analyze how ultranatural laws aid in the 
production of Natural and UN-event sequences$,$ and develop an
external unification of all physical theories. \pars
\hrule
\smallskip
\hrule
\smallskip

Various subdevelopmental (or subgeneral) paradigms $\rm d_i$ are obtained by considering the 
actual descriptive content (i.e. events) of specific theories $\rm\Gamma_i$ that 
are deduced from hypotheses $\rm\eta_i,$ usually$,$ by finitary consequence operators 
$\rm S_i$ (the inner logics) that are compatible with $\r S.$ In this case$,$ 
$\rm d_i \subset S_i(\eta_i).$ It is also possible to include within $
\{\rm d_i\}$ and $\{\rm \eta_i\}$ the assumed descriptive {chaotic behavior} that 
seems to have no apparent set of hypotheses except for that particular 
developmental paradigm itself and no apparent deductive process 
except for the identity consequence operator. In this way$,$ such scientific 
nontheories can still be considered as a formal theory produced by a finitary 
consequence operator applied to an hypothesis. Many of these hypotheses 
$\rm \eta_i$ contain the so-called natural laws (or first-principles) peculiar to 
the formal theories $\rm \Gamma_i$  and the theories language$,$ where it is assume 
that such languages are at least closed under the informal conjunction and 
conditional.\pars
Consider each $\rm \eta_i$ to be a general paradigm. For the appropriate M type 
set constructed from the denumerable set $\r B = \{\bigcup\{\rm d_i\} 
\cup(\bigcup \{\eta_i\}),$ redefine $\rm M_B$ to be the smallest subset 
of $\rm P_0$ containing B and closed under finite $(\geq 0)$ conjunction. (The 
usual type of inductively defined $\rm M_B.$) Then there exist ultrawords $w_i 
\in \Hyper {\bf M_B} - \Hyper {\b B}$ such that ${\bf \eta_i} \subset \Hyper {\b 
S}(\{w_i\})$ (where due to parameters usually {\it ultranatural laws} exist in $\Hyper {\b S}(\{w_i\})- {\bf \eta_i}$) and ${\bf d_i} \subset \Hyper {\b S}(\{w_i\}).$ Using methods 
such as those in Theorem 7.3.4$,$ it follows that there exists some $w" \in 
\Hyper {\bf M_B} - {\bf M_B}$ such that $ w_i \in \Hyper {\b  S}(\{w"\})$ and$,$ 
consequently$,$ ${\bf \eta_i \cup d_i} \subset \Hyper {\b S}(\{w"\}).$  
Linguistically$,$ it is hard to describe the ultraword $w".$ Such a $w"$ might 
be called an {{\it ultimate  ultranatural hypothesis}} or {{\it the ultimate 
building plain}}. \pars
{\it Remark.} It is not required that the so-called Natural laws that appear in some 
of the $\rm \eta_i$ be either cosmic time or universally applicable. They could 
refer only to local first-principles. It is not 
assumed that those first-principles that display themselves in our local 
environment are universally space-time valid.\pars
Since the consequence operator $\r S$ is compatible with each $\rm S_i,$ 
it is useful to proceed in the following manner. First$,$ apply the IUN-process
$\Hyper {\b S}$ to $\{w"\}.$ Then ${\bf d_i \cup \eta_i} \subset \Hyper {\b 
S}(\{w"\}).$ It now follows that ${\bf d_i \cup \eta_i} \subset \Hyper {\b 
S}(\{w_i\}) \subset \Hyper {\bf S_i}(\Hyper {\b S}(\{w_i\})) \subset \Hyper 
{\bf S_i}(\Hyper {\bf S_i}(\{w_i\})) = \Hyper {\bf S_i}(\{w_i\}).$ Observe 
that for each $\rm a \in \Gamma_i$ there exists some finite $\rm F_i \subset 
\eta_i$ such that $\bf a \in S_i(F_i).$ However$,$ $\bf F_i \subset \eta_i$ for 
each member of $F(\rm \eta_i)$ implies that $\b a \in \Hyper {\bf S_i(F_i)} 
\subset \Hyper {\bf S_i}(\Hyper {\b S}(\{w_i\})).$ Consequently$,$ ${\bf 
\Gamma_i} \subset \Hyper {\bf S_i}(\Hyper {\b S}(\{w_i\})).$ The {ultimate 
ultraword} suffices for the descriptive content and inner logics associated 
with each theory $\rm \Gamma_i$.\pars
We now make the following observations relative to ``rules'' and deductive 
logic. It has been said that science is a combination of empirical data$,$ 
induction and deduction$,$ and that you can have the first two without the last. 
 That this belief is totally false should be 
self-evident since the philosophy of science requires its own general rules 
for observation$,$ induction$,$ data collection$,$ proper experimentation and the 
like. All of these general rules require logical deduction for their 
application to specific cases --- the metalogic. Further$,$ there are specific 
rules for linguistics that also must be properly applied prior to scientific 
communication. Indeed$,$ we cannot even open the laboratory door --- or at least 
describe the process --- without application of deductive logic. The concept 
of deductive logic as being the patterns our ``minds'' follow and its use 
exterior to the inner logic of some theory should not be dismissed for even 
the (assumed?) mental methods of human choice that occur prior to 
communicating various scientific statements and descriptions. \pars 
Finally$,$ with respect to the {hypothesis rule} in [9]$,$ it 
might be argued that we can easily analyze the specific composition of all 
significant ultrawords$,$ as has been previously done$,$ and the composition of 
the nonstandard extension of the general paradigm. Using this assumed analysis 
and an additional alphabet$,$ one {\it might} obtain specific information about  
pure NSP-world ultranatural laws or refined behavior.  Such an argument would 
seem to invalidate the cautious hypothesis rule and lead to appropriate 
speculation. However$,$ such an argument would itself be invalid.\pars 
 Let ${\cal W}_1$ be an infinite set of meaningful readable sentences for some 
description and assume that ${\cal W}_1$ does not contain any infinite subset of 
readable sentences each one of which 
contains a mathematically interpreted entry such as a real number or the like. 
Since ${\cal W}_1 \subset {\cal W}$ and the totality $\r T_i = \{\r X
\r W_i \mid \r X \in {\cal W} \}$ is denumerable$,$ the subtotality 
${\r T}_i^\prime = \{\r X{\r W}_i \mid \r X \in {\cal W}_1 \}$ is also 
denumerable. Hence$,$ the external cardinality of $\Hyper {\bf T_i^\prime} \geq 
2^{\vert{\cal M}\vert}.$ \pars
Consider the following sentence \parm
\line{\hfil $ \forall z(z \in i[{\cal W}_1] \to 
\exists y\exists x((y \in 
i[{\cal W}]^{[0,1]}) \land (x \in {\bf T_i^\prime}) \land (y \in x) \land 
 $\hfil}\pars
\line{(10.2.1)\hfil$((0,i({\r W_i})) \in y)\land
((1,z) \in y )))).$\qquad\hfil }\parm
By *-transfer and letting  ``z'' be an element in $\Hyper (i[{\cal W}_1])-i[{\cal W}_1]$ it 
follows that we can have little knowledge about the remaining and what must be 
unreadable portions that take the ``X'' position. If one assumes 
that members of ${\cal W}_1$ are possible descriptions for possible NSP-world 
behavior at the time $t_i,$ then it may be assumed that at the time $t_i$ the 
members of$ \Hyper {\bf T_i^\prime} - {\bf T_i}$ describe NSP-world behavior at NSP-
world (and N-world) time $t_i.$ Now as i varies over $\hypernat,$ pure 
nonstandard subdevelopmental paradigms (with or without the time index 
statement $\r W_i$) exist with members in $\Hyper {\cal T}$ and may be 
considered as descriptions for time refined NSP-world behavior$,$ especially for 
a NSP-world time index $i \in \nat_\infty.$\pars   
\bigskip
\centerline{\bf CHAPTER 10 REFERENCES}
\medskip
\noindent {\bf 1} Beltrametti$,$ E. G.$,$ Enrico$,$ G. and G. Cassinelli$,$ {\it The 
Logic of Quantum Mechanics,} Encyclopedia of Mathematics and Its Applications$,$ 
Vol. 15$,$ Addison-Wesley$,$ Reading$,$ 1981.\pars
\noindent {\bf 2} d'Espagnat$,$ B.$,$ The quantum theory and realism$,$ Scientific 
America$,$ 241(5)(1979)$,$ 177. \pars
\noindent {\bf 3} {\it Ibid.} \pars
\noindent {\bf 4} {\it Ibid.}$,$ 181. \pars
\noindent {\bf 5} {\it Ibid.}$,$ \pars
\noindent {\bf 6} {\it Ibid.}$,$ 180.\pars
\noindent {\bf 7} Feinberg$,$ G.$,$ Possibility of faster-than-light particles$,$ 
Physical Review$,$ 159(5)(1976)$,$ 1089---1105.\pars
\noindent {\bf 8} Hanson$,$ W. C.$,$ The isomorphism property in nonstandard 
analysis and its use in the theory of Banach Spaces$,$ J. of Symbolic Logic$,$ 
39(4)(1974)$,$ 717---731.\pars
\noindent {\bf 9} Herrmann$,$ R. A.$,$ D-world evidence$,$ C.R.S. Quarterly$,$ 
23(2)(l1986)$,$ 47---54.\pars
\noindent {\bf 10} Herrmann$,$ R. A.$,$ The Q-topology$,$ Whyburn type filters and 
the cluster set map$,$ Proceedings Amer. Math. Soc.$,$  59(1)(1975)$,$ 161---
166.\pars
\noindent {\bf 11} Kleene$,$ S. C.$,$ {\it Introduction to  Metamathematics}$,$ D. 
Van Nostrand Co.$,$ Princeton$,$ 1950.\pars
\noindent {\bf 12} Prokhovnik$,$ S. J.$,$ {\it The Logic of Special Relativity}$,$ 
Cambridge University Press$,$ Cambridge$,$ 1967.\pars
\noindent {\bf 13} Stroyan$,$ K. D. and W. A. J. Luxemburg$,$ {\it Introduction to 
the Theory of Infinitesimals}$,$ Academic Press$,$ New York$,$ 1976. \pars
\noindent {\bf 14} Tarski$,$ A.$,$ {\it Logic$,$ Semantics$,$ Metamathematics}$,$ 
Clarendon Press$,$ Oxford$,$ 1969.\pars
\noindent {\bf 15} Thurber$,$ J. K. and J. Katz$,$ Applications of fractional 
powers of delta functions$,$ {\it Victoria Symposium on Nonstandard Analysis},
Springer-Verlag$,$ New York$,$ 1974.\pars
\noindent {\bf 16} Zakon$,$ E.$,$ Remarks on the nonstandard real axis$,$ {\it 
Applications of Model Theory to Algebra$,$ Analysis and Probability}$,$ Holt$,$ 
Rinehart and Winston$,$ New York$,$ 1969. \pars
\noindent {\bf 17} Note that the NSP-world model is not a local hidden 
variable theory. \par

\medskip
\line{\leaderfill}
\medskip
\hrule\smallskip \hrule\smallskip
In the next section$,$ it is shown how the mathematical structure predicts
the possible existence of the ultra-properton and how$,$ by the very simple process
of ultrafinite combinations$,$ all of the Natural universe assumed 
fundamental entities can be produced. Also$,$ this allows for the construction 
of infinitely many distinct collections of fundamental entities for distinct 
universes. Further$,$ it allows the virtual particle concept to be replaced by 
ultrafinite combinations.\par
\smallskip
\hrule\smallskip\hrule\bigskip 

\leftline{\bf 11.1 Propertons (Subparticles)).}\par 
\medskip
What is a properton? Or$,$ what is an infant? Or$,$ better still$,$ what is a 
thing? I first used the name {infant} for these strange objects. I then 
coined the term properton. It would have been better to have simply called 
them ``things.'' As stated in [9]$,$ these objects are not to be described in terms 
of any geometric configuration. These {multifaceted things}$,$ these propertons$,$ 
are not to be construed as either particles nor waves nor quanta nor anything 
that can be represented by some fixed imagery. propertons are to be viewed only 
operationally. Propertons are only to be considered as represented by a 
*-finite sequence $\{a_i\}_{i=1}^n,\ n \in \hypernat,$ of hyperreal numbers. Indeed$,$ the idea of the n-tuple 
$(a_1,a_2,\ldots,a_i,\ldots)$ notation is useful and we assume that 
$n$ is a fixed member of $\nat_\infty.$ The language of coordinates 
for this notation is used$,$ where the i'th coordinate means the i'th value of the 
sequence. Obviously$,$ 0 is not a domain member for our sequential representation. 
\pars
The first coordinate 
$a_1$ is a ``naming'' coordinate. The remaining coordinates are used to 
represent various real numbers$,$ complex numbers$,$ vectors$,$ and the like physical
qualities needed for different physical theories. For example$,$ $a_2 = 1$ might 
be a counting coordinate. Then $a_i,\ 3\leq i\leq 6$ are hyperreal numbers that
represent NSP-world 
coordinate locations of the properton named by $a_1$ --- $a_7,\ a_8$  
represent the positive or negative charges that can be assigned to every 
properton --- $a_9,\ a_{11},\ a_{13}$ {hyperreal representations} for the inertial$,$ 
gravitational and intrinsic (rest) mass. For vector quantities$,$ continue this 
coordinate assignment and assign specific coordinate locations for the vector 
components. So as not to be biased$,$ include as other coordinates hyperreal 
measures for qualities such as energy$,$ apparent momentum$,$ and all other physical  
qualities required within theories that must be combined in order to produce a 
reasonable description for N-world behavior. For the same reason$,$ we do not 
assume that such N-world properties as the uncertainty principle hold for 
the NSP-world. (See note (2) on page 60.)\pars

It is purposely assumed that the qualities represented by the coordinate 
$a_i, \ i \geq 3$ are not inner-related$,$ in  their basic construction$,$ by any 
mathematical relation since it is such {inner-relations} that are assumed to 
mirror the N-world laws that govern the development of not only our present 
universe but previous as well as future developmental alterations. The same 
remarks apply to any possible and distinctly different universes that may or 
not occur. Thus$,$ for these reasons$,$ we view the properton as being totally 
characterized by such a sequence $\{a_i\}$ and always proceed cautiously when 
any attempt is made to describe all but the most general properton 
behavior. Why have we chosen to presuppose that propertons are 
characterized by sequences$,$ where the coordinates are hyperreal 
numbers?\pars
For chapters 11$,$ 12 assume EGS. Let $r$ be a positive real number. The number $r$ can be represented by a 
decimal number$,$ where for uniqueness$,$ the repeated 9s case is used for all 
terminating decimals. From this$,$ it is seen that there is a sequence $S_i$ of 
natural numbers such that $S_i/10^i \to r.$ Consequently$,$ for any $ \omega \in 
\nat_\infty=\hypernat-\nat,$ it follows that $\pm \hyper S_\omega/10^\omega \in  \monad {\pm r},$  
where $\hyper S_\omega \in \hypernat$ and $\monad {\pm r}$ is the {monad} 
about $r.$ In [9]$,$ it is assumed that each coordinate $a_i,\ i \geq 3$ is 
characterized by the numerical quantity $\pm 10^{-\omega},\ \omega \in 
\nat_\infty.$ Obviously$,$ we need not confine ourselves to the number $10^{-
\omega}.$  Indeed$,$ the next theorem has interesting applications relative to 
such a selection.\parm
{\bf Theorem 11.1.1} {\sl Let $\omega \in \nat_\infty.$ Then for each $ r \in 
\real$ there exists an $x \in \{m/\omega \mid( m \in \Hyper {\b Z}) \land
(\vert m\vert < \omega^2)\}$ such that $x \approx r$ (i.e. $ x \in \monad 
r.)$}\pars
Proof. Let $r \in \real.$ Then there exists some integer $n  \in \b Z$ such 
that $n\leq r< n+1.$ Partition $[n,n+1]$ as follows: let $ 0\not= m \in \nat$ and 
consider $n \leq n+1/m \leq \cdots \leq n+ (m-1)/m \leq n+1.$ Then there 
exists a unique $a \in \{0,1,\ldots, m-1\}$ such that $r \in [n 
+a/m,n+(a+1)/m).$ For $ m = 1,$ let $S_1 = n = f_1/1.$ For $ m \geq 2,$ if $a 
\in \{0,1,\ldots, m-2 \},$ then let $S_m = n + (a+1)/m = (nm + a + 1)/m = 
f_m/m$; if $ a = m-1,$ then let $S_m = n +(m-1)/m = (nm + m -1)/m = f_m/m.$ 
This yields two sequences $S\colon \nat \to \b Q$ and $ f\colon \nat \to \b 
Z.$ It follows easily that $S_i \to r.$ Hence$,$ for each $\omega \in 
\nat_\infty, \ \hyper S_\omega =\hyper f_\omega/\omega \approx r$ and $\hyper f_\omega \in \Hyper {\b 
Z}.$ Observe that $\hyper f_\omega/\omega$ is a finite (i.e. limited) number.  
Hence$,$ $ \vert \hyper f_\omega/\omega \vert < \omega$ entails that $\vert 
\hyper f_\omega \vert < \omega^2.$ Therefore$,$ $\hyper f_\omega/\omega \in \{m/\omega 
\mid( m \in \Hyper {\b Z}) \land (\vert m\vert < \omega^2)\}.$ \qed
{\bf Corollary 11.1.1.1} { \sl Let $ \omega \in \nat_\infty$.
Then there exists a sequence $f\colon \nat \to \b Z$ such that $\hyper f_\omega/\hyper m_\omega\in \monad r$ for each $r \in \real.$ Also$,$ there exists a sequence $g\colon \nat \to \nat$ such that for each real $r \geq 0,\ (\pm \hyper g_\omega/\omega) \approx \pm r.$
}\parm
For the {{\it ultra-properton}}$,$ each coordinate $a_i = 1/10^\omega \ i 
\geq 3$ and odd, $a_i =- 1/10^\omega \ i 
\geq 4$ and even$,$ $\omega \in \nat_\infty.$ From the above theorem$,$ the 
choice of $10^{-\omega}$ as the basic numerical quantity is for 
convenience only and is not unique accept in its infinitesimal character. 
Of course$,$ the sequences chosen to represent the ultra-properton are pure 
internal objects and as such are considered to directly or indirectly affect 
the N-world. Why might the *-finite ``length''
 of such propertons (here is 
where we have replaced the NSP-world entity by its corresponding sequence)  
be of significance? \pars
First$,$ since our N-world languages are formed from a finite set of alphabets$,$ 
it is not unreasonable to assume that NSP-world ``languages'' are composed from 
a *-finite set of alphabets. Indeed$,$ since it should not be presupposed that there is 
an upper limit to the N-world alphabets$,$ it would follow that the basic 
NSP-world set of alphabets is an infinite *-finite set. Although the 
interpretation method that has been chosen does not require such a restriction 
to be placed upon NSP-world alphabets$,$ it is useful$,$ for consistency$,$ to assume 
that descriptions for substratum processes that affect$,$ in either a directly 
or indirectly detectable manner$,$ N-world events be so restricted. For the 
external NSP-world viewpoint$,$ all such infinite *-finite objects have a very 
significant common property.  [Note relative to the basic book "The Theory of Ultralogics." In what follows ${\cal M}_1$ is the standard superstructure constructed on page 76 and not the object defined on page 57.]\parm
{\bf Theorem 11.1.2} {\sl All infinite *-finite members of 
our (ultralimit) model $\Hyper {\cal M}_1$ have the same external cardinality which is
$\geq \vert {\cal M}_1\vert.$} \pars
Proof. Hanson [8] and Zakon [16] have done all of the difficult work for this 
result to hold. First$,$ one of the results shown by Henson is that all
infinite *-finite  members 
of our ultralimit model have the same external cardinality. Since our model 
is a comprehensive enlargement$,$ Zakon's theorem 3.8 in [16] applies. Zakon 
shows that there exists a *-finite set$,$ $A,$ such that $\vert A \vert \geq 
\vert {\cal M}_1 \vert= \vert {\cal R}\vert.$ 
Since $A$ is infinite$,$ Hanson's result now implies that all 
infinite *-finite members of 
our model satisfy this inequality.\qed
For an extended infinite standard set $\hyper A$ it is well-known that 
$\vert \hyper A \vert \geq 2^{\vert{\cal M}_1\vert}.$  One may use these
various results and establish easily that there exist more than enough 
propertons to obtain all of the cardinality statements relative to the three 
substratum levels that appear in [9]  even if we assume that there are a 
continuum of finitely many properton qualities that are needed to create all 
of the N-world. \pars
Consider the following infinite set of statements expressed in an extended 
alphabet. \parm
\line{\hfil${\rm G_A = \{An\sp elementary\sp particle\sp}
\r k^\prime(\r i^\prime,\r j^\prime)\sp{\rm with\sp}$\hfil}\pars
\line{\hfil${\rm total\sp energy\sp}\r c^\prime{+}1/(\r n^\prime).\mid ((i,j,n) \in $\hfil}\pars
\line{\qquad\hfil$\nat^+ \times 
\nat^+ \times \nat^+)\land (1\leq k\leq m) \},$\hfil(11.1.1)}\parm
\noindent where $\nat^+$ is the set of all nonzero  natural numbers and $m \in \nat^+.$ 
Applying the same procedure that appears in the proof of Theorem 9.3.1 and 
with a NSP-world alphabet ${\cal W}'$, we obtain\parm
\line{\hfil${\rm G_A^\prime = \{An\sp elementary\sp particle\sp} 
\r k^\prime(\r i^\prime,\r j^\prime)\sp{\rm with\sp}$\hfil}\pars
\line{\hfil${\rm total\sp energy\sp}\r c^\prime{+}1/(\r n^\prime).\mid ((i,j,n) \in $\hfil}\pars
\line{\qquad\hfil$\hypernat^+ \times 
\hypernat^+ \times \hypernat^+)\land (1\leq k\leq m )\},$\hfil(11.1.2}\parm
Assume that there is at least one type of elementary particle with the properties 
stated in the set ${\rm G_A}.$ It will be shown in the next section 
that within the NSP-world there may be simple properties that lead to  
N-world energy being a manifestation of mass. For $c = 0,$ we have  another 
internal set of descriptions that forms a subset of $\rm G_A^\prime.$ \parm
\line{\hfil${\rm \{An\sp elementary\sp particle\sp} 
\r k^\prime(\r i^\prime,\r j^\prime)\sp{\rm with\sp}$\hfil}\pars
\line{\hfil${\rm total\sp energy\sp}\r c^\prime{+}1/(10^{\gamma^\prime}).\mid 
((i,j,\gamma) \in $\hfil}\pars
\line{\qquad\hfil$\hypernat^+ \times 
\hypernat^+ \times \hypernat^+)\land (1\leq k\leq m )\},$\hfil(11.1.3)}\parm
For our purposes$,$ (11.1.3) leads immediately to the not ad hoc concept of 
propertons with {infinitesimal proper mass}. As will be shown$,$ such 
infinitesimal proper mass can be assumed to characterize any possible zero 
proper mass N-world entity. The set $\rm G_A^\prime$ has meaning if there 
exists at least one natural entity that can possess the energy expressed by 
$\rm G_A,$ where this energy is measured in some private unit of measure.\pars
Human beings combine together finitely many sentences to produce 
comprehensible descriptions. Moreover$,$ all N-world human construction requires 
the composition of objectively real N-world objects. We model the idea of {{\it
finite composition}} or {{\it finite combination}} by an N-world process. 
This produces a corresponding 
NSP-world intrinsic ultranatural process {{\it ultrafinite composition}} or 
{{\it ultrafinite combination}} that 
can either directly or indirectly affect the N-world$,$ where its effect is 
{indirectly inferred}. \pars 
Let the index $j$ vary over a hyperfinite interval and fix the other indices. 
Then the set of sentences\parm
\line{\hfil${\rm G_A^{\prime\prime} = \{An\sp elementary\sp particle\sp} 
\r k^\prime(\r i^\prime,\r j^\prime)\sp{\rm with\sp}$\hfil}\pars
\line{\hfil${\rm total\sp energy\sp}\r c^\prime{+}1/(\r n^\prime).
\mid (j \in \hypernat^+)$\hfil}\pars
\line{\hfil\qquad$\land (1\leq j\leq \lambda) \},$\hfil(11.1.4)}\parm
\noindent where $\lambda \in \hypernat^+,\ 3\leq i \in \hypernat,\ n \in\nat_\infty$ 
and $1 \leq 
k\leq m,$ forms an internal linguistic object that can be assumed to describe a 
hyperfinite collection of ultranatural entities. Each member of $\rm 
G_A^{\prime\prime}$ has the $i$'th coordinate that  measures the proper 
mass and  is infinitesimal (with respect to NSP-world private units of 
measure). In the {N-world$,$ finite combinations yield an event.} Thus$,$  with 
respect to such sets as $\rm G_A^{\prime\prime},$ one can say that there are 
such N-world events iff there are ultrafinite combinations of NSP-world entities. And 
such ultrafinite combinations yield a  NSP-world event that 
is an ultranatural entity. \pars
Associated with such ultrafinite combinations for the entities described in 
$\rm G_A^{\prime\prime}$ there is a very significant procedure that yields 
the i'th coordinate value for the entity obtained by such ultrafinite combinations. 
Such entities are called {{\it intermediate propertons}}. Let $m_0\geq 0$ be the 
N-world proper mass for an assumed elementary particle denoted by $\rm 
k^\prime.$ If $m_0 = 0,$ then let $\lambda = 1.$ Otherwise$,$ from Theorem 11.1.1$,$ we know that there is a $\lambda \in\hypernat
$ such that $\lambda /(10^\omega) \in \monad {m_0},$  where $\omega \in 
\nat_\infty$ and since $m_0 \not=0,$ $\lambda \in \nat_\infty.$ 
Consequently$,$ for $b_n = 10^{-\omega},$ the *-finite sum
$$\sum_{n=1}^\lambda b_n=\sum_{n=1}^\lambda {{1}\over{10^\omega}} = {{\lambda}\over {10^\omega}}\leqno 
(11.1.5)$$
has the property that $\st {\sum_{n=1}^\lambda 1/(10^\omega)} = m_0.$ (Note the special summation notation for a constant summand.) The 
{standard part operator st is an important external operator that is a continuous 
[11] NSP-world process that yields N-world effects.} The appropriate 
interpretation is that\parm 
{\leftskip 0.5in \rightskip 0.5in \noindent {\it ultrafinite combinations of 
ultra-propertons yield an intermediate properton that$,$ after application 
of the standard part operator$,$ has the same effect as an elementary particle 
with proper mass $m_0.$}\par}\parm
An additional relevant idea deals with the interpretation that the *-finite 
set $\rm G_A^{\prime\prime}$ exists at$,$ say$,$ nonstandard time$,$ and that such a 
set is manifested at standard time when the operator st is applied. The 
standard part operator is one of those external operators that can be 
indirectly detected by the presence of elementary particles with proper mass 
$m_0.$ \pars
The above discussion of the creation of {intermediate propertons} yields a 
possible manner in which ultra-propertons are combined within the NSP-world
to yield appropriate energy or mass coordinates for the multifaceted 
propertons. But is there an indication that all standard world physical 
qualities that are denoted by qualitative measures begin as infinitesimals?   
\pars 
Consider the infinitesimal methods used to obtain such things as the charge on 
a sphere$,$ charge density and the like. In all such cases$,$ it is assumed that 
charge can be infinitesimalized. In 1972$,$ it was shown how a classical theory 
for the electron$,$ when infinitesimalized$,$ leads  to the point charge concept 
of quantum field theory and 
then how the *-finite many body problem produced the quasi-particle. [15] 
Although this method is not the same as the more general and less ad hoc 
properton approach$,$ it does present a procedure that leads to an 
infinitesimal charge density and then$,$  in a very ad hoc manner$,$ it is assumed 
that there are objects that when *-finitely combined together entail a real 
charge and charge density. Further$,$ it is the highly successful use of the 
modeling methods of infinitesimal calculus over hundreds of years that has
lead to our additional presumption that all coordinates of the basic 
sequential properton representation are a $\pm$ fixed infinitesimal. \pars
In order to retain the general independence of the coordinate representation$,$ 
{{\it independent *-finite coordinate summation}} is allowed$,$ recalling that 
such objects are to be utilized to construct many possible universes. [This 
is the same idea as *-finitely repeated simple affine or linear transformations.] Thus$,$ 
distinct from coordinatewise addition$,$ *-finitely many such sequences can be 
added together by means of a fixed coordinate operation in the following sense.
Let $\{a_i\}$ represent an ultra-properton. Fix the coordinate $j,$ then the 
sequence $\{c_i\},\ c_i = a_i ,\ i \not= j$ and $c_j = 2a_j$ forms an 
intermediate properton.  As will be shown$,$ it is only after the formation of 
such intermediate propertons that the customary coordinatewise 
addition is allowed and this yields$,$ after the standard part operator is 
applied$,$ representations for elementary particles. Hence$,$ from our previous 
example$,$ we have that ultrafinite combinations of ultra-propertons yield 
propertons with ``proper mass'' $\lambda/(10^\omega) \approx m_0$ while all 
other coordinates remain as $\pm 10^\omega.$ This
physical-like process is not a speculative ad hoc construct$,$ but$,$ rather$,$
it is modeled  after what occurs in our observable natural world.
Intuitively, this  type of summation is modeled after the process of
inserting finitely  many pieces of information (mail) into a single
``postal box$,$'' where  these boxes are found in rectangular arrays in post
offices throughout the  world.\pars 
Now other ultra-propertons are ultrafinitely combined and yield for a specific 
coordinate the $\pm$ unite charge or$,$ if quarks exist$,$ other N-world charges$,$ 
while all other coordinates remain fixed as $\pm 1/(10^\omega)$$,$ etc. 
Rationally$,$ how can one conceive of a combination of these intermediate 
propertons$,$ a combination that will produce entities that can be 
characterized in a standard particle or wave language?\pars
Recall that a finite summation is a *-finite summation within the NSP-world. 
Therefore$,$ a finite combination of intermediate propertons is an allowed 
internal process. [Note that external processes are always allowed but with 
respect to our interpretation procedures we always have direct or indirect knowledge 
relative to application of internal processes. Only for very special 
and reasonable external processes do we have direct or indirect knowledge that 
they have been applied.]  Let $\gamma_i \in \monad 0,\ i= 1,\ldots,n.$ Then 
$\gamma_1 + \cdots \gamma_n \in \monad 0.$ The {final stage in properton 
formation for our universe} --- the final stage in particle or wave 
substratum formation --- would be {\it finite coordinatewise} summation of 
finitely many intermediate propertons. This presupposes that the N-world 
environment is characterized by but finitely many qualities that can be 
numerically characterized. This produces the following type of coordinate 
representation for a specific coordinate $j$ after $n$ summations with $n$ 
other intermediate propertons that have only infinitesimals in the $j$ 
coordinate 
position.
$$\sum_{i=1}^{\lambda}(1/(10^\omega) + \sum_{i=1}^n\gamma_i. \eqno (11.1.6)$$
Assuming $\lambda$ is one of those members of $\nat_\infty$ or equal to 1 as 
used in (11.1.5)$,$ 
then the standard part operator can now be applied to (11.1.6) and the result 
is the same as $\st {\sum_{i=1}^{\lambda}(1/(10^\omega)}.$\pars

The process outlined in (11.1.6) is then applied to finitely many distinct 
intermediate propertons --- those that characterize an elementary particle. 
The result is a properton each coordinate of which is infinitely close to 
the value of a numerical characterization or an infinitesimal. When the 
standard part operator is applied under the  usual coordinatewise procedure$,$ 
the coordinates are either the specific real coordinatewise characterizations 
or zero. Therefore$,$ N-world formation of particles$,$ the dense substratum 
field$,$ or even gross matter may be accomplished by a ultrafinite combination
of ultra-propertons that leads to the intermediate 
properton; followed by finite combinations of intermediate propertons that 
produce the N-world objects. Please note$,$ however$,$ that prior to application 
of the standard part operator such propertons retain infinitesimal nonzero
coordinate characterizations in other noncharacterizing positions. (See note (1) on 60.)\pars  
             
We must always keep in mind the hypothesis law [9] and {avoid unwarranted 
speculation}. We do not speculate whether or not the formed particles have 
{point-like} or  ``{spread out}'' properties within our space-time environment. 
These additional concepts may be pure {catalyst} type statements within some 
standard N-world theory and could have no significance for either the N-world 
or NSP-world. \pars
With respect to field effects$,$ the cardinality of the set of all 
ultra-propertons clearly implies that there can be ultrafinite combinations   
of ultra-propertons  ``located'' at every   ``point'' of any 
finite dimensional continuum. Thus the field effects yielded by propertons 
may present a {completely dense continuum} type of pattern within the N-world 
environment although from the monadic viewpoint this is not necessarily how 
they   ``appear'' within the NSP-world.\pars
There are many scenarios for {quantum transitions} if such occur in objective 
reality. The simplest is a re-ultrafinite combination of the ultra-propertons 
present within the different objects. However$,$ it is also possible that this 
is not the case and$,$ depending upon the preparation or scenario$,$ the 
so-called   ``conservation'' laws do not hold in the N-world. \pars
As an example$,$ the {neutrino could be a complete fiction}$,$ only endorsed as a 
type of catalyst to force certain laws to hold under a particular scenario. 
Consider the set of sentences \parm
\line{\hfil${\rm G_B = \{An\sp elementary\sp particle\sp}
\r k^\prime(\r i^\prime,\r j^\prime)\sp{\rm with\sp}$\hfil}\pars
\line{\hfil${\rm total\sp energy\sp}\r c^\prime{+}\r n^\prime.\mid ((i,j,n) \in $\hfil}\pars
\line{(11.1.7)\hfil$\nat^+ \times 
\nat^+ \times \nat^+)\land (1\leq k\leq m )\}.$\qquad\hfil}\parm
It is claimed by many individuals that such objects as being described in $\rm 
G_B$ exist in objective reality. Indeed$,$ certain well-known scenarios for a 
possible cosmology require$,$ at least$,$ one  ``particle'' to be characterized by 
such a collection $ \rm G_B.$ By the usual method$,$ these statements are *-
transferred to \parm 
\line{\hfil${\rm G_B^\prime = \{An\sp elementary\sp particle\sp}
\r k^\prime(\r i^\prime,\r j^\prime)\sp{\rm with\sp}$\hfil}\pars
\line{\hfil${\rm total\sp energy\sp\r c^\prime{+}\r n^\prime.\mid ((i,j,n) \in }$\hfil}\pars
\line{(11.1.8)\hfil$\hypernat^+ \times 
\hypernat^+ \times \hypernat^+)\land (1\leq k\leq m )\}.$\qquad\hfil}\parm
\noindent Hence$,$ letting $n \in \nat_\infty$ then various   {``infinite'' 
NSP-world energies} emerge from our procedures. With respect to the total 
energy coordinate(s)$,$ ultra-propertons may also be ultrafinitely combined to 
produce such possibilities. Let $\lambda = 10^{2\omega}$ [ resp. $\lambda = 
\omega^2$] and $\omega \in \nat_\infty.$ Then 
$$\sum_{n=1}^\lambda {{1}\over{10^\omega}} =10^\omega \  {\rm 
[resp.\ }\sum_{n=1}^\lambda {{1}\over {\omega}} = \omega] \in 
\nat_\infty.\leqno (11.1.9)$$\par
Of course$,$ these numerical characterizations are external to the N-world. 
Various distinct {``infinite'' qualities can exist} rationally in the NSP-world 
without altering our interpretation techniques. The behavior of the infinite 
hypernatural numbers is very interesting when considered as a model for 
NSP-world behavior. A {transfer of finite energy}$,$ momentum and$,$ indeed$,$ all other 
N-world characterizing quantities$,$ back and forth$,$ between these two worlds is 
clearly possible without destroying  {NSP-world infinite conservation 
concepts}. 
\pars

Further$,$ observe that  various intermediate propertons carrying nearstandard 
coordinate values could be present at nearstandard space-time coordinates$,$ and 
application of the continuous and external standard part operator would 
produce an apparent not conserved N-world effect. These concepts will be 
considered anew when we discuss the Bell inequality.\pars

Previously$,$ ultrawords were obtained by application of certain concurrent 
relations. Actually$,$ basic ultrawords exist in any elementary nonstandard 
superstructure model$,$ as will now be established for the general paradigm. 
\pars
Referring back to $\rm G_A$ equation (11.1.1)$,$ for some fixed $k,\ 1\leq k\leq 
m,$ let 
$\r h_k \colon \nat^+ \times \nat^+ \times \nat^+ \to  \rm G_A$ be 
defined as follows: $\r h_k(i,j,n) = {\rm An\sp elementary\sp particle\sp}$ $\r k^\prime 
(\r i^\prime,\r j^\prime){\rm \sp with\sp total\sp}$ ${\rm energy\sp} 
\r c^\prime{+} 1/(\r n^\prime).$ Since the set $F(\nat^+ \times \nat^+ \times 
\nat^+)$ is denumerable$,$ there exists a bijection $\r H\colon \nat \to$\break $ 
F(\nat^+ \times \nat^+ \times \nat^+).$ For each $1\leq \lambda \in \nat$ and fixed 
$i,\ n \in \nat^+,$ 
let $\rm G_A(\lambda) = \{{\rm An\sp elementary\sp particle\sp} \r k^\prime 
(\r i^\prime,\r j^\prime) {\rm \sp with\sp total\sp energy\sp} 
c^\prime{+}1/(n^\prime).\mid (1\leq j\leq \lambda)\land$ $(j \in \nat^+) \}.$ 
Let $p \in \nat.$ If $\vert \r H(p) \vert \geq 2,$ define finite $\r M(\r h_k[\r H(p)]) = 
{\rm \{A_1\sp and\sp A_2\sp and\sp\cdots \sp and\sp} \r A_m\},$ where ${\rm A_j} \in 
\r h_k[\r H(p)], m = \vert \r H(p)\vert.$ If $\vert \r H(p) \vert \leq 1,$ then define
$\r M(\r h_k[\r H(p)]) = \emptyset.$ Let $\r M^0 = \bigcup\{\r M(\r h_k[\r H(p)])\mid p \in \nat 
\}.$ Please note that the $\r k^\prime$ represents the  ``type'' or name of the 
elementary particle$,$ assuming that only finitely many different types exist,
$\r i^\prime$ is reserved for other purposes$,$ and the $\r j^\prime$ the number of such elementary particles of 
type $\r k^\prime.$ \parm
{\bf Theorem 11.1.3} {\sl For any $i,\ n,\ \lambda \in \hypernat^+,$ such that $2 
\leq \lambda,$ there exists $w \in \Hyper {\bf M^0 - G_A},\ \Hyper{\bf 
G_A(\lambda)} \subset \Hyper{\b S(\{w\})}$ and if $A \in \Hyper {\bf G_A} - 
\Hyper{\bf G_A(\lambda)},$ 
then $A \notin \Hyper{\b S(\{w\})}.$} \pars
Proof. Let $i,\ j,\ \lambda \in \nat^+$ and $ 2 \leq \lambda.$ Then there 
exists some $ r\in \nat$ such that $\r h_k[\r H(r)] = {\rm G_A(\lambda)}.$ From 
the construction of $\r M^0,$ there exists some $r^\prime \in \nat$ such that 
$\r w(r^\prime) = {\rm An\sp elementary\sp 
particle\sp}\r k^\prime(\r i^\prime,1^\prime){\rm \sp with\sp total\sp energy \sp} 
\r c^\prime{+}1/(\r n^\prime).\break {\rm \sp and\sp An\sp elementary\sp 
particle\sp}\r k^\prime(\r i^\prime,2^\prime){\rm \sp with\sp total\sp energy 
\sp}
\r c^\prime{+}1/(\r n^\prime).\sp {\rm and\sp \cdots\sp and\sp}$\break${\rm An\sp elementary\sp 
particle\sp}\r k^\prime(\r i^\prime,\lambda^\prime){\rm \sp}{\rm with\sp total\sp 
energy \sp} \r c^\prime{+}1/(\r n^\prime). \in \r M[\r h_k\r H(r^\prime)].$ Note that 
$\r w(r^\prime) \notin {\rm G_A},\ \r h_k\r [\r H(r^\prime)] \subset \r S(\{\r 
w(r^\prime)\})$ and if ${\rm A \in G_A} -\r h_k[\r H(r)],$ then $\rm A \notin \rm 
S(\{\r w(r^\prime)\}).$ The result follows by our embedding and *-transfer. \qed
The ultrawords utilized to describe various propertons$,$ whether obtained as 
in Theorem 11.1.3 or by concurrent relations$,$ are called {{\it ultramixtures}} 
due to their applications. The ultrafinite choice operator $\bf C_1$ can 
select them$,$ prior to application of $\Hyper {\b S}.$ Moreover$,$ application of 
the ultrafinite combination operator entails a specific intermediate 
properton with the appropriate nearstandard coordinate characterizations. 
Please notice that the same type of sentence collections may be employed to 
infinitesimalize all other quantities$,$ although the sentences need not have 
meaning for certain popular N-world theories. Simply because substitution of 
the word ``charge'' for  ``energy'' in the above sentences  $\rm G_A$ does not 
yield a particular modern theory description$,$ it does yield the infinitesimal 
charge concept prevalent in many older classical theories. \pars
Using such altered $\rm G_A$ statements$,$ one shows that there does exist 
ultramixtures $w_i$ for each intermediate properton and$,$ thus$,$ a single
ultimate ultramixture $w$ such that $\Hyper {\b S}(\{w_i\}) \subset 
\Hyper {\b S}(\{w\}).$ Each particle or elementary particle may$,$ thus$,$ be assumed to 
originate from $w$ through application of the ultralogic $\Hyper {\b S}.$\par

\bigskip
\leftline{{\bf 11.1.2 More on Propertons}}\par
\medskip

The general process for construction of all N-world fundamental entities from 
propertons can be improved upon or achieved in an alternate fashion. One of 
the basic assumptions of subatomic physics is that in the Natural-world two 
fundamental subatomic objects$,$ such as two electrons$,$ cannot be differentiated 
one from another by any of its Natural-world properties. One of the 
conclusions of what comes next is that in the NSP-world this need not be the 
case. Of course$,$ this can also be considered but an auxiliary result and need 
have no applications.
At a 
particular instant of (universal) time$,$ it is  possible to associate with 
each entity a distinct ``name'' or identifier through properton 
construction. This is done through application of the properton naming 
coordinate  $a_1.$ As will be shown$,$ the concept of independent *-finite 
coordinate summation followed by n-tuple vector addition can be accomplished 
by means of a simple linear transformation. However$,$ by doing so$,$ the concept 
of the *-finite combinations or the gathering together of propertons  
as a NSP-world physical-like process is suppressed. Further$,$ a simple method 
to identify each N-world entity or Natural-system is not apparent. Thus$,$ we 
first keep the above two processes so as to adjoin to each entity constructed 
an 
appropriate identifier. \pars

There is a coding $i$ used in the original foundations, that codes each member of ${\cal W}'$ via the natural numbers. (Herrmann, 1979-1993). This coding can be dropped and has no effect upon and the results in the new foundations that include a general language such as ${\cal W}'$ as a subset of the ground set. This embedding is now removed as allowed by the constructions in Herrmann (1979-1993). Further, a general language ${\cal W}'$ is now considered as a subset of the ground set for the standard model ${\cal M}_1.$ As mentioned in Herrmann (1997-93), the members of ${\cal W}'$ can be expressed in a different color than all other symbols used. (Apparently, Robinson (1963) was the first to use only such a ground set, (i.e. not using $\cal E$). He differentiates between members of ${\cal W}'$ by simply stating that they are individuals and are different relative to all the symbols that appear in another set.)  \pars 

There is a bijection $w \colon {\cal W}'\to {\cal E},$ such that, for $a \in {\cal W}',$ $\ w(a) = [g].$ The only difference between whether $i$ is included or not is that for an $[k] \in {\cal E},$ the range of the partial sequence $k$ is a subset of $i[{\cal W}']$, a set of natural numbers, or the range is a subset of ${\cal W}'$, respectively. It is but a matter of ``style'' whether $i$ is included or not.  There are two special unique members in each $[g], f_0,\ f_m.$ Intuitively, via the  Markov join operator, $f_m \in [g],\ f_m \in {{\cal W}'}^{[0,m]}$ and the ``ordered'' $a = f_m(m)\cdots f_m(0),$ and each $f_m(j) \in {\cal A}'$ the extended alphabet. Then $f_0 \in [g]$ and $f_0(0)= a.$ Of course, relative to a physical or non-physical world, all such mathematically modeled objects but ``represent'' entities or processes. \pars 

{\bf Definition 1.1.} For fixed even $K >2$ and $n \in \nat'$, consider the sets of K+2-tuples $\r C_{n} = \{(\r h(j), 1, -1/10^n, 1/10^n,\ldots,-1/10^{n}, 1/10^n)\mid (i \in \nat)\land(n\in \nat) \}$.   \parm

From definition 1.1, via *-transform, we have the hyperfinite set
$$\Hyper C =\{ (\Hyper {h}(j), 1, -1/10^n, 1/10^n,\ldots,-1/10^{n}, 1/10^n)\mid (j \in \hypernat)\land(n\in \hypernat)\}.$$
 Let $\omega \in \hypernat' - \nat'.$ \pars

{\bf Definition 1.2, Ultra-propertons}. The set of all ultra-propertons is (represented by) the hyperfinite ${C} =\{ (\Hyper {h}(j), 1, -1/10^\omega, 1/10^\omega,\ldots,-1/10^\omega, 1/10^\omega)\mid (j \in \hypernat)\}.$ \parm

For Definition 1.2, it is assumed that there is no more than $K$ physical or physical-like numerical or coded descriptive characteristics for the any elementary entity. \parm

Let $r_1 \in \real.$ By Theorem 11.1.1 in Herrmann (1979-1993), there is a $\lambda_1 \in \nat_\infty$ such that $\lambda_1/10^\omega \in \monad {|r|}.$ Hence, $\st {(\lambda_1/10^\omega)} = |r|.$ Then there are $K$, $\lambda_i, \ i \in [1,K]$ that yield the $K$ characteristics. For an elementary entity $\r e_j$, some characteristics can be 0, meaning that the measure has value 0. Throughout the combining processes, if a coordinate retains its infinitesimal value $\pm 1/10^\omega,$ this indicates that the characteristic has no meaning for $\r e_j.$ In order to indicate these differences, any characteristic that has measure 0 is obtained from a combination of two ultra-propertons. The standard part physical realization operator $\st {\cdot}$ is only applied to coordinates of the intermediate properton representations with the form $\pm \lambda/10^\omega,$ where $\lambda \geq 2.$\pars

There are other characteristics such as spin, where the 0 takes on a different meaning. However, such coding is rather arbitrary and can be replaced with non-zero numbers or non-zero codings for the characteristics so as to not confuse them with a 0 measurement. For the needed intermediate properton $e_1$, with a third coordinate  characteristic  under independent coordinate addition, the hyperfinite set of ultra-propertons $\{(\Hyper h(j),1,-1/10^\omega, \ldots, 1/10^\omega)\mid j \in [1,\lambda_1]\}$ is employed. Hence, the first intermediate properton is $(\Pi_{1}^{\lambda_1}, \lambda_1, -\lambda_1/10^\omega,1/10^\omega, \ldots, 1/10^\omega).$ For a forth coordinate intermediate properton for value $r_2,$ consider $\{(\Hyper h(i),  \lambda_2,-1/10^\omega, \lambda_2/10^\omega,\ldots, 1/10^\omega)\mid i \in [\lambda_1 +1,\lambda_1 +\lambda_2]\}.$ Continue these definition for each member of $[1,K].$ Thus the entire collection of ultra-propertons used to obtain one of the $e_1$ entities is $\lambda_1 + \cdots +\lambda_K = \delta_1 \in \nat_\infty.$ \pars

It is assumed that there are a nonempty countable (i.e non-zero finite or denumerable) collection of $\{e_i\}$ needed. Thus there is a non-zero finite or denumerable set $\{\delta_i\}$ and in the finite case, consider $\sum \delta_i \in \nat_\infty.$ Next consider $\{\delta_i\mid i \in \nat'\}.$ Then $\{\delta_i\mid i \in \nat'\} \subset \hypernat.$ The $|\{\delta_i\mid i \in \nat'\}| < |{\cal M}_1|^+$. Hence, there is a $\gamma_1 \in \nat_\infty$ such that $\{\delta_i\mid i \in \nat'\} \subset [1,\gamma_1]$ by application of Corollary 1.1.1 and Theorem 1.2. Thus, in both cases, there is a $\Gamma_1 \in \nat_\infty$ such that $\{\delta_i\} \subset [1,\Gamma_1]$. This shows that there are ``enough'' ultra-propertons to produce the set $\{e_i\}.$ For another type of elementary particle, simply repeat this for the identifiers $h(j),\ j > \Gamma_1].$ Then continue by induction. \pars

For this application, it appears unnecessary to consider more than $H$, where  $1 \leq H \in \nat$, different types of elementary entities. The set of ultra-propertons $\{(\Hyper h(j), 1,-1/10^\omega, \ldots, 1/10^\omega)\mid j \in \hypernat'\}= C$ is an internal set and as such the hyperfinite operator $\hyper {\cal F}$ is defined for it. For properton generation, a universe can be considered as a collection of physical-systems. Hence application of a finite ($>0$) iteration $\hyper {\cal F}^i$ to $C$ yields ${\cal C}=\bigcup \{\hyper {\cal F}^i({C})\mid (1\leq i \leq n)\land (i \in \nat)\},$ an internal collection that is sufficient to generate the physical-systems for any of the presently considered cosmologies. To accommodate the formation of the physical-like systems, infinite hyperfinite set $X$ of internal sets that is disjoint from $\bigcup \{\hyper {\cal F}^i({C})\mid (0\leq i \leq n)\land (i \in \nat)\}$ is adjoined to $\bigcup \{\hyper {\cal F}^i({C})\mid (0\leq i \leq n)\land (i \in \nat)\}$ Then $\Pi^+ = {\cal C}\cup X.$\pars 

 Rather than the $\r f^q(i,j)$ being a general description, one considers instructions or rules $\r I^q(i,j)$ - a nonempty finite subset of ${\cal W}'$, which is equivalent to a single word in ${\cal W}'$. These sets of instructions - instruction-sets - (also called instruction-information) are indexed in the same way as the general descriptions and determine the {\it instruction paradigm} ${\cal I}_q$. Indeed, there is an injection $\r H$ on  $\r d_q$  onto ${\cal I}_q$, where $\r H(\r f^q(i,j)) = \r I^q(i,j)$ and $(i,j)$ varies over the same set of integers and natural numbers, respectively. (See appendix.) There is one instruction paradigm for each pre-designed universe and there can be a vast collection of such universes. Rather than simply applying this bijection as a means to reproduce each of the instruction paradigm results, the instruction paradigm is analyzed directly.  \pars

As symbol strings, the set $\dag \{\rm There\sp are\sp n'\sp ultra-propertons\sp combined\sp to\sp$ $\rm produce\sp an\sp intermediate\sp properton. \mid n\in \nat\}$ is a member of $\cal N.$ But the terms ``ultra-properton'' and ``properton'' have no physical meanings within the physical world. Further, it corresponds under $w$ to a specific subset of $\cal E.$ Using the method of Theorem 9.3.1 in Herrmann (1979-93) and *-transform, this set corresponds to a subset of $\Hyper {{\cal W}}'$ as well as a subset of $\Hyper {\cal E} - {\cal E},$ where the $n'$ now varies over members of $\nat$ for one subset and $\nat_\infty$ for the other. But the language ${\cal W}'$ does not have alphabet symbols that correspond to members of $\nat_\infty.$ For specific members of $\nat_\infty,$ representative symbols, constructed or otherwise, that are not members ${\cal W}'$ are employed. There are objects in $\Hyper {{\cal W}} - {\cal W}$ that do correspond to symbols in the extend alphabet $\Hyper {\cal A}'.$ One such object is denoted in the following display by the $\lambda.$ This leads to a specific ordered string of symbols via the corresponding member of $\Hyper {\cal E}$ that is intuitively represented by 

$$\dagger \rm There\sp are\sp \lambda\sp ultra-propertons\sp combined\sp to\sp$$ $$\rm produce\sp an\sp intermediate\sp properton.$$

Such $\dagger$ *instructions have ``interpretations'' in terms of the GGU-model language as do other members of $\Hyper {\cal M}$ or $\Hyper {\cal M}_1.$ They can have additional symbol strings taken from ${\cal W}'$ that have no meanings until interpreted for the GGU-model. \pars
\pars

Each physical-system is actually a physical-like system since their construction includes *instructions for, at present,   unknowable processes or ``things,'' as represented by the $X,$ that seem to ``force'' these combinations to occur.  Notice that symbol-strings, diagrams, images and sensor information represented by a standard language ${\cal W}'$ are comprehensible only when they carry an additional component  - meanings. "Meanings" are understood by the mind and cannot lead to mere circular thinking. \pars

For human endeavors, there is a five-step process. (1) A standard meaningful instruction-set. (2) The instructions are mentally comprehended. (3) This mental comprehension is transmitted, via electro-chemical actions, to other human physical locations. (4) At these physical  locations actions are performed. (5) These actions produce a physical entity that corresponds to the original  instruction-set. Errors can occur along this entire path. If these processes are performed in an errorless manner and the physical entity produced does not correspond to a desired physical object, then this alone cannot change the instruction-set. The instruction-set (1) would need to be altered in a meaningful way as required by (2). This alteration is done by an intelligent physical entity. For the GGU-model, a standard meaningful instruction-set contains operative statements.\pars 

Aside from such human endeavors, Nature neither displays physical laws nor numerical parameters. We display such laws and parameter values as representations for behavior we cannot otherwise comprehend. By definition, such objects as quantum fields or strings are not displayed by Nature as independent agents. Again they are used as representations for behavior we cannot, at present, otherwise comprehend. Hence, distinct from human endeavors, steps 1 - 5 represent behavior we cannot, at present, otherwise comprehend. \pars

 From the modeling viewpoint, processes within the substratum are non-physical. Thus, step (1) is non-physical and steps (2) and (3) are consider as statements relative to a  non-physical medium and  a non-physical mode of transmission, respectively. For each universe-wide frozen-frame, step (4) corresponds to  a subsets of $\cal C$ - the substratum info-fields. Step (5) is the application of the $\st {\cdot}$ operator and yields that physical products of the entire process. In order to be operative, the above displayed *instruction as well as similar ones have non-physical components.\pars

The process of applying the standard part operator does not erase the 
original coordinates of the constructed objects. The process only yields the 
physical-world codes. Such alterations of the fundamental subatomic or field 
entities is  modeled by
vector space subtraction and  this would be followed by 
independent *-finite subtraction. This represents a ``breaking apart'' of 
the fundamental entity into its original ultra-properton constituents. 
The original identifiers are restored. An altered constituent is then obtained by 
repeating the construction process. This can be used to eliminate the virtual 
particle or process concept within reality models. The models that need such concepts 
to predict 
behavior are now just that  ``models'' for predicted behavior and would not
require the virtual ``stuff'' to exist in reality.\pars

As mentioned$,$ the formation of fundamental entities from ultra-propertons 
through the process of the intermediate properton formation can be modeled 
by means of  a very simply (diagonal styled) *-continuous 
linear transformation. Let 
a fundamental entity have the required finitely many coordinate numerical 
characteristics as discussed previously and let nonempty finite $A \subset 
\hypernat$ be the specific coordinate names. Then there is a finite set of
hypernatural numbers $\{\lambda_i\mid I \in A\}$ that represent the 
number of summands in the respective
*-finite summation process. (The process that yields the intermediate 
propertons.) Consider an $n \times n$ hypermatrix with the diagonal elements
$\lambda_i = b_{ii},\ i \in A,$ and $b_{jj} = 1,\  j \in (\hypernat - A),\ j \leq 
n$ and
$b_{ij} = 0$ otherwise. Then letting $(a_j)$ be a coordinate representation for an ultra-properton, consider $(b_{ij})(a_j)^{\r T}.$ This hypermatrix represents an internal function
that will take a single {\it representation} for an ultra-properton and  
will yield$,$ after application of the  standard part operator to the 
appropriate coordinate values (or all of them if one wishes)$,$ a {\it 
representation} for a fundamental entity. BUT note that identifier coordinate 
would not be that of the intermediate properton. 
Unless we use the sum of the $\lambda_i$s in the second 
diagonal place$,$ the counting coordinate would not be the appropriate value. 
Of course$,$ the 
 matrix inverse would yield a representation of an original ultra-properton 
with its identifier. 
Thus using either 
method$,$ the identifier would be lost to the N-world.  
But using one of the favorite simple mathematical models$,$ the matrix approach,
the fundamental entity identifier would be lost to the NSP-world as well. 
This shows how two 
different mathematical procedures can lead to equivalent N-world results$,$ but 
yield considerably different NSP-world ramifications. \pars  
The next step in a properton re-formulation of particle physics would be to 
consider specific NSP-world predictions for what is claimed are random or$,$ at 
least$,$ probabilistic in character actual observed events. In this regard$,$ I 
mention the known fact that a probably density functions behavior can be 
predicted by a specific well-defined sequence. For example$,$ it is 
known that there is a real number $x$ and a specifically defined sequence
$x2^n$ such that if one takes the fractional part of each $x2^n$$,$ less than 
1/2 and correspond it to H$,$ and the fractional part greater or equal to 1/2 and 
correspond it to a T$,$ then the predictive results will pass every statistical 
test that implies that H and T are randomly obtained by a sequence of H = 
heads$,$ T = tails coin tosses. Clearly$,$ this process$,$ as far as the sequence 
$x2^n$ 
is concerned is a predictable process and contradicts scientific randomness.
But the specific numbers $x$ need  not be known by entities within 
the Natural universe.\par\bigskip

\noindent {\bf 11.2 An Ultimate GGU-model Conclusion.}\parm

In the book ``Ultralogics and More'' other properton properties are 
discussed. These include the ultraenergetic and ultrafast propertons. The 
ultrafast propertons are of significance in that they can be used to provide 
seemingly instantaneous informational transmissions throughout the Natural 
universe. This is significant for the ultralogic, ultraword (and ultra-logic-logic) generation of a 
universe.  Clearly the research presented in this monograph is not complete 
but is only designed to present the beginning concepts in what is hoped will 
be a continuing research activity. One important question is relative to the 
fact that an ultimate ultraword such as $w'$ generates a preselected or 
theory generated ideal 
universe. How can this be made to correspond to the actual universe in which 
we live where each Natural-system is perturbed or altered within certain 
limits dictated by Natural law from this ideal case? Such alterations can also be produced by 
partially independent agencies such as biological entities. 
There are different ways to attack this 
problem but the existence of the UN-events would require that speculation be 
restrained.\pars 

One method is slightly similar to the Everett-Wheeler-Graham many-worlds 
interpretation (parallel universes) but is much less esoteric in 
character. One can consider countably many
$w_t'$ from Corollary 7.3.4.1 each one containing the allowable alterations in 
Natural-system behavior starting from the moment $t.$ Denote the 
corresponding set of universes by $\cal U.$ At any instant of 
substratum time $t,$ all the 
universes but one are ``covirtual'' universes. This does not necessary mean that 
they actually exist in some type of NSP-world reality. This simply means that 
they exist potentially. When any one or more Natural-systems is perturbed 
then the UN-events react first so to speak. Information relative to this 
reaction is transmitted within $\{w_t'\}$ by means of ultrafast propertons. 
This information causes a  $w'_{U_t}$ to be selected such that 
$U_t \subset \Hyper {\b S(\{w'_{U_t}\})},\ U_t \in \cal U$. For the  ``next'' 
instant, this 
generated $U_t$ becomes the Natural universe objective reality. Now technically there 
would be a ``next'' instant since there would be an NSP-world time interval 
over 
which all of the Natural-systems would have a fixed frozen segment segments 
that would not be altered. The only possible alterations would be in the 
UN-events. This process then continues throughout all of Natural universe 
time.\pars 
There is a second approach that can solve this perturbation problem but 
requires a complete restructuring of ultraword generation. This method is 
similar 
to the concept of ``parallel'' logic. Time in this case is modeled as previously done but on interval
$(-\infty,+\infty).$ At a particular observer beginning instant $t,$ a time 
slice in the mathematical 
sense is obtained. This slice will yield a frozen segment describing a
Natural event for 
each of the present Natural-systems. This will lead as in Chapter 7 to an 
ultraword $x$ that will by means of $\Hyper {\b S}$ generate this slice.  
Let ${\cal NA}$ denote the set of all potential Natural-systems. This means 
that some of these at the instant $t$ contain only frozen segments from 
$\Hyper {\b T_i} - \b T_i.$\pars 
 Assume that ${\cal NA}$ is at the least 
denumerable. Then each $x_t$ (the ultraword associated with the instant $t$) 
also contains UN-events. Again the alterations 
in Natural-system behavior within the limits of Natural law lead to changes  
in the frozen segment descriptions for the very ``next'' Natural event. These 
changes are incorporated within the NSP-world time interval into the 
``next'' ultraword $x_{t_1}$. Application of $\Hyper {\b S}$ to $x_{t_1}$ now 
yields the ``next'' slice. This process continues throughout all of the Natural 
universe time frame. Of course the ``time'' being referred to here is not 
N-world measured time but the time concept previously discussed that is 
concerned with  
``before and 
after.''\pars 
The first of the above models appears to be the most significant and the one 
that is probably the closest to objective reality. This portion of this model leads to the moment-to-moment re-generation of our universe since its behavior most certainly is perturbed moment-to-moment from the ideal.\pars 

As asked previously, is there something within Nature that ``forces'' physical behavior within our universe to follow expressed laws as well as to satisfy expressed parameter values? GGU-model descriptions represent non-physical substratum entities and behavior that provide an answer to this question. 
\parm
\centerline{\bf CHAPTER 11 REFERENCES}
\smallskip
\noindent {\bf 1} Beltrametti E. G. Enrico G. and G. Cassinelli {\it The 
Logic of Quantum Mechanics} Encyclopedia of Mathematics and Its Applications 
Vol. 15 Addison-Wesley Reading 1981.\pars 
\noindent {\bf 2} d'Espagnat B. The quantum theory and realism Scientific 
America 241(5)(1979) 177. \pars 
\noindent {\bf 3} {\it Ibid.} \pars 
\noindent {\bf 4} {\it Ibid.} 181. \pars 
\noindent {\bf 5} {\it Ibid.} \pars 
\noindent {\bf 6} {\it Ibid.} 180.\pars 
\noindent {\bf 7} Feinberg G. Possibility of faster-than-light particles 
Physical Review 159(5)(1976) 1089--1105.\pars 
\noindent {\bf 8} Hanson W. C. The isomorphism property in nonstandard 
analysis and its use in the theory of Banach Spaces J. of Symbolic Logic 
39(4)(1974) 717--731.\pars 
\noindent {\bf 9} Herrmann R. A. D-world evidence 
C.R.S. Quarterly 23(1986) 47--53.\pars 
\noindent {\bf 10} Herrmann R. A. The Q-topology Whyburn type filters and 
the cluster set map Proceedings Amer. Math. Soc.  59(1)(1975) 161--166.\pars 
\noindent {\bf 11} Kleene S. C. {\it Introduction to  Metamathematics} D. 
Van Nostrand Co. Princeton 1950.\pars 
\noindent {\bf 12} Prokhovnik S. J. {\it The Logic of Special Relativity} 
Cambridge University Press Cambridge 1967.\pars 
\noindent {\bf 13} Stroyan K. D. and W. A. J. Luxemburg {\it Introduction to 
the Theory of Infinitesimals} Academic Press New York 1976. \pars 
\noindent {\bf 14} Tarski A. {\it Logic Semantics Metamathematics} 
Clarendon Press Oxford 1969.\pars 
\noindent {\bf 15} Thurber J. K. and J. Katz Applications of fractional 
powers of delta functions {\it Victoria Symposium on Nonstandard Analysis}
Springer-Verlag New York 1974.\pars 
\noindent {\bf 16} Zakon E. Remarks on the nonstandard real axis {\it 
Applications of Model Theory to Algebra Analysis and Probability} Holt 
Rinehart and Winston New York 1969. \pars 
\noindent {\bf 17} Note that the NSP-world model is not a local hidden 
variable theory.\pars 
\noindent {\bf 18} Herrmann R. A. Fractals and ultrasmooth microeffects
J. Math. Physics 30(4) April 1989 805--808.\pars 
\noindent {\bf 19} Davis M. {\it Applied Nonstandard Analysis} John Wiley 
\& Sons New York 1977. \pars
\noindent {\bf 20} http://www.raherrmann.com/glossary.htm  \parm 
(1) For propertons, only two possible intrinsic properties for elementary particle formation are here considered. Assuming that there are such things as particles or elementary particles, then they would be differentiated one from the other by their intrinsic properties that are encoded within properton coordinates. When there are particle interactions, these intrinsic properties can be altered or even changed to extrinsic properties. How the alteration from intrinsic to extrinsic occurs probable cannot be known since it most likely is an ultranatural event. For new results on properton formations, see http://arxiv.org/abs/quant-ph/9909078 \pars
(2) To conceive of subparticles properly, quantum theory is viewed as an approximation. Moreover, in terms of physically determined units, the numerical characteristics produced by applications of the standard part operator are  considered as exact. \parm\vfil\eject
\centerline{\bf Appendix-Ultra-logic-systems}\parm

\noindent{\bf 1. Logic-System Generation for Instructions}\parm

As is customary, the nonstandard model used in all of the articles on the GGU-model is a polysaturated polyenlargements (Lobe and Wolff, 2000; Stroyan and Bayod, 1986). In this paper, $q =1,2,3,4$. These numbers denote the four primitive-time intervals (Herrmann 2006) employed for the GGU-model.  The ultraword approach to generate a universe is replaced with an ultra-logic-system. This is a hyperfinite logic-system where, after application of the extended logic-system algorithm, generates each member of the hyperfinite instruction paradigm $d^q_x$ in the proper $\leq_{d^q_x}$ order such that $\b d_q \subset d^q_x \subset \hyper {\b d_q},$ where $q = 1,2,3,4$ and $x = \lambda, \nu\lambda, \mu\lambda,\nu\gamma\lambda$ respectively. Finally, in this article, the term "properton" was previously used. To prevent incorrect mental images as to models for propertons, the term "properton" replaces the term "subparticle." Without visualizing, a properton is an entity characterized only by a list of properties.\pars

The primitive entity that yields physical reality for any GGU-model generated universe is dense collection of ultra-propertons. 
 When first conceived this author had not investigated quantum field theory and did not base proprtins upon any quantum theoretic approach. All of the GGU-model entities and processes can be considered as existing in a {\bf background universe} or {\bf substratum world}. This world can be considered as a physical-like world, where the rules that govern universe formation are distinct from those processes and rules that govern the development of \underbar{any} physical universe. They are simple rules that only refer to counting. This substratum world is also interpreted philosophically in other ways. \pars

If necessary for a specific physical theory, any continuity requirement is satisfied by the properton field (Herrmann, 1983, 1989). For our universe, a collection of propertons has been shown to be closely associated with relativistic effects (Herrmann, 2003). No other known primitive entities, such as superstrings, will have any effect upon the application of propertons as the primitive entities that generate a universe. The processes used to obtain particles and all other physical entities from ultra-propertons need not correspond to the rules of quantum field theory or any additional rules like how quarks combine to form particles. \pars
For our universe, quantum field theory contains descriptions (rules or instructions) that produce such particles from  immaterial fields. Such fields are quantum  mechanical systems and, when represented, have various degrees of freedom. These are but parameters that contribute to the overall state of the system. For various particles, parameters for physical measures or states are the characterizing features of propertons. The physical appearance  and disappearance of particles are trivial applications of properton processes. For quantum field theory, one has the ``creation'' and ``annihilation'' operators that mathematically yield the same results.\pars

For the GGU-model, quantum theory does not produce steps in a development since the method of production must be universe and physical law independent. For our universe, the development ``satisfies'' the predictions of accepted physical theories. I personally consider quantum theory as mostly a product of human imagination that predicts behavior, behavior that we cannot otherwise comprehend. That is, it is a model that mimics. \pars

The GGU-model can be based upon observable human behavior and the mathematics predicts, for our universe, behavior that satisfies the behavior predicted by accepted physical theories. There is a vast amount of evidence for the predicted GGU-model processes. Whether such processes exist in some sort of reality is a philosophic choice. One can make this choice based upon various factors. One can choose to accept properton existence based upon the same philosophy expressed by those that accept that entities postulated in quantum field and particle theory exist.\pars  

The concept of instructions or rules is generalized to instructions that yield a physical reality from combinations of propertons. They are substratum laws. (So as not to confuse these with physical laws, they are called instructions. Further, in what follows, the events that correspond to each $\r f^q(i,j)$ are denoted by $\r E^q(i,j). $) This does \underbar{not} mean that the rules used in quantum theory (QT) actually yield each $\r E^q(i,j).$ As mentioned, what this signifies is that the QT rules are verified via the production of event sequences that yield our universe. For the GGU-model, the physical realization of each $\r f^q(i,j)$ is not the result of any of these physical theories. These theories are but verified by each realized $\r f^q(i,j)$ and they allow us to predict what behavior occurred in or will occur within other realized $\r f^q(p,k).$ For the GGU-model, the ``instructions'' are rather simple ones that lead to all the characteristics that allow one to identify any material entity for any of the presently known cosmologies. \pars 

 Rather than the $\r f^q(i,j)$ being a general description, one considers instructions or rules $\r I^q(i,j)$ - a nonempty finite subset of L, which is equivalent to a single word in L. These sets of instructions - instruction-sets - (also called instruction-information) are also indexed in the same way as the general descriptions and determine the {\it instruction paradigm} ${\cal I}_q$. Indeed, there is an injection $\r H$ on  $\r d_q$  onto ${\cal I}_q$, where $\r H(\r f^q(i,j)) = \r I^q(i,j)$ and $(i,j)$ varies over the same set of integers and natural numbers. There is one instruction paradigm for each pre-designed universe and there can be a vast collection of such universes. Rather than simply applying this bijection as a means to reproduce each of the instruction paradigm results from the developmental paradigm results, what follows is a duplicate of these results and how they are obtained in terms of instruction paradigm notation. \pars

Relative to the GGU-model and generation of a universe, a hyperfinite $\Hyper {\b I}^q(i,j)$ yields a universe-wide frozen-frame. Each instruction $x \in \Hyper {\b I}^q(i,j)$, yields a physical or physical-like system. The physical-systems are disjoint. Each collection of ultra-propertons that yields a specific physical-system is distinct from the set of ultra-propertons that yields any other physical-system. Hence, each physical-system within a universe-wide frozen-frame has a distinct identifier via the collection of all of the identifiers for the ultra-propertons or the intermediate propertons employed to produce the physical-system.\parm

\noindent{\bf 2.  Logic-System Generation for the Type-1 Interval.}\parm

The notation in all that follows is from Herrmann (2006). Notice that there are two different $t$ sequence notations. One $\r t $ is in the informal world, while another $t$ is in the formal standard superstructure. These two sequence are, of course, consider as equivalent since the set of objects that informally yield the informal $\r t$ are also formally present within the standard superstructure. The informal composition $\r f^q =\r I^q \circ \r t^q,$ when embedded relative to $\cal E$  is denoted by $\b f^q = \b I^q \circ t^q$ since the $\r t^q$ is not embedded relative to $\cal E$ and it merely generates a rational number sequence for the embedded informal paradigm.  Usually, these different notations are eliminated and only the math-italics font is employed. This is the customary practice throughout Herrmann (1979 - 1993). Notation for  informal natural, rational and real numbers, if applicable, is usually the same for the informal and more formal superstructure objects. Each $t^q(i,j)$ is a rational number. Each $\b f(i,j)$ is a nonempty instruction-set.\pars

Each member of ${\cal I}_q$ is now considered as determined by a function defined on a set $R_q$ of rational numbers, \b Q. The members of $R_q$ carry the restricted rational simple order and the order $\leq_{{\cal I}^q}$ for the members of ${\cal I}_q$ (the lexicographic order) is order isomorphic  to $R_q$ in the obvious way. Each interval partition is of the form $[c_i, c_{i+1})$ (with a closed interval in two cases), where $i \in \b Z$ and $\b Z$ is the set of integers, and $t^q(i,0) = i,\ t^q(i+1,0) =i +1.$  Then each member of $(c_i,c_{i+1})$ is a defined rational number $t^q(i,j),$ where $i < j < i+1.$ For example, consider $[c_2,c_3)$. Then $t^q(2,1) = 3- 1/2,\ t^q(2,2) = 3 - 1/4,\ t^q(2,3) =  3 - 1/8,$ then, in general, $t^q(2,j) = 3 -1/2^j.$ Hence, $\r f^q(2,0) <_{{\cal I}^q} \r f^q(2,1) <_{{\cal I}^q} \r f^q(2,2) <_{{\cal I}^q} \cdots  <_{{\cal I}^q} \r f(3,0).$ (The order $\leq_{\underline{\cal I}^q}$ is  lexicographic and is isomorphic to the rational number order for a specific set of rational numbers.)\pars

Let ${\cal I}_1$ be the standard instruction paradigm.  An instruction paradigm is defined mathematically in the exact same manner as that of the developmental paradigm in Herrmann (2006) and is equivalent to the range of a sequence $\rm g' \colon \nat\to \power{{\cal W}'},$ where ${\cal W}'$ is our denumerable general language. The first case illustrated for the GGU-model is for a developing universe starting with a frozen segments (frame) instruction-set $\r g'(0)$. For the other three GGU-model cases, this sequence is appropriately modified. In all cases, the $(\r f^q(i,j),\r f^q(p,k))$ is equivalent to ``If $\r f^q(i,j),$ then $\r f^q(p,k)).$ This notation will be simplified later. \pars

For the type-1 case $[0,b],\ b >0$, as indicated above, a denumerable instruction paradigm displays a refined form. 
For $1 < m \in \nat,\  {\cal I}_1 = \{\r f^1(i,j)\mid (0 \leq i \leq m)\land(i \in \b Z)\land (j \in \nat).$ (For each of the types, $\r f(i,j) \in {\cal W}'.$ Using ${\cal I}_1,$ consider the following logic-system.\pars

Due to the simplicity and special nature of the logic-systems used, a simplified algorithm is employed. 
The basic logic-system algorithm is re-defined for sets of two distinct objects $\{ \r A, \r B \}$. If a deduction yields $\r C$ and $\r C$ is a member of $\{ \r A, \r B \}$, then the  ``other'' member is a deduction. Hence, if A is deduced, then from $\{ \r A, \r B\}$, B is deduced. This can be written as $\{\r A, \r B \} - \{\r A\}$ is deduced. In general, this approach is only valid for these special collections of two element sets. This process mimics the proposition-logic modus ponens rule of inference $\{(\r X\to \r Y, \r X, \r Y) \mid \r X,\ \r Y \ {\rm are\ propositions}\}.$ However, for both logic-systems only one member of any two element set is deducible. \parm

{\bf Definition 2.1} Let $i \in \b Z.$ For each $n \in \nat,$ let $\r k^1_i(n) = \{ \{ \r f^1(i,j), \r f^1(i,j+1)\}\mid (0\leq j \leq n-1)\land (j\in \nat)\},\ \r K^1(n) =\bigcup \{\r k^1_i(n)\mid (0\leq i < m)\land(i \in \b Z)\}.$ Finally, let finite $\Lambda^1(n) = \{\r f^1(0,0)\} \cup \r K^1(n) \cup \{ \{ \r f^1(p-1,n),\r f^1(p,0) \}\mid (0< p \leq m)\land (p \in \b Z)\}$ and ${\cal L}^1 =\{\Lambda^1(x)\mid x \in \nat\}.$  The set $\{ \{\r f^1(p-1,n),\r f^1(p,0)\}\mid (0< p \leq m)\land (p \in \b Z)\}$ is called the ``jump elements." Also, each $\Lambda^1(n)$ is a finite set. \parm 

In general, members in ${\cal L}^q$ can be characterized by a first-order sentence.  When the deduction algorithm is applied to $\Lambda^1(n)$ the result is an ordered set of words from ${\cal W}'$ -  the ordered instruction paradigm. In accordance with the juxtaposition join operator that yields words in ${\cal W}'$, this ordered instruction paradigm is a word in ${\cal W}'$L. It can be obtained using the spacing symbol where each member of this paradigm is considered a sentence. For a  multi-universe theory, each such universe is a portion of each of the original members of the instruction paradigm.  \pars

In order to make the notation as simple as possible for the next construction, notice that ${\cal L}^1$ is denumerable.  Let $\nat -\{0\} = \nat'.$ Thus, there is a bijection $\r D^1 \colon \nat' \to {\cal L}^1.$ We use the subscript notation for this bijection. Thus, consider ${\cal L}^1 = \{\r D^1_i\mid i \in \nat'\}$. For each $n \in \nat'$, define $\r M^1_n = \{\{\r D^1_1,\ldots,\r D^1_n\}\}.$ Let ${\cal M}^1 =\{ M^1_n\mid n \in \nat'\}.$ The set $\r M^1_n = \{\{\r D^1_1,\ldots,\r D^1_n\}\}$, as before, can be considered as a single word-like object. \pars

(There are a few typographic errors in Herrmann (2006) and (2006a). For example, in Theorem 4.1, $m >0$ should read $m >1,$ and $\Hyper {\b D}$, should read $\Hyper {\b D}_1.$ In Herrmann (2006a), page 12, in the first (4), the $\nu \in \Hyper {\b Z}^{\geq 0} - \b Z$ should be replaced with $\nu \in \Hyper {\b Z}^{\leq 0} - \b Z$, $\gamma \in \Hyper {\b Z}^{\leq 0} - \b Z$ should be replaced with $\gamma \in \Hyper {\b Z}^{\geq 0} - \b Z$.)\pars
 
A finite consequence  operator S is defined in Herrmann (1979 - 1993, p. 65).  However, a new simplified logic-system ${\cal S}^q,\ q = 1,2,3,4$ is defined. When a logic-system is applied, it generates a specific finite consequence operator. It is the logic-system algorithm that does this.  In this article, this algorithm is explicitly noted since only logic-systems are used.  In general, logic-systems are stated in terms of metamathematics n-tuples. If a set  $\{\r A,\r B,\r C,\ldots,\r D\}$ is used as an hypothesis, then it is word-like since the objects the logical deduction models via the algorithm yields words or word-like objects. \pars

Define ${\cal M}^q, \ q = 2,3,4,$ in the same manner as ${\cal M}^1,$ from members of  ${\cal L}^q$. For each $\r G^q \in {\cal M}^q$, there exists a unique $n \in \nat'$ such that $\r G^q \in \r M^q_n.$  
This $\r G^q = \{\r D^q_1,\ldots,\r D^q_n\},\ \r D^q_i \in {\cal L}^q, \ 1\leq i\leq n.$  \pars 

Define the logic-system that generates $\r S^q$ as
 ${\cal S}^q = \{\{x,y\}\mid (\exists n (n \in \nat')) \land (x \in \r M^q_n) \land (y \in {\cal L}^q )\land (y \in x)\}.$ (This definition can be further described in order to  characterize the doubleton set notion and can include all necessary bounds for the quantifiers.) Further, under the simplification used here, each member of ${\cal S}^q$ is a propositional tautology. Notice that $\r M^q$ is a function with values a singleton set containing an n-set (i.e. a set of ``n'' members).\pars 
 Usually, such a logic-system would use ordered pairs  to model the rules of inference. Within these rules, finite conjunctions are displayed as first coordinates via n-sets. Again the simplified doubleton-set approach is  used here, where one of these sets is $\{\{D_1\}, D_1\}$.   \pars

Hypotheses are considered as members of a set (a 1-ary relation), when part of a logic-system. They are, usually, considered as a list of the members of this set.   In general, a logic-system, when considered as an operator,  is defined on subsets of the language employed. \pars

From the definitions employed for the logic-systems used here, the properties of the logic-system algorithm ${\cal A}$ 
can be explicitly described in set-theoretic notation. For these applications, $\cal A$ is a function defined on various defined logic-systems and a set of hypotheses. For example, the entire set of deductions or the order in which the deductions are made, among a few other characteristics. In our application to a logic-system, the notation used signifies all of the ``deduced'' results the algorithm produces when the logic-system is applied to a set of hypotheses. This yields the same results as a corresponding finite consequence operator. What the notation indicates is that the finite consequence operator is being displayed in a more refined and explicit manner. Hence, the algorithm and its relation to the logic-system can be embedded into the formal structure via formalizable characteristics.\pars

When the application characteristics are *-transferred, then the notation $\hyper {\underline{\cal A}}$ is employed. The process of applying the algorithm to the logic-system ${\cal S}^q,$ that is applied it to a set of hypotheses Y, is denoted by ${\cal A}(({\cal S}^q, \r Y)).$ Hence, $\cal A$ is defined upon a set of ordered pairs. The result of ${\cal A}(({\cal S}^q,\r Y))$ is a set. An additional step can be included for this specific algorithm, where $\r Y$ is removed. When this is done the algorithm is denoted by ${\cal A}'$. The necessary informally and, hence, formally described properties are specifically displayed. In general, the $q$ notion is not included as part of the $\cal A$ notation unless confusion would result. \pars

For the denumerable set ${\cal L}^1$, notice that for any $\Lambda^1(k),\ k \in \nat$  there exists an $k' \in \nat$ and $\r X^1_{k'} \in {\r M}^1_{k'},$ such that $\Lambda^1(k) \in {\cal A}'(({\cal S}^1,\{\r X^1_{k'}\}))$ and, in this case, finite choice yields the $\Lambda^1(k)$ logic-system. Notice that the logic-system $\Lambda^1(k)$ is considered as a set-theoretic set. 
Then the logic-system algorithm $\cal A$ is applied to  $(\Lambda^1(k),\{\r f^1(0,0)\}),$ where $\r f^1(0,0)$ is the only hypothesis contained in the logic-system. This yields $\r f^1(i,j)\in {\cal I}_1$ as a deduction from $\r f^1(0,0)$. Conversely, if $\r f^1(i,j) \in {\cal I}_1,$ then there is an $\r X^1_{k'} \in {\r M}^1_{k'}$ and a logic-system $\Lambda(k) \in {\cal A}'({\cal S}^1,\{\r X^1_{k'}\})$ such that application of the logic-system algorithm $\cal A$ to  $(\Lambda^1(k),\{\r f^1(0,0)\})$ yields $\r f^1(i,j)$ as a deduction from $\r f^1(0,0)$.  \pars

The informal algorithm $\cal A$ is defined on any logic-system that contains an hypothesis and, in this paper, such a logic-system is $\Lambda^q (x)$ and application is on $(\Lambda^q(x), \r Y)$ where Y is an hypothesis contained in the logic-system and containing but one member. Due to the construction of the $\Lambda^q (x),$ this yields a partial sequence of members of ${\cal I}_q$. This sequence is denoted by ${\cal A}[(\Lambda^q,\r Y)].$ This sequence represents the steps in the deduction and satisfies the $\leq_{{\cal I}^q_x}$ order. Also, for this case, ${\cal A}((\Lambda^q(x), \r Y)) = {\cal I}^q_{x} \subset {\cal I}_q.$ Significantly, for $n,\ k \in \nat,\ n \leq k, {\cal A}((\Lambda^1(n), \r Y)) \subset {\cal A}((\Lambda^1(k), \r Y))$ and ${\cal A}[(\Lambda^1(k),\r Y)]|[1,n] = {\cal A}[(\Lambda^1(n),\r Y)].$ \pars

In the usual way, all of the above informally defined objects are embedded relative to  $\cal E.$ When the informal set-theoretic expresses are considered as embedded into the standard superstructure, all of the bold font conventions defined in Herrmann (1979-1993) are observed. All other embedded symbols retain their math-italics form. Where script notation is used, an underline is used in place of the bold face font. All the following results are relative to our nonstandard model $\hyper {\cal M} = \langle \Hyper {\b Q}, \in , = \rangle$ or $\hyper {\cal M} = \langle \Hyper {\real }, \in ,= \rangle$ (Herrmann, (1979 -  1993)).   \parm 

\noindent{\bf Theorem 2.1} {\it Consider primitive time interval $1 = [0,b],  b>0.$ It can always be assumed that interval 1 is partitioned into two or more intervals $[c_0,c_1),\ldots$ $ [c_{m-1}, c_m], \ c_m = b, \ m >1,\ m \in \b Z.$ Let $\underline{\cal I}_1$ be an instruction paradigm order isomorphic to the rational numbers $R_1 \subset [0,b].$ For any $\lambda \in \nat_\infty,$ there exists a unique hyperfinite $\hyper {\b \Lambda^1}(\lambda) \in \Hyper {\underline{\cal L}}^1$ and a $\lambda' \in \hypernat$ such that the ultra-word-like $X^1_{\lambda'} \in  \Hyper {\b M}^1_{\lambda'}$ and ultra-logic-system $\hyper {\b \Lambda^1}(\lambda) \in  
\hyper { {\underline{\cal A}}'} ((\Hyper {\underline{\cal S}^1},\{X^1_{\lambda'}\}))$ and $\sig{\underline{\cal I}_1} \subset 
\hyper { {\underline{\cal A}}}(({\hyper {\b \Lambda^1}}(\lambda),\{\Hyper {\b f}^1(0,0)\})) = {\cal I}^1_\lambda\subset \Hyper {\underline{\cal I}}_1$. 
 Also the $\hyper { {\underline{\cal A}}}[(\hyper{\b \Lambda^1}(\lambda),\{\Hyper {\b f}^1(0,0)\})]$ *steps satisfy the $\leq_{{\cal I}^1_\lambda}$ order  and  $(\Hyper {\underline{\cal I}_1} - \sig{\underline{\cal I}_1})\cap \hyper { {\underline{\cal A}}}((\hyper {\b \Lambda^1}(\lambda),\{\Hyper {\b f}^1(0,0)\}))= $ an infinite set.} 
\pars 
Proof. This follows in the same manner as Theorem 4.1 in Herrmann (2006) by *-transfer of the appropriate first-order statements that precede this theorem statement. Also note that since for every $n \in \nat'$, the $\Lambda(n)$ is finite, then, via the identification process, $\sig {\b \Lambda(n)} =\b \Lambda(n).$ It also follows that $\Hyper {\b \Lambda(n)} = \b \Lambda(n)$ under the customary conventions. Since for any $n,\ k \in \nat',\ n \leq k$, $ {\cal A}((\b \Lambda(n),\{{\b f}^1(0,0)\})) \subset  {\cal A}((\b \Lambda(k),\{ \b f^1(0,0)\})),$  from the  above and, via *-transfer, it follows that  $^\sig{\underline{\cal I}_1} \subset  \hyper {\underline{\cal A}}((\hyper {\b \Lambda}^1(\lambda).\{\Hyper {\b f}^1(0,0)\})) = {\cal I}^1_\lambda \subset \Hyper {\underline{\cal I}}_1.$  From the definition of $\Lambda^1(n),$ these steps numbers are order isomorphic the set of rational numbers $R_1.$ Hence, $\hyper {\underline{\cal A}}((\hyper {\b \Lambda}^1(\lambda),\{\Hyper {\b f}^1(0,0)\}))$ is *order isomorphic to a hyperfinite subset of $\Hyper {\b Q}.$  
Since there are infinitely many $i < \lambda$ and $i \in \nat_\infty$,  there are infinitely many $\Hyper {\b f}(i,j) \in \hyper {\underline{\cal A}}((\hyper{\b \Lambda^1}(\lambda),\{\Hyper {\b f}^1(0,0)\}))\subset \Hyper {\underline{\cal I}_1},$ where $ \Hyper {\b f}(i,j) \in \Hyper {\underline{\cal I}_1} - \sig{\underline{\cal I}_1}.$ These are interpreted as ultranatural events but in some cases may differ from physical events only in their primitive time identifications. This completes the proof. \qed\parm 
  
By considering the definition of ${\cal L}^1$, it follows that the given $1 < m \in \nat,\ \hyper {\b \Lambda^1}(\lambda)$ is precisely 
$\{\Hyper {\b f}^1(0,0)\} \cup \{\bigcup\{\Hyper{\b k}^1_i(\lambda)\mid 0 \leq i <m\}\} \cup \{\{\Hyper {\b f}^1(p-1,\lambda),\Hyper {\b f}^1(p,0)\}\mid (0< p\leq m) \land (p \in \Hyper {\b Z})\}.$ Of significance is the fact that the steps in the *-deduction $\hyper { {\underline{\cal A}}}((\hyper{\b \Lambda^1}(\lambda),\{\Hyper {\b f}^1(0,0)\}))$ preserve the order $\leq_{\Hyper {\underline{\cal I}}_1}$. Notice that $\hyper {\b \Lambda^1}(\lambda)$ is obtained by hyperfinite choice.  Further, any $\Hyper {\b f}^1(i,j) \in \{\Hyper {\b f}^1(x,y)\mid (0\leq x <m)\land(0\leq y \leq \lambda)\land(x \in \Hyper {\b Z})\land( y \in \hypernat)\} \cup \{\Hyper {\b f}^1(m,0)\}$ is a hyperfinite *-deduction from $\b f^1(0,0) = \Hyper {\b f}^1(0,0).$ And, it also follows that the set of all such *deductions yields a hyperfinite set ${\cal I}^1_\lambda$ such that $\sig{\underline{\cal I}_1} \subset {\cal I}^1_\lambda \subset \Hyper {\underline{\cal I}_1}.$\pars

\noindent{\bf 3. Logic-System Generation for the Type-2 Interval}\parm

For the type-2 case $[0,+\infty)$, a denumerable instruction paradigm displays a refined form. 
For this case, ${\cal I}_2 = \{\r f^2(i,j)\mid (0 \leq i )\land(i \in \b Z)\land (j \in \nat).$  Using ${\cal I}_2,$ consider the following logic-system.\parm\vfil\eject

{\bf Definition 3.1} Let $0 \leq i \in \b Z.$ For each $n \in \nat,$ let $\r k^2_i(n) = \{\{\r f^2(i,j), \r f^2(i,j+1)\}\mid (0\leq j \leq n-1)\land (j\in \nat)\}.$ For $0< m \in \b Z,$ let $\r K^2(m,n) =\bigcup \{\r k^2_i(n)\mid (0\leq i < m)\land(i \in \b Z)\}.$ Finally, let $\Lambda^2(m,n) = \{\r f^2(0,0)\} \cup \r K^2(m,n) \cup \{\{\r f^2(p-1,n),\r f^2(p,0)\}\mid (0< p\leq m)\land (p \in \b Z)\}\cup \{\{\r f^2(m,j), \r f^2(m,j+1)\}\mid (0\leq j < n)\land (j \in \nat)\},$ and ${\cal L}^2 = \{\Lambda^2(x,y)\mid (0 \leq x \in \b Z)\land (y \in \nat)\}.$ Notice that if $0 \leq i < k, \ i,\ k \in \b Z$, then ${\cal A}((\Lambda^2(i,j), \{\r f^2(0,0)\})) \subset {\cal A}((\Lambda^2(k,n),\{\r f^2(0,0)\}))$ for any $j,\ n \in \nat.$ 
Also, each $\Lambda^2(m,n)$ is a finite set. (Notice that members in ${\cal L}^2$ can be characterized by a first-order sentence.)\parm

Consider any $\Lambda^2(q,k).$ Then there exists an $q'k'\ \in \nat'$ ($q'k'$ is a natural number in $\nat'$) and the 
$q'k'$-set $\r X^2_{q'k'} \in \r M^2_{q'k'},$ such that $\Lambda^2(q,k) \in {\cal S}^2(\{\r X^2_{q'k'}\})$ and, in this case, finite choice yields the $\Lambda^2(q,k)$ logic-system. Then the logic-system algorithm $\cal A$ applied to $(\Lambda^2(q,k),\{\r f^2(0,0)\})$ yields $\r f^2(q,k)$ as a deduction from $\r f^2(0,0).$ Further, $\r f^2(q,k) \in {\cal I}_2.$ Conversely, if $\r f^2(q,k) \in {\cal I}_2,$ then there exists an $q'k' \in \nat'$ and an $\r X^2_{q'k'} \in {\r M}^2_{q'k'}$ and a logic-system $\Lambda(q,k) \in { {\cal A}'}(({\cal S}^2,\{\r X^2_{q'k'}\}))$ such that application of the logic-system algorithm $\cal A$ to  $(\Lambda^2(q,k),\{\r f^2(0,0)\})$ yields a deduction of $\r f^2(q,k)$ from $\r f^2(0,0)$. \parm

\noindent{\bf Theorem 3.1} {\it Consider primitive time interval $2 = [0,+\infty).$ It can always be assumed that interval 2 is partitioned  into intervals $[c_0,c_1),\ldots [c_{m-1}, c_m), \ m >1,\ m \in \b Z.$ Let $\b d_2$ be an instruction paradigm order isomorphic to the rational numbers $R_2 \subset [0,+\infty).$ For any $\lambda \in \nat_\infty$ and $\nu \in 
\Hyper {\b Z} - \b Z, \ \nu >0,$ there exists a unique hyperfinite $\hyper {\b \Lambda^2}(\nu,\lambda) \in \Hyper {\underline{\cal L}^2}$ and $\nu', \lambda'  \in \hypernat$ such that the ultra-word-like $X^2_{\nu'\lambda'} \in \Hyper {\b M}^2_{\nu'\lambda'}$ and ultra-logic-system $\hyper {\b \Lambda^2}(\nu,\lambda) \in 
\hyper { {\underline{\cal A}}'}((\Hyper {{\cal S}^2,\{X^2_{\nu'\lambda'}\}}))$ and $\sig{\underline{\cal I}_2} \subset \hyper { {\underline{\cal A}}}((\hyper {\b \Lambda^2}(\nu,\lambda),\{\Hyper {\b f}^2(0,0)\})) = 
{\cal I}^2_{\nu\lambda}\subset \Hyper {\underline{\cal I}}_2$.
Also the $\hyper { {\underline{\cal A}}}[(\hyper {\b \Lambda^2}(\nu,\lambda),\{\Hyper {\b f}^2(0,0)\}))]$ *steps satisfy the $\leq_{{\cal I}^2_{\nu\lambda}}$ order  and  $(\Hyper {\underline{\cal I}_2} -  \sig{\underline{\cal I}_2})\cap \hyper { {\underline{\cal A}}}(\hyper {\b \Lambda^2}(\nu,\lambda),\{\Hyper {\b f}^2(0,0)\})) = $ an infinite set.} 
\pars
Proof. As in Theorem 2.1, the proof follows by *-transfer of the appropriate formally presented material that appears above in this section 3.\parm

By considering the definition of ${\cal L}^2$, it follows that the $\hyper {\b \Lambda^2}(\nu,\lambda)$ is precisely 
$ \{\Hyper {\b f}^2(0,0)\} \cup \{\bigcup\{\Hyper {\b k}^2_i(\lambda)\mid 0 \leq i< \nu\}\} \cup \{(\{\Hyper {\b f}^2(p-1,\lambda),\Hyper {\b f}^2(p,0)\}\mid (0 < p\leq\nu)\land (p \in \Hyper {\b Z})\} \cup \{\{\Hyper {\b f}^2(\nu,j),\Hyper {\b f}^2(\nu,j+1)\}\mid (0\leq j < \lambda)\land(j \in \hypernat)\}$. Of significance is the fact that the steps in the *-deduction $\hyper { {\underline{\cal A}}}[\hyper {\b \Lambda^2}(\nu,\lambda),\{\Hyper {\b f}^2(0,0)\}))]$ preserve the order $\leq_{\Hyper {\underline{\cal I}}_2}$.  Notice that $\hyper {\b \Lambda^2}(\nu,\lambda)$ is obtained by hyperfinite choice. Further, any $\Hyper {\b f}^2(i,j) \in \{\Hyper {\b f}^2(x,y)\mid (0\leq x \leq \nu)\land(0\leq y \leq \lambda)\land(x \in \Hyper {\b Z})\land (y \in \hypernat)\}$ is a hyperfinite *-deduction from ${\b f}^2(0,0).$ And, it also follows that the set of all such *deductions yield a hyperfinite set ${\cal I}^2_{\nu\lambda}$ such that $\sig{\underline{\cal I}_2} \subset {\cal I}^2_{\nu\lambda} \subset \Hyper {\underline{\cal I}_2}.$\parm

\noindent{\bf 4. Logic-System Generation for the Type-3 Interval}\parm

For the type-3 case $(-\infty,0]$, a denumerable instruction paradigm displays a refined form. 
For this case, ${\cal I}_3 = \{\r f^3(i,j)\mid (i\leq 0 )\land(i \in \b Z)\land (j \in \nat).$  Using ${\cal I}_3,$ consider the following logic-system.\parm

{\bf Definition 4.1} Let $i \in \b Z,\ i\leq 0.$ For each $n \in \nat,$ let $\r k^3_i(n) = \{\{\r f^2(i,j), \r f^1(i,j+1)\}\mid (0\leq j \leq n-1)\land (j\in \nat)\}.$ For $m \in \b Z\, m<0,$ let $\r K^3(m,n) =\bigcup \{\r k^3_i(n)\mid (m \leq i < 0)\land(i \in \b Z)\}.$ Finally, let $\Lambda^3(m,n) = \{\r f^3(m,0)\} \cup \r K^3(m,n) \cup \{\{ \r f^3(p-1,n),\r f^3(p,0)\}\mid (m < p \leq 0)\land (p \in \b Z)\},$ and ${\cal L}^3 =\{\Lambda^2(x,y)\mid (0 \leq x \in \b Z)\land (y \in \nat)\}.$ Notice that if $ i < k \leq 0, \ i,\ k \in \b Z$, then ${\cal A}((\Lambda^3(i,j)),\{\r f^3(m,0)\})) \subset {\cal A}((\Lambda^3(k,n), \{\r F^3(m,0)\}))$ for any $j,\ n \in \nat.$ 
 Also, each $\Lambda^3(m,n)$ is a finite set. (Notice that members in ${\cal L}^3$ can be characterized by a first-order sentence.)\parm

Consider any $\Lambda^3(q,k).$ Then there exists an $q'k' \in \nat$ and $\r X^3_{q'k'} \in \r M^3_{q'k'},$ such that $\Lambda^3(q,k) \in {\cal A}'(({\cal S}^3,\{\r X^3_{q'k'}\}))$ and, in this case, finite choice yields the $\Lambda^3(q,k)$ logic-system. Then the logic-system algorithm $\cal A$ applied to $(\Lambda^3(q,k),\{\r f^3(q,0)\}))$ yields $\r f^3(q,k)$ as a deduction from $\r f^3(q,0).$ Further, $\r f^3(q,k) \in {\cal I}_3.$ Conversely, if $\r f^3(q,k) \in \r {\cal I}_3,$ then there is an $\r X^3_{q'k'} \in \r {\cal M}^3_{q'k'}$ and a logic-system $\Lambda(q,k) \in \r S^3(\{\r X^3_{q'k'}\})$ such that application of the logic-system algorithm $\cal A$ to  $(\Lambda^3(q,k),\{\Hyper {\b f}^3(q,0)\}))$ yields $\r f^3(q,k)$ as a deduction from $\r f^3(q,0)$. \parm

\noindent{\bf Theorem 4.1} {\it Consider primitive time interval $3 = (-\infty, 0].$ It can always be assumed that interval 3 is partitioned  into intervals $ \ldots, [c_{-2},c_{-1}), [c_{-1},c_0]$. Let $\b d_3$ be an instruction  paradigm order isomorphic to the rational numbers $R_3 \subset (-\infty,0].$ 
For any $\lambda \in \nat_\infty,\ \mu \in \Hyper {\b Z} - \b Z,\ \mu < 0,$
 there exists a unique hyperfinite $\hyper {\b \Lambda^3}(\mu,\lambda) \in \Hyper {\underline{\cal L}^3}$ and $\mu', \lambda'  \in \hypernat$ such that the ultra-word-like $X^3_{\mu'\lambda'} \in \Hyper {\b M}^3_{\mu'\lambda'}$ and ultra-logic-system 
$\hyper {\b \Lambda^3}(\mu,\lambda) \in 
\hyper { {\underline{\cal A}}'}((\Hyper {{\cal S}^3,\{X^3_{\mu'\lambda'}\}}))$ and $\sig{\underline{\cal I}_3} \subset \hyper { {\underline{\cal A}}}(\hyper {\b \Lambda^3}(\mu,\lambda),\{\Hyper {\b f}^3(\mu,0)\})) = {\cal I}^3_{\mu\lambda}\subset \Hyper {\underline{\cal I}}_3$.
Also the $\hyper { {\underline{\cal A}}}[(\hyper {\Lambda^3}(\mu,\lambda),\{\Hyper {\b f}^3(\mu,0)\}))]$ *steps 
satisfy the $\leq_{{\cal I}^3_{\mu\lambda}}$ order  and  $(\Hyper {\underline{\cal I}_3} - 
^{\underline{\cal I}_3})\cap \hyper { {\underline{\cal A}}}((\hyper {\b \Lambda^3}(\mu,\lambda)) ,\{\Hyper {\b f}^3(\mu,0)\}))= $ an infinite set.} 
\pars

Proof. As in Theorem 3.1, the proof follows by *-transfer of the appropriate formally presented material that appears above in this section 3.\parm

By considering the definition of ${\cal L}^3$, it follows that the $\hyper {\b \Lambda^3}(\mu,\lambda)$ is precisely 
$ \{\Hyper {\b f}^3(\mu,0)\} \cup \{\bigcup\{\Hyper {\b k}^3_i(\lambda)\mid \mu \leq i < 0\}\} \cup \{\{\Hyper {\b f}^3(p-1,\lambda),\Hyper {\b f}^3(p,0)\}\mid (\mu < p \leq 0)\land (p \in \Hyper {\b Z})\} $. Of significance is the fact that the steps in the *-deduction $\hyper {\underline{\cal A}}[(\hyper {\b \Lambda^3}(\mu,\lambda), \{\Hyper {\b f}^3(\mu,0)\}]$ preserve the order $\leq_{\Hyper {\underline{\cal I}}_3}$. Notice that $\hyper {\b \Lambda^3}(\mu,\lambda)$ is obtained by hyperfinite choice.  Further, any $\Hyper {\b f}^3(i,j) \in \{\Hyper {\b f}^3(x,y)\mid (\mu \leq x < 0)\land(0\leq y \leq \lambda)\} \cup \{{\b f}^3(0,0\}$ is a hyperfinite *-deduction from $\b f^3(\mu,0).$ And, it also follows that the set of all such *deductions is a hyperfinite set ${\cal I}^3_{\nu\lambda}$ such that $\sig{\underline{\cal I}_3} \subset {\cal I}^3_{\nu\lambda} \subset \Hyper {\underline{\cal I}_3}.$\parm

\noindent{\bf 5. Logic-System Generation for the Type-4 Interval}\parm

\noindent{\bf Theorem 5.1} {\it Consider primitive time interval $4 = (-\infty, +\infty).$ It can always be assumed that interval 4 is partitioned  into intervals $ \ldots, [c_{-2},c_{-1}), [c_{-1},c_0), \ldots$. Let $\b d_4$ be a instruction paradigm order isomorphic to the rational numbers $R_4 \subset (-\infty,+\infty).$ For any $\lambda \in \nat_\infty,\ \nu, \gamma \in \Hyper {\b Z} - \b Z,$ such that $\nu \leq 0, \ \gamma\geq 0,$ 
 there exists a unique hyperfinite $\hyper {\b \Lambda^4}(\nu,\gamma,\lambda) \in \Hyper {\underline{\cal L}^4}$ and $\nu', \gamma', \lambda'  \in \hypernat$ such that the ultra-word-like $X^4_{\nu'\gamma'\lambda'} \in \Hyper {\b M}^4_{\nu'\gamma'\lambda'}$ and ultra-logic-system $\hyper {\b \Lambda^4}(\nu,\gamma,\lambda) \in 
\hyper { {\underline{\cal A}}'}((\Hyper {{\cal S}^4,\{X^4_{\nu'\gamma'\lambda'}\}}))$ and $\sig{\underline{\cal I}_4} \subset \hyper { {\underline{\cal A}}}((\hyper {\b \Lambda^4}(\nu,\gamma,\lambda),\{\Hyper {\b f}^4(\nu,0)\})) = {\cal I}^4_{\nu\gamma\lambda}\subset \Hyper {\underline{\cal I}}_4$. Also the $\hyper {\underline{\cal A}}[(\hyper {\b \Lambda^4}(\nu,\gamma,\lambda),\{\Hyper {\b f}^4(\nu,0)\})]$ *steps  
satisfy the $\leq_{{\cal I}^4_{\nu\gamma\lambda}}$ order  and $(\Hyper {\underline{\cal I}_4} - \sig{\underline{\cal I}_4})\cap \hyper {\underline{\cal A}}((\hyper {\b \Lambda^4}(\nu,\gamma,\lambda),\{\Hyper {\b f}^4(\nu,0)\}))= $ an infinite set.} \parm

By considering the definition of ${\cal L}^4$, it follows that the $\hyper {\b \Lambda^4}(\nu,\gamma,\lambda)$ is precisely 
$ \{\Hyper {\b f}^4(\nu,0)\} \cup  \{\bigcup\{\Hyper {\b k}^4_i(\lambda)\mid (\nu \leq i < \gamma) \land (i \in \Hyper {\b Z})\}\} \cup  \{\{\Hyper {\b f}^4(p-1,\lambda),\Hyper {\b f}^4(p,0)\}\mid (\nu<p\leq \gamma)\land(p \in \Hyper {\b Z})\} \cup\{\{\Hyper {\b f}^4(\gamma,j),\Hyper {\b f}^4(\gamma,j+1)\}\mid (0\leq j < \lambda)\land (j \in \hypernat)\}.$ Of significance is the fact that the steps in the *-deduction $\hyper {\underline{\cal A}}[(\hyper {\b \Lambda^4}(\nu,\gamma,\lambda),\{\Hyper {\b f}^4(\nu,0)\})]$ preserve the order $\leq_{\Hyper {\underline{\cal I}}_4}$. Notice that $\hyper {\b \Lambda^4}(\nu,\gamma,\lambda)$ is obtained by hyperfinite choice.  Further, any $\Hyper {\b f}^4(i,j) \in \{\Hyper {\b f}^4(x,y)\mid (\nu \leq x \leq \gamma)\land(0\leq y \leq \lambda)\}$ is a hyperfinite *-deduction from $\b f^4(\nu,0).$ And, it also follows that the set of all such *deductions is a hyperfinite set 
${\cal I}^4_{\nu\gamma\lambda}$ such that $\sig{\underline{\cal I}_4} \subset {\cal I}^4_{\nu\gamma\lambda} \subset \Hyper {\underline{\cal I}_4}.$\parm

\noindent{\bf 6. Necessary refinements.}\parm

For the GGU-model, a universe is a nonempty collection of empty-systems, physical-systems, physical-like systems or other-systems. In general, an infinite hyperfinite set $X$ of internal sets, disjoint from $\bigcup \{\hyper {\cal F}^i({\cal C})\mid (1\leq i \leq n)\land (i \in \nat)\},\ n > 0$, is adjoined when info-fields are employed. Let ${\cal K} = \power{ \bigcup \{\hyper {\cal F}^i({\cal C})\mid (1\leq i \leq n)\land (i \in \nat)\} \cup X}.$ \pars

 It is now necessary that a more refined definition for each $\b f^q(i,j)$, which yields each $\Hyper {\b f}^q(i.j),$ be given. The notion of the ``non-operative'' instruction is used. Using the alphabet symbol $\r y\r X\in {\cal W}'$, consider the meaningless ``word'' $\r y\r X$. If this word is considered an instruction, then it has no operative content and $\bf yX$ yields neither properton combinations of any form nor any entities that require the adjoined set $X$. Its application yields an empty-system. This word is introduced in order to simplify the following refinement. \pars

Although thus far $\Hyper {\b f}^q(i,j)$ has been considered as an *instruction-set and any further refinements as to how it is constructed were unnecessary, this is no longer the case. None of the previous results are altered by this refinement. All members of $\Hyper {\cal E}'$ being considered in this section are *instructions. Let $\b Z_q,\  q = 1,2,3,4,$ be as  employed to define the ${\cal I}_q.$  Consider, as with $q$ and defined in the same manner, $\b Z_r \subset \b Z,$ where $
r = 1,2,3,4.$ Notationally, let $\b T = \b Z_r \times \nat.$ Both $\b Z_q \times \nat$ and $\b T$ carry the simple lexicographic 
order $\preceq$ (Herrmann, 2006). \pars

In what follows, given a particular $q$, then $r$ is fixed for all universe-wide frozen-frames. There exists a function $\r g^{(q,r)}\colon \b Z_q \times\nat \to ({\cal F}'({\cal W}'))^{\b T}$ with the following properties. For $(i,j) \in \b Z_r \times \nat,$ $\r g^{(q,r)}(i,j) = \r v.$ Then for each $(k,s) \in \b Z_r \times \nat$, $\r v(k,s)$ is an appropriate instruction-set. Notationally, the instruction-set is $(\r g^{(q,r)}(i,j))(k,s) = \r v(k,s) = \r g^{(q,r)}(i,j;k,s).$ \pars

Using this notation, for a fixed $(i,j)$, the $\r g^{(q,r)}(i,j;k,s)$ has the same type of primitive time ordering (lexicographic) as does the universe-wide frozen-frames.  For each $(i,j,k) \in \b Z_q \times \nat \times \b Z_r,$ let $\r h^{(q,r)}(i,j;k) =\bigcup \{ \r g^{(q,r)}(i,j;k,s)\mid s \in \nat\}$ and, for each $(i,j) \in \b Z_q \times \nat$, let the instruction-set $\r f^q(i,j) = \bigcup \{\r h^{(q,r)}(i,j;k)\mid k \in \b Z_r\}.$ By embedding and *-transfer, $\Hyper (\b g^{(q,r)}(i,j))(k,s) = \hyper {\b v}(k,s) = \hyper {\b g}^{(q,r)}(i,j;k,s).$ 
Then, for each $(i,j,k) \in \Hyper {\b Z}_q \times \hypernat \times \Hyper {\b Z_r},$ $\Hyper {\b h}^{(q,r)}(i,j;k) =\bigcup \{ \Hyper {\b g}^{(q,r)}(i,j;k,s)\mid s \in \hypernat\}$ and the *instruction-set $ \Hyper {\b f}^{(q,r)}(i,j) = \bigcup \{\Hyper {\b h}^{(q,r)}(i,j,k)\mid k \in \Hyper {\b Z_r}\}.$ The $\hyper {\preceq}$ *ordering is also satisfied. \pars

With respect to the original method used to obtained words in ${\cal W}'$,  note that from the definitions and when considered as restrictions, for fixed $(i,j,k),$ each $\r g^{(q,r)}(i,j;k,r)$ is an instruction-set, each $\r h^{(q,r)}(i,j;k)$ is an instruction-set (of instruction-sets) as is each $\r f^{(q,r)}(i,j)$. As previously mentioned, for fixed $(i,j),\ (k,s)$ there is a single word $\r W^{(q,r)}_{(i,j;k,s)}$ in ${\cal W}'$  that corresponds to  $\r g^{(q,r)}(i,j;k,s).$ Then for each $k$ there is a word $\r W^{(q,r)}_{(i,j;k)} \in {\cal W}'$  corresponding to $\r h^{(q,r)}(i,j;k)$ that is formed by ordering intuitively from right-to-left, via the word theory join operator, the  individual symbol-strings, diagrams, images or sensory information determined by each word $\r g^{(q,r)}(i,j;k,s)$ as $s$ varies via the linear portion of the lexicographic order. This gives a word in ${\cal W}'$ composed of the individual words. This word yields in a written or thought-order (left-to-right or right-to-left depending upon the language) set of instructions for each of the finite $k$ systems. However, the speaking-order is fixed. \pars 

In like manner for fixed $(i,j)$, there is a member of ${\cal W}'$ that yields, in written or thought word-order form, via the lexicographic order, a single word $\r W^{(q,r)}_{(i,j)} \in {\cal W}'.$ Then there is a single word $\r W^{(q,r)}(n)$ that yields in written or thought word-order form, again via the lexicographic order, a word that corresponds to a complete development for a finitely generated universe. But, notice that the speaking-order or corresponding thought-.order is fixed. These results are immediately extended to the hyperfinite cases. Then the finite cases for the $\r W^{(q,r)}(n)$ immediately yield the existence of an ultraword $W^{(q,r)}(\lambda)$ with the same first-order properties as $\r W^{(q,r)}(n)$. This can be considered as a higher-form of an hyper-ordered "spoken" or "thought" word. This has theological applications.\pars

For each $(i,j) \in \Hyper {\b Z}_q \times \hypernat$ and $k \in \Hyper {\b Z}_r,$ $\Hyper {\b h}^{(q,r)}(i,j;k)$ yields an empty-system, physical-system, physical-like system or an other-system. (Note that physical-like properties include physical-like behavior relative to non-physical entities, where the entities have no other known properties.)  Of course, each $\Hyper {\b f}^{(q,r)}(i,j)$ is an *instruction-set as is each $\Hyper {\b h}^{(q,r)}(i,j;k)$. By application of the word $\bf yX$ and the GGU-model construction of each universe-wide frozen frame, there are only finitely or hyperfinitely many physical or physical-like systems. Further more, various physical-like systems can be physical-systems in that only physical events exist. Which physical-like systems are but physical depends upon choice since the GGU-model is not dependent upon what science classifies as ``physical.''  Such a choice is dependent upon a chosen interpretation. The same holds for what are considered as empty-systems or other-systems. \pars

For each $(i,j) \in \Hyper {\b Z}_q \times \hypernat$ that yields a universe-wise frozen-frame, $\Hyper {\b g}^{(q,r)}(i,j;k,s)$ produces an info-field in the same manner as an entire universe is considered associated with a *developmental paradigm. These info-fields can also be considered in system form via the $\Hyper {\b h}^{q,r)}(i,j;k)$ *instruction-sets. If empty-systems are used, only a type-r = 4 ultra-logic-system need be considered. In this case, this yields for each such $(i,j),$ and fixed $\{\mu, \gamma, \lambda\} \subset \hypernat$, a generating ultra-logic-system is 
$$ \Hyper {\b F}^{(q,4)}(i,j) = \{\Hyper {\b g}^{(q,4)}(i,j;\nu,0)\} \cup  \{\bigcup\{\Hyper {\b k}^{(q,4)}_x(\lambda)\mid (\nu \leq x< \gamma) \land (x \in \Hyper {\b Z})\}\} \cup$$ 
$$\{\{\Hyper {\b g}^{(q,4)}(i,j;k-1,\lambda),\Hyper {\b g}^{(q,4)}(i,j;k,0)\}\mid (\nu<k\leq \gamma)\land(k \in \Hyper {\b Z_4})\} \cup$$ $$\{\{\Hyper {\b g}^{(q,4)}(i,j;\gamma,s),\Hyper {\b g}^{(q,4)}(i,j;\gamma,s+1)\}\mid (0\leq s < \lambda)\land (s \in \hypernat)\}.$$\pars

For this refined approach, the algorithm $\cal A$ has an extended definition. It is applied to these special logic-systems where the members are themselves logic-systems. This is how modified ${\cal A}'$ is applied. As each logic-system is obtained in the indicated ordered manner, $\cal A$ is applied to it. This is what would occur in the standard definition for application of $\cal A$ to a collection of logic-systems in n-tuple form except that the logic-systems are obtained deductively in a specific order and then as each is deduced the deduction is completed for the deduced logic-system.  This is a rather natural way one would preceed. Under this extended definition for $\cal A$ all of the previous theorems in sections 2, 3, 4, 5 hold where F [resp. $\b F$, $\Hyper {\b F}$] replaces f [resp. $\b f$, $\Hyper {\b f}$]. \pars

Let $D_i \subset \real$  be a countable set of non-zero numerical or coded values for a particular ultra-properton coordinate $i \in K$. Let $D = \bigcup \{D_i\mid i \in K\}.$ The set $G(0)$ is the set of all finite numbers. For each $r \in D,$ let $\Lambda_p = \{ x \mid (x \in G(0))\land (\st {x/10^\omega} = p)\}.$ The sets $\Lambda_p$ are disjoint and by choice consider distinct $\lambda_p \in \Lambda_p$ for each $p \in D.$ For any $p \in D,$ a set $P_p$ of ultra-propertons is $\lambda_p$-finite, if $P_p$ is hyperfinite and there exists a bijection from $[1,\lambda_p]$ onto $P_p$. From the definition of the set of all ultra-propertons, for each $i',j'  \in [1,K],$ and each $p,y \in D,$ there exist a $\lambda_p$-finite $x(p,i')  \in \hyper {\cal F}({\cal C})$ and a $\lambda_y$-finite $x(y,j') \in \hyper {\cal F}({\cal C})$ such that if $i \not = j$, then $x(p,i') \cap x(p,j') = \emptyset$. Of course, there are distinct $\lambda_y$-finite sets that yield physical or physical-like events. \pars  
  
{\leftskip=0.5in \rightskip=0.5in \noindent All of the previous results hold for the case of a corresponding developmental paradigm by substituting corresponding developmental paradigm $\r f^{(q,r)},\ \r g^{(q,r)}(i,j;k,s)$, and the like for an instruction paradigm notation.\par}\pars

There is a set-function 
$$G_z^{(q,r)} \colon \Hyper {\cal P}(\{ \Hyper {\b g}^{(q,r)}(i,j;k,s)\mid (\alpha \leq i \leq \beta)\land(i\in \Hyper {\b Z}_q)\land (j \in \hypernat)\land$$ $$(k \in \Hyper {\b Z}_r) \land (s \in \hypernat)\}) \to \Pi^+$$
where $\alpha, \beta$ depend upon $q.$ 
 For this approach, one considers $\Hyper {\cal P}(\{ \Hyper {\b g}^{(q,r)}(i,j;k,s)\mid (\alpha \leq i \leq \beta)\land(i\in \Hyper {\b Z}_q)\land (j \in \hypernat)\land(k \in \Hyper {\b Z}_r) \land (s \in \hypernat)\}) \subset \Hyper {\cal P}(\Hyper {\cal E}').$ Of course, an image can be the empty set.\pars

The process of applying the standard part operator does not erase the 
original coordinates of the constructed objects. The process only yields the 
physical-world objects. Such alterations of the fundamental subatomic or field 
entities is  modeled by
vector space subtraction and  this would be followed by 
independent *-finite subtraction. This represents a ``breaking apart'' of 
the fundamental entity into its original ultra-properton constituents. 
The original identifiers are restored. An altered constituent is then obtained by 
repeating the construction process. This can be used to eliminate the virtual 
particle or process concept within reality models. The models that need such concepts 
to predict 
behavior are now just that  ``models'' for predicted behavior and would not
require the virtual ``stuff'' to exist in reality.\pars

Of course, for each $(i,j)$, there is the important $G_z^{(q,r)}[\{\Hyper {\b h}^{(q,r)}(i,j;k))\mid k \in \b Z_r\}]$ that yields  empty-systems, physical-systems, physical-like systems or other-systems. The set $G^{q,r)}_z[\{\Hyper {\b h}^{(q,r)}(i,j;k)\mid k \in \b Z_r\}]$ is an info-field. The $G^{(q,r)}_z$ represents a substratum medium and processes that yield this action. \pars

Each member of the domain of $G_z^{(q,r)}$ can be replaced  by a single member of $\Hyper {\cal E}'$ since any nonempty finite subset set of ${\cal W}'$ can be replaced with a string of symbols connected by an $\&$ or a similar ``conjuction'' symbol. This new form, in this interpretation, has the same meaning as the set of finitely many members of ${\cal W}'$. This yields a single member of ${\cal E}'$ that can replace the finite instruction-set. Using strings of mathematical symbols that are members of $\Hyper {\cal E}'$ the same can be done for each member of the range of $G_z^{(q,r)}$. This yields a new binary relation ${\cal G}_z^{(q,r)} \subset \Hyper {\cal E}' \times \Hyper {\cal E}'$.  (Or by application of $\hyper {w}^{-1}$, $\Hyper {\cal W}' \times \Hyper {\cal W}'$\pars

 The binary relation ${\cal G}_z^{(q,r)}$ can be interpreted as a logic-system and as such the entire non-physical process can be interpreted as representing an intelligent design. Further, since the domain elements are so designed, then by implication so are the info-fields. \parm

\noindent{\bf 7. The Complete GGU-model Scheme}\parm

We assume that the previous results, not so generalized, have been generalized in order to accommodate  the above necessary refinement . For the $({\tt St}G^{(q,r)}_z)$ and the $'{\tt St}$ defined in Herrmann (2006), the following scheme is not in composition notational form due to one application of choice and the step-by-step application of $({\tt St}G^{(q.r)}_z).$ It represents an ordered application of the GGU-model operators. For $q = 1,2,3,4,$ the $a,b,c$ take the appropriate value for a specific $q.$

$$({\tt St}G^{(q,4)}_z)(\hyper {\underline{\cal A}}((\hyper {\b \Lambda^{(q,4)}}(a),\{\Hyper {\b F}^{(q,r)}(b,c)\})))(\hyper {\underline{\cal A}'}((\Hyper {{\cal S}^{(q,r)},\{X^{(q,r)}_{a'}\}}))).$$

The ${\b \Lambda^{(q,4)}}(a)$ is obtained, as previously, except that $\Hyper {\b f}^q(i,j)$ is replaced with $\Hyper {\b F}^{(q,4)}(i,j).$  The operators $\underline{\cal A}$ and $\underline{\cal A}'$ have characterizing first-order statements. These statements need not capture all of  of the intuitive statements that describe the algorithms. The results of application of $\underline{\cal A}'$ as formalized can show major aspects of the algorithm's selection process. For example, in terms of the corresponding $(q,r)$ definitions,
$$\forall x \forall y\forall z\forall w((w \in \underline{\cal M}^{(q,r)}\land(y \in {\cal F}(\underline{\cal L}^{(q,4)})) \land(y\in w)\land (x \in  \underline{\cal A}'((\underline{\cal S}^{(q,r)},y)) \to$$ 
$$\exists p((p \in \underline{\cal S}^{(q,r)})\land (y \in p)\land (x \in p)\land (y \not= x)).$$
\pars

For the instruction-information behavior (1 - 5), the function $G_z^{(q,r)}$ represents non-physical (2) and (3) and yields (4) the non-physical info-fields.  For a particular universe and for each corresponding $\lambda_{p}$-finite $x(p,i')$ there is a set of *instructions $I(p,i')$ such that $G^{(q,r)}_z((I(p,i')) = x(p,i').$ These collections are considered as ``bound.'' This binding is represented by application of $\lambda_p$-finite independent coordinate addition. The *instruction-sets that yield physical-systems contain various standard statements and these *instruction-sets also contain statements that refer to the non-physical. \pars

 The realization operator $\st{\cdot}$ is only applied to the $\lambda_{p} $-finite collections of ultra-propertons and, when applied, an entity represented by an intermediate properton is produced.  From the modeling viewpoint, processes within the substratum are non-physical. Thus in summary, relative to the GGU-model and for comprehension, steps (2) and (3) are consider as statements relative to a  non-physical medium and  a non-physical mode of transmission, respectively. For each universe-wide frozen-frame, step (4) corresponds to  a non-physical info-field. Step (5) is the application of the $\st {\cdot}$ operator and yields that physical products of the entire process. In order to be operative, the above displayed *instruction as well as similar ones have non-physical components.\pars

The latest refinements and alterations, if any, to the GGU-model can be found in the Herrmann (2014, 2013) references. \parm

\noindent{\bf 8. The Participator Universe.}\parm

(The following is not presented in the generalized refined form.) For the GGU-model, one of the most difficult requirements is to include the concept of the ``participator'' universe. As stated at the May 1974 Oxford Symposium in Quantum Gravity, Patton and Wheeler describe how existence of human beings alter the universe to various degrees. ``To that degree the future of the universe is changed. We change it. We have to cross out that old term `observed' and replace it with the new term `participator.' In some strange sense the quantum principle tells us that we are dealing with a participator universe." (Patton and Wheeler (1975, p. 562).) This aspect of the GGU-model is  only descriptively displayed in section 4.8 in Herrmann (2002). It is now possible to obtain formally the collection of pre-designed universes that satisfies this participator requirement. \pars

The previous notation is modified for finitely many ($> 0$) instruction paradigms as previously denoted by ${\cal I}_q.$ From the construction of each instruction paradigm using  ${\cal W}'$, it follows that there is, at least, a sequence of possible alterations. An instruction paradigm is a nonempty subset of the subsets ${\cal W}'$. Hence, the collection of all such instruction paradigms is a member of $\power{\power {{\cal W}'}}$. There can be infinitely many basic universes. These are universes prior to participator alterations. For each of these, there is an collection of ultra-word-like objects of the appropriate type. What follows next is for an arbitrary member of  this collection of ultra-world-like objects and, hence, an arbitrary basic universe. \pars

So as to include the type of universe being considered, let $q \colon  \nat' \times [1,4] \to \power{\power {{\cal W}'}}.$ Then for a specific $p \in [1,4]$ an expression $\{x \mid (x = \r q(n,p)) \land (n \in \nat')\} = {\cal I}_{p}$ represents this denumerable set of instruction paradigms for type-p universes. (If $n' \not= m ,$ then $q(n',p) \not= q(m,p).$)  Let  $\{x \mid \exists p (p \in [1,4] \land(x =  {\cal I}_{p})\} =   {\cal I}.$  Then, a  specific $\r f^1(0,0)$ is further identified relative to the sequences. As an example, $\r f^{\r q(3,1)}(0,0) \subset {\cal W}'$ represents a specific $n =3$ type-1  instruction-set. Thus, as embedded into the formal structure this last expression reads $\b f^{\b q(3,1)}(0,0) \subset {\cal E}'$.\pars

An original alteration can be miniscule and made in one or more of the necessary parameters that are satisfied by a specific cosmology. This can be done in such a way that only miniscule alterations in physical-system satisfy  the alterations. On the other hand, a highly altered cosmology can also occur. An alteration is local prior to it being propagated during a universe's development. Although not specifically included, and indeed the definition would need to be altered slightly, other sequences $\r q'$ can be used where various values can be empty or repeated. Also, for a universe with infinitely many local alterations at the same moment, one can include the obvious change in the $\r q$ sequence, where the sequence $q$ is a sequence of type-4 except that each image is a ``universe.'' \pars

For the GGU-model, the various members of ${\cal I}_p$ satisfy the ``participator" requirements, when participators exist, for each of the known suggested cosmologies. When embedded into the formal structure, properties of $q$ can be easily characterized, using various forms, in a first order language. For example, 
$$\forall x \forall p((x \in \nat') \land(p \in [1,4]) \to \exists y\exists z((z \in \underline{{\cal I}})\land (y \in z) \land (\b q(x,p) = y)).$$ Consider a specific $p' \in [1,4]$. Then 
$$\forall y((y \in \underline{{\cal I}}_{p'})) \to \exists x((x \in \nat') \land (\b q(x,p') =y)).$$ 
In *-transfer form, these two sentences read
$$\forall x \forall p((x \in \hypernat') \land(p \in [1,4]) \to \exists y\exists z((z \in \Hyper{\underline{{\cal I}}})\land (y \in z) \land (\hyper{\b q(x,p)} = y)).$$
$$\forall y((y \in \Hyper {\underline{{\cal I}}}_{p'})) \to \exists x((x \in \hypernat') \land (\hyper{\b q}(x,p') =y)).$$ 

The previous Theorems 2.1, 3.1, 4.1, 5.1 are all relative to a specific instruction paradigm and each holds for a collection of these instruction paradigms. Thus, the notion can be added and the additional statement that the results hold for each $n \in \hypernat'$ and each $p \in [1,4].$ The special processes noted in the scheme in section 6 are applied to each set of instruction paradigms.  \pars 

Each member of In Herrmann (2002), hyperfast propertons are mentioned as mediators for the automatic selection of a particlar member of $\Hyper {\cal I}_{p}.$ Certain not realized members of $\Hyper \{{\cal I}\}_p$ are termed as covirtual. Notice that from the transformed formal statements above, there exist a member of  $\Hyper {\underline{{\cal I}}}_{p'}$ for each $\gamma \in \hypernat' - \nat'.$ These can be used for various interpretations using either a $\r q,\ \r q',\ q$ type of sequence. Further, GGU-model predicted processes and entities can aid in comprehending the notion of the non-temporal and its relation to the temporal.  \pars

 An important refinement to this participator model can be found in Herrmann (2014).\parm

\centerline{\bf References}\par\bigskip 
\id{H}errmann, R. A., (2014), The Participator Model, 
http://vixra.org/abs/1410.0181 

\id{H}errmann, R. A., (2014a), An Alteration to the Foundations of the Theory of Ultralogics, 
http://vixra.org/abs/1406.0100

\id{H}errmann, R. A., (2013), General Logic-Systems that Determine Significant Collections of Consequence Operators, http://vixra.org/abs/1309.0013

\id{H}errmann, R. A., (2013a), GGU-model Ultra-Logic-Systems Applied to Developmental\hfil\break  Paradigms, http://vixra.org/abs/1309.0004 

\id{H}errmann, R. A., (2013b), The GGU-model and Generation of Developmental Paradigms, \hfil\break http://vixra.org/abs/1308.0145 

\id{H}errmann, R. A., (2013c), Nonstandard Ultralogic-Logic-Systems Applied to the GGU-model, \hfil\break
http://vixra.org/abs/1308.0125 

\id{H}errmann, R. A., (2006), The GGU-model and Generation of Developmental Paradigms, http://arxiv.org/abs/math/0605120 
\id{H}errmann, R. A., (2006a), General Logic-systems that Determine Significant Collections of Consequence Operators, \hfil\break http://arxiv.org/abs/math/0603573 

\id{H}errmann, R. A., (2003), The Theory of Infinitesimal Light-Clocks, \hfil \break http://arxiv.org/abs/math/0312189 
\id{H}errmann, R. A., (2002), Science Declares that Our Universe IS Intelligently Designed, Xulon Press, Longwood, FL. (and elsewhere)
\id{H}errmann, R. A., (1979 - 1993). The Theory of Ultralogics, \hfil\break  
http://raherrmann.com/books.htm 
http://arxiv.org/abs/math/9903082 http://arxiv.org/abs/math/9903081 
\id{H}errmann, R. A., (1989), Fractals and ultrasmooth microeffects, J. Math. Physics, 30(4), :805-808. (Note that there are typographical errors in this paper. In the proof of Theorem 4.1, in equations $h(x,c,d), \ G_j(x),$ the ) + )) should be )) + ). In $G_j(x),$ the second $c$ should be replaced with $a_j.$ On page 808, the second column, second paragraph, line six, $\st {D}$ should read $\st {\hyper D}$ and, trivially, $x \in \mu (p)$, should read $x \in \mu(p) \cap \hyper {D}.$ In the proof of Theorem 3.1, first line $\Hyper {\bf R}^m$ and should read ${\b R}^m.$
\id{H}errmann, R. A., (1983). Mathematical philosophy and developmental processes, Nature and System 5(1/2):17-36.
\id{L}oeb, P. A. and M. Wolff, (2000). Nonstandard Analysis for the Working Mathematician, Kluwer Academic Publishers, Boston.
\id{M}endelson, E., (1987), Introduction to Mathematical Logic, Wadsworth \& Brooks/Cole  Advanced Books \& Software, Monterey, CA. 
\id{P}aton, C. M. and J. A. Wheeler, (1975). Is physics legislated by a cosmogony? In {\it Quantum Gravity} ed. Isham, Penrose and Sciama, Clarendon Press, Oxford. 
\id{R}obinson, A. (1993). On languages which are based on non-standard arithmetic." Nagoya Math. J. 22:83-117. (See also
Geiser, J. A., (1968). Nonstandard Logic, J. Symbolic Logic, 33(2):236-250.)
\id{S}troyan, K. D. and J . M. Bayod, (1986). Foundations of Infinitesimal Stochastic Analysis, North Holland, N.Y.
\end